\documentclass[journal]{IEEEtran}
\usepackage{amsmath,amssymb,amsfonts}
\usepackage{amsmath}
\usepackage{glossaries}
\usepackage{algorithm}
\usepackage{algorithmicx}
\usepackage{array}
\usepackage[caption=false,font=normalsize,labelfont=sf,textfont=sf]{subfig}
\usepackage{textcomp}
\usepackage{stfloats}
\usepackage{url}
\usepackage{verbatim}
\usepackage{graphicx,color}
\usepackage{cite}
\usepackage{xcolor}
\usepackage{bm}
\usepackage{algpseudocode}
\usepackage{siunitx}
\usepackage{hyperref}
\usepackage{subfig}
\usepackage{booktabs}
\usepackage{cuted}

\algdef{SE}[DOWHILE]{Do}{doWhile}{\algorithmicdo}[1]{\algorithmicwhile\ #1}

\newcommand{\ed}{\color{black}} 

\newcommand\norm[1]{\left\lVert#1\right\rVert}

 \usepackage{scalerel}
 \usepackage{tikz}
\usetikzlibrary{svg.path}

\definecolor{IEEEblue}{RGB}{0 98 155}
\definecolor{IEEElightblue}{RGB}{0 181 226}
\definecolor{IEEEdarkblue}{RGB}{0 98 155}
\definecolor{IEEEturquoise}{RGB}{0 156 166}
\definecolor{IEEEred}{RGB}{186 12 47}
\definecolor{IEEEdarkred}{RGB}{134 32 65}
\definecolor{IEEEgreen}{RGB}{0 132 61}
\definecolor{IEEElightgreen}{RGB}{120 190 32}
\definecolor{IEEEorange}{RGB}{225 163 0}
\definecolor{IEEEdarkorange}{RGB}{232 119 34}
\definecolor{IEEEyellow}{RGB}{255 209 0}
\definecolor{IEEEslategray}{RGB}{112 128 144}

\definecolor{xcolorMahogany}{rgb}{0.6627 0.2039 0.1216}
\floatname{algorithm}{Alg.}

\usepackage{pgfplots}
\usetikzlibrary{pgfplots.groupplots} 
\usetikzlibrary{pgfplots.patchplots}
    \usepgfplotslibrary{colormaps}
    \pgfplotsset{
        compat=1.18,
    }\usepackage{textcomp}

\newlength{\plotWidth}		
\newlength{\plotHeight}		

\usepackage{etoolbox}
\usetikzlibrary{tikzmark}
\usetikzlibrary{calc}  
\newcommand{\ALGtikzmarkcolor}{black}
\newcommand{\ALGtikzmarkextraindent}{4pt}
\newcommand{\ALGtikzmarkverticaloffsetstart}{-.5ex}
\newcommand{\ALGtikzmarkverticaloffsetend}{-.5ex}
\makeatletter
\newcounter{ALG@tikzmark@tempcnta}

\newcommand\ALG@tikzmark@start{%
    \global\let\ALG@tikzmark@last\ALG@tikzmark@starttext%
    \expandafter\edef\csname ALG@tikzmark@\theALG@nested\endcsname{\theALG@tikzmark@tempcnta}%
    \tikzmark{ALG@tikzmark@start@\csname ALG@tikzmark@\theALG@nested\endcsname}%
    \addtocounter{ALG@tikzmark@tempcnta}{1}%
}

\def\ALG@tikzmark@starttext{start}
\newcommand\ALG@tikzmark@end{%
    \ifx\ALG@tikzmark@last\ALG@tikzmark@starttext
    \else
        \tikzmark{ALG@tikzmark@end@\csname ALG@tikzmark@\theALG@nested\endcsname}%
        \tikz[overlay,remember picture] \draw[\ALGtikzmarkcolor] let \p{S}=($(pic cs:ALG@tikzmark@start@\csname ALG@tikzmark@\theALG@nested\endcsname)+(\ALGtikzmarkextraindent,\ALGtikzmarkverticaloffsetstart)$), \p{E}=($(pic cs:ALG@tikzmark@end@\csname ALG@tikzmark@\theALG@nested\endcsname)+(\ALGtikzmarkextraindent,\ALGtikzmarkverticaloffsetend)$) in (\x{S},\y{S})--(\x{S},\y{E});%
    \fi
    \gdef\ALG@tikzmark@last{end}%
}

\apptocmd{\ALG@beginblock}{\ALG@tikzmark@start}{}{\errmessage{failed to patch}}
\pretocmd{\ALG@endblock}{\ALG@tikzmark@end}{}{\errmessage{failed to patch}}
\makeatother

\definecolor{orcidlogocol}{HTML}{A6CE39}
\tikzset{
  orcidlogo/.pic={
    \fill[orcidlogocol] svg{M256,128c0,70.7-57.3,128-128,128C57.3,256,0,198.7,0,128C0,57.3,57.3,0,128,0C198.7,0,256,57.3,256,128z};
    \fill[white] svg{M86.3,186.2H70.9V79.1h15.4v48.4V186.2z}
                 svg{M108.9,79.1h41.6c39.6,0,57,28.3,57,53.6c0,27.5-21.5,53.6-56.8,53.6h-41.8V79.1z M124.3,172.4h24.5c34.9,0,42.9-26.5,42.9-39.7c0-21.5-13.7-39.7-43.7-39.7h-23.7V172.4z}
                 svg{M88.7,56.8c0,5.5-4.5,10.1-10.1,10.1c-5.6,0-10.1-4.6-10.1-10.1c0-5.6,4.5-10.1,10.1-10.1C84.2,46.7,88.7,51.3,88.7,56.8z};
  }
}

\newcommand\orcidicon[1]{\href{https://orcid.org/#1}{\mbox{\scalerel*{
\begin{tikzpicture}[yscale=-1,transform shape]
\pic{orcidlogo};
\end{tikzpicture}
}{|}}}}

\newcommand{\RML}{\text{\tiny RML}}
\newcommand{\ML}{\text{\tiny ML}}
\newcommand{\LoS}{\text{\tiny LoS}}
\newcommand{\NLoS}{\text{\tiny NLoS}}

\DeclareMathOperator*{\argmin}{arg\,min}

\newcommand{\eqdef}{\stackrel{\textrm{\tiny def}}{=}}

\usepackage{titlesec}
\titlespacing*{\subsection}
{0pt}{10pt plus 2pt minus 2pt}{4pt plus 2pt minus 2pt}

\newacronym{AOA}{AOA}{Angle-of-Arrival}
\newacronym{AWGN}{AWGN}{Additive White Gaussian Noise}
\newacronym{CRLB}{CRLB}{Cramér-Rao Lower Bound}
\newacronym{CNN}{CNN}{Convolutional Neural Network}
\newacronym{DEB}{DEB}{Distance Error Bound}
\newacronym{DM-LED}{DM-LED}{Dual-Mode LED}
\newacronym{DOP}{DOP}{Dilution of Precision}
\newacronym{FIM}{FIM}{Fisher Information Matrix}
\newacronym{FoV}{FoV}{Field of View}
\newacronym{IID}{IID}{Indipendent and Identically Distributed}
\newacronym{ILS}{ILS}{Iterative Least Squares}
\newacronym{IPS}{IPS}{Indoor Positioning System}
\newacronym{IRS}{IRS}{Intelligent Reflective Surface}
\newacronym{IWLS}{IWLS}{Iterative Weighted Least Squares}
\newacronym{OIRS}{OIRS}{Optical Intelligent Reflective Surface}
\newacronym{LED}{LED}{Light Emitting Diode}
\newacronym{LLS}{LLS}{Linear Least Square}
\newacronym{LoS}{LoS}{Line-of-Sight}
\newacronym{ML}{ML}{Maximum Likelihood}
\newacronym{MCRB}{MCRB}{Misspecified Cramér–Rao Bound}
\newacronym{MLE}{MLE}{Maximum Likelihood Estimation}
\newacronym{MSE}{MSE}{Mean Squared Error}
\newacronym{NLoS}{NLoS}{Non-Line-of-Sight}
\newacronym{PD}{PD}{Photo Detector}
\newacronym{PDF}{PDF}{Probability Density Function}
\newacronym{PEB}{PEB}{Position Error Bound}
\newacronym{PSO}{PSO}{Particle Swarm Optimization}
\newacronym{QRX}{QRX}{Quadrant Receiver}
\newacronym{RIS}{RIS}{Reconfigurable Intelligent Surface}
\newacronym{RMSE}{RMSE}{Root Mean Squared Error}
\newacronym{RSS}{RSS}{Received Signal Strength}
\newacronym{RF}{RF}{Radio-Frequency}
\newacronym{SNR}{SNR}{Signal-to-Noise Ratio}
\newacronym{SPAO}{SPAO}{Simultaneous Positioning and Orientating}
\newacronym{TOA}{TOA}{Time-Of-Arrival}
\newacronym{TDOA}{TDOA}{Time-Difference-Of-Arrival}
\newacronym{ULA}{ULA}{Uniform Linear Array}
\newacronym{VLC}{VLC}{Visible Light Communication}
\newacronym{VLP}{VLP}{Visible Light Positioning}

\begin{document}

\title{Visible Light Indoor Positioning with a Single LED and Distributed Single-Element OIRS: An Iterative Approach with Adaptive Beam Steering}

\author{\IEEEauthorblockN{
Daniele Pugliese~\orcidicon{0000-0001-6436-2359},~\IEEEmembership{Student Member,~IEEE}, 
Giovanni Iacovelli
~\orcidicon{0000-0002-3551-4584}, \IEEEmembership{Member,~IEEE}, 
Alessio~Fascista~\orcidicon{0000-0001-6645-6391},~\IEEEmembership{Member,~IEEE}}, 
Domenico Striccoli
~\orcidicon{0000-0003-2904-6961}, 
Oleksandr Romanov
~\orcidicon{0000-0002-8683-3286}, \IEEEmembership{Member,~IEEE}, \\
Luigi Alfredo Grieco
~\orcidicon{0000-0002-3443-6924}, \IEEEmembership{Senior Member,~IEEE}, and~Gennaro~Boggia
~\orcidicon{0000-0002-0883-3045}, \IEEEmembership{Senior Member,~IEEE}
\thanks{D. Pugliese, A. Fascista, D. Striccoli, L.A. Grieco, and G. Boggia are with the Department of Electrical and Information Engineering (DEI), Politecnico di Bari, Italy (e-mail: name.surname@poliba.it) and with Consorzio Nazionale Interuniversitario per le Telecomunicazioni (CNIT). 

O. Romanov is with Institute of Telecommunication Systems of the National Technical University of Ukraine “Igor Sikorsky Kyiv Polytechnic Institute”, Ukraine (email: a\_i\_romanov@ukr.net). 

G. Iacovelli is with the Signal Processing and Communications (SIGCOM) Research Group at Interdisciplinary Centre for Security, Reliability and Trust (SnT), University of Luxembourg, (email: giovanni.iacovelli@uni.lu).

{\ed This work was supported by the EU under Italy NRRP (NextGenerationEU): RESTART (PE00000001, CUP: D93C22000910001), MOST (CN00000023, CUP: D93C22000410001), and SERICS (PE00000007, CUP: D33C22001300002, project ISP5G+). Additional support came from PRIN projects INSPIRE (2022BEXMXN 01) and HORUS (2022P44KA8) funded by MUR, and the HORIZON MSCA project BRIDGITISE (grant 101119554).}}
}

\markboth{\tiny This work has been submitted to the IEEE for possible publication. Copyright may be transferred without notice, after which this version may no longer be accessible.}
{Shell \MakeLowercase{\textit{et al.}}: A Sample Article Using IEEEtran.cls for IEEE Journals}

\maketitle

\begin{abstract}
The integration of Optical Intelligent Reflective Surfaces (OIRSs) into Visible Light Communication (VLC) systems is gaining momentum as a valid alternative to RF technologies, harnessing the existing lighting infrastructures and the vast unlicensed optical spectrum to enable higher spectral efficiency, improved resilience to Line-of-Sight (LoS) blockages, and enhanced positioning capabilities. This paper investigates the problem of localizing a low-cost Photo Detector (PD) in a VLC-based indoor environment consisting of only a single Light Emitting Diode (LED) as an active anchor, and multiple spatially distributed single-element OIRSs. We formulate the problem within an indirect, computationally efficient localization framework: first, the optimal Maximum Likelihood (ML) estimators of the LoS and Non-Line-of-Sight (NLoS) distances are derived, using a suitable OIRS activation strategy to prevent interferences. To overcome the grid-based optimization required by the ML NLoS estimator, we devise a novel algorithm based on an unstructured noise variance transformation, which admits a closed-form solution. The set of estimated LoS/NLoS distances are then used within a low-complexity localization algorithm combining an Iterative Weighted Least Squares (IWLS) procedure, whose weights are set according to the inverse of the Cramér-Rao Lower Bound (CRLB), with an adaptive beam steering strategy that allows the OIRSs network to dynamically steer the reflected NLoS signals toward the PD, without any prior knowledge of its position. Accordingly, we derive the CRLB for both LoS/NLoS distance and PD position estimation. Simulation results demonstrate the effectiveness of our approach in terms of localization accuracy, robustness against OIRSs misalignment conditions, and low number of iterations required to attain the theoretical bounds.
\end{abstract}


\section{Introduction}
\IEEEPARstart{T}{he} 
growing scarcity of the \gls{RF} spectrum has driven the development of \gls{VLC} as a promising alternative for wireless data transmission \cite{CommMagaz1}. Recently standardized under IEEE 802.11bb \cite{80211bb}, the \gls{VLC} technology offers the potential to provide high-speed broadband internet services by harnessing \glspl{LED}. Leveraging the existing widespread lighting infrastructures --- commonly found in residential, industrial, and urban environments --- \gls{VLC} reduces deployment costs while simultaneously enhancing energy efficiency, the latter being a critical requirement for next-generation communication systems \cite{Barbarossa6G}. This can be achieved through the inherent advantages of \glspl{LED}, including their extended operational lifespan and lower power consumption.
Moreover, the unlicensed VLC spectrum, spanning approximately 400–800 THz, is thousands of times wider than the RF spectrum and enables data rates up to 10 Gbps, offering a promising opportunity for high-capacity communication \cite{petrosino2023light}.

\glspl{IPS} utilizing \gls{VLC} have attracted growing interest from academia and industry due to their immunity to multipath interference and inter-room leakage, stemming from the high propagation loss and inability of visible light to penetrate walls \cite{sun2022joint}. These features offer enhanced spatial confinement compared to RF-based systems (e.g., WiFi, Bluetooth), motivating the development of dedicated VLC positioning algorithms \cite{zhuang2018survey}.

Several signal processing techniques have been explored for \gls{VLC}-based positioning. Among them, \gls{RSS}-based methods are widely adopted due to their simplicity and effectiveness, taking advantage from the strong correlation between received optical power and distance under low noise conditions \cite{Bozanis2023Indoor,Ma2023Centimeter,Liu2022Machine,RSS2022,RSSFing2021,RobustRSS_2020}. Time-based approaches, such as \gls{TOA} and \gls{TDOA}, use propagation delays from multiple anchors but face practical challenges such as strict synchronization requirements, need for high-speed \glspl{PD}, and vulnerability to multipath and \gls{NLoS} \cite{TOA1,TDOA1,TDOA2}. Angular methods, such as \gls{AOA}, exploit direction of incoming rays but require specialized hardware like directional \glspl{LED} or arrays, and are impaired in presence of significant obstructions \cite{AOA1,AOA2,AOA3}.

Recent studies have explored the integration of \gls{VLC} with \glspl{OIRS} --- also referred to as \glspl{IRS} or \glspl{RIS} --- with the term \gls{OIRS} adopted here to highlight their optical domain distinction \cite{aboagye2023ris}. {\ed In line with the benefits brought by \glspl{RIS} in the RF domain in terms of localization \cite{RIS_mio}, channel estimation \cite{Rev3_1}, and enhanced wireless coverage \cite{Rev3_2}, the integration of the \gls{OIRS} technology into \gls{VLC} systems offers numerous advantages}, including enhanced channel capacity \cite{10198213,10190313}, improved robustness to \gls{LoS} blockages  \cite{9543660,Guzman2025}, and optimized spectrum and energy utilization \cite{sun2022joint,10526227}.
In \gls{VLC} applications, two \gls{OIRS} hardware architectures are generally available: \emph{mirror array-based}, which rely on geometric optics with tiltable mirrors, and \emph{metasurface-based}, which use subwavelength dielectric structures to induce programmable phase shifts \cite{Abdelhady2021Visible,9681888}. In this work, we focus on mirror array-based \glspl{OIRS}, for which practical implementations already exist \cite{Zhang:24}.

\subsection{Related Work} \label{subsec:rel_wrks}
{\ed
To exploit the potential of \glspl{OIRS} in enhancing optical communications, the reflected light must be accurately steered toward the user by tuning either the tilt angles (for mirror-array \glspl{OIRS}) or the phase shifts (for metasurface-based ones). This typically requires integration with \gls{VLC}-based \glspl{IPS} using multiple \glspl{LED} as anchors.
In \cite{Wu:25}, an integrated \gls{VLC}/\gls{VLP} framework is proposed, where the user position is first estimated via sequential \gls{LoS} \gls{RSS} measurements from individual \glspl{LED}, and then used to configure the \gls{OIRS} for communication enhancement. The role of \gls{NLoS} reflections is also investigated to mitigate \gls{LoS} blockages, supported by a \gls{CRLB}-based analysis under imperfect position knowledge. A number of other works focused on OIRS-assisted localization using RSS-based methods. In \cite{11161078}, an RSS fingerprinting approach is considered, involving an offline training phase, where OIRSs are exploited in a multi-LED scenario to enhance the distinguishability of RSS measurements across adjacent spatial locations. An OIRS-assisted single-LED localization system is instead presented in \cite{Wang23}, where a sparse Bayesian learning (SBL) algorithm jointly estimates user position, noise variance, and sparsity under \gls{LoS} blockage conditions. In \cite{11096556}, an RSS-based simultaneous positioning and orientation (SPAO) problem is formulated for a single-LED, single-OIRS, and single-PD setup, and solved via a differential evolution algorithm due to the highly nonconvex nature of the optimization.

From a system setup and specific problem perspective, the most closely related works employ a direct localization paradigm, in which the user position is inferred directly from the raw received signals.
In \cite{Kokdogan24}, a direct \gls{MLE}-based positioning method is introduced using multiple \glspl{LED} and \glspl{OIRS}, adopting a mixed diffuse/specular model with reflections assumed near the center of each element. An orientation adjustment mechanism estimates the user location using multi-\gls{LED} signals. However, the \gls{ML} estimator requires numerical optimization due to the lack of a closed-form solution.
The work in \cite{IDDRISU2025109867} extended \cite{Kokdogan24} by analyzing the impact of orientation mismatches in the \gls{OIRS} setup, quantified through a \gls{MCRB}. Similarly, \cite{TARHAN2025104799} derived a direct ML estimator for an OIRS-assisted single-LED system using RSS measurements, under the assumption that the PD is always illuminated by all the OIRSs reflected paths.
Moreover, \cite{10669070} proposed a joint position and orientation estimation algorithm using multiple \glspl{LED} and suitable signal separation techniques. A particle swarm optimization is employed, with beam steering confined to a single axis with phase shifts pre-aligned to maximize channel gain, implying some prior knowledge of the PD location.

Finally, alternative localization approaches using  different measurements have also been explored. In \cite{10974611}, the authors introduced an OIRS-assisted TDOA-based localization framework. Despite requiring advanced hardware capabilities at the PD to support high sampling rates, the method achieves centimeter-level accuracy via cross-correlation-based delay estimation. The framework, however, relies on a priori knowledge of which reflecting element generates each delayed signal path. To overcome this limitation, \cite{Shi:25} addressed the challenge of separating NLoS paths, using a suitable OIRS element activation protocol. Specifically, NLoS power measurements are collected by deactivating one element at a time and computing differences of RSS, leading to a least-squares problem that can be efficiently solved for the user position.}

\subsection{Contributions}\label{subsec:contrib}
{\ed The aforementioned works tackle the \gls{OIRS}-assisted \gls{VLC} positioning problem from diverse and insightful perspectives, but generally leverage  multiple \glspl{LED}. Moreover, those sharing a similar setup and problem perspective, adopt a direct localization framework, often requiring some prior knowledge or an initial estimate of the user position. Despite their superior performance \cite{ClosasDirect}, direct  methods frequently lack closed-form solutions and rely on intensive numerical optimization, limiting their applicability in low-cost \gls{VLC} devices.}

In this work, we take a different path and investigate the problem of localizing a \gls{PD} in an \gls{IPS} setup that leverages only a single \gls{LED} combined with mirror array-based \glspl{OIRS}, spatially distributed in the environment. Our method relies on an indirect, computationally efficient localization framework based on suitable closed-form estimators that do not require any prior knowledge on the user location.
Moreover, we focus on the case of single-element \glspl{OIRS}, which serve as the fundamental building blocks of larger mirror arrays. As shown in \cite{Abdelhady2021Visible}, a multi-element \gls{OIRS} can be indeed approximated as the superposition of its individual elements, making the single-element case both challenging and worth investigating due to its limited degrees of freedom.
Specifically, the main contributions of this paper are as follows:
\begin{itemize}
    \item We derive the \gls{ML} estimators of the \gls{LoS} distance (between the \gls{LED} and the \gls{PD}) and \gls{NLoS} distances (between each \gls{OIRS} and the \gls{PD}), leveraging a structured activation protocol to avoid interference among \glspl{OIRS}, followed by a successive subtraction that isolate the sole \gls{NLoS} components. To overcome the need of a grid-based optimization required by the \gls{ML} \gls{NLoS} distance estimator, we propose a suitable relaxed \gls{ML} estimator based on an unstructured transformation of the noise variance, which admits a closed-form solution.
    \item An iterative localization algorithm is developed following an indirect estimation paradigm, using the closed-form estimates of the \gls{LoS} and \gls{NLoS} distances as input for a low-complexity \gls{IWLS} algorithm.  Remarkably, the algorithm operates without any prior knowledge of the \gls{PD} position, leveraging an adaptive beam steering mechanism to dynamically orient the distributed \glspl{OIRS} toward the \gls{PD}, and converges to the theoretical bounds within a few iterations. {\ed In addition, a theoretical complexity analysis is conducted to quantify the computational advantages of the proposed approach.}
    \item {\ed We derive the \gls{CRLB} for \gls{LoS}/\gls{NLoS} distance estimation and  position estimation, in a setup involving a single \gls{LED} and  single-element \glspl{OIRS}, and conduct an theoretical analysis of the achievable performance. As a byproduct, the weights of the \gls{IWLS} algorithm are set based on the inverse of the \gls{CRLB}, evaluated at the estimated distances to account for path-dependent uncertainty.}
    \item A simulation campaign is carried out to assess the performance of the proposed distance and position estimation algorithms in comparison with the corresponding \gls{DEB} and \gls{PEB}, {\ed as well as against the state-of-the-art direct ML estimator, under different system parameters and operating conditions, including scenarios with multi-element \glspl{OIRS}. In addition, we  evaluate the computational complexity of all the considered algorithms in terms of their average execution times.} 
\end{itemize}

The remainder of the paper is organized as follows. Sec.~\ref{sec:sys_chan} introduces the system geometry, outlines the \gls{LoS} and \gls{NLoS} channel models, and defines the main problem. Sec.~\ref{sec:los_est} and \ref{sec:nlos_est} present the proposed \gls{LoS} and \gls{NLoS} distance estimators, including both \gls{ML} and relaxed ML approaches. In Sec.~\ref{sec:bounds}, we conduct a Fisher information analysis and derive the theoretical bounds for distance and position estimation. {\ed The complete localization algorithm and its theoretical complexity are detailed in Sec.~\ref{sec:algorithm}, whereas} its performance is assessed in Sec.~\ref{sec:results}. Conclusions are drawn in Sec.~\ref{sec:conclusion}.


\section{System and Channel Models} \label{sec:sys_chan}
We consider an \gls{OIRS}-assisted \gls{VLC} system deployed in an indoor environment and consisting of (i) a single \gls{LED} serving as the primary light source, (ii) $N$ single-element \glspl{OIRS} \cite{Zhang:24,aboagye2023ris} distributed on the walls of the room, and (iii) a \gls{PD} placed on the ground at an unknown location, as shown in Fig.~\ref{fig:sysmodel}.

\subsection{System Geometry}\label{subsec:sys_geom}
\begin{figure*}
    \centering
    \subfloat[]{%
        \includegraphics[width=0.63\linewidth]{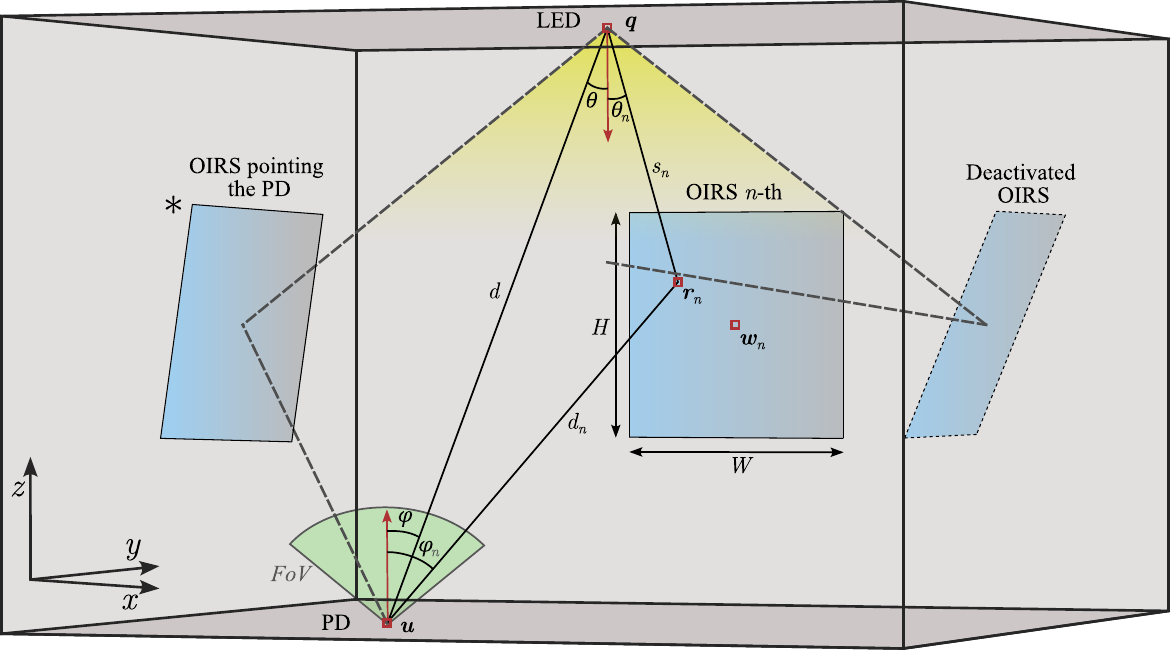}%
        \label{fig:sysmodel_full}
    }
    \hspace{1.3cm}
    \subfloat[]{%
\includegraphics[width=0.2\linewidth]{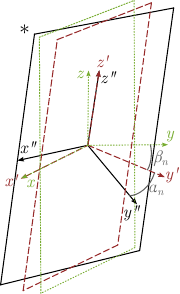}%
        \label{fig:sysmodel_tilt}
    }    
    \caption{(a) Representative scenario of \gls{VLC}-based indoor positioning supported by a single \gls{LED} and a network of $N$ distributed \glspl{OIRS}; (b) Orientation control of the $n$-th \gls{OIRS} via its vertical tilt angle $\beta_n$ and horizontal tilt angle $\alpha_n$ (the green dotted contour indicates the initial orientation of the \gls{OIRS}).}
   \label{fig:sysmodel}
\end{figure*}

The \gls{LED} emitting incoherent light is located at a known position $\bm{q} = \left[ q_x \ q_y \ q_z\right]^\mathsf{T}$. The $n$-th \gls{OIRS}, described by its width $W$ and height $H$ parameters (assumed the same for all \glspl{OIRS}), has its center located at $\bm{w}_n = \left[ w_{n,x} \ w_{n,y} \ w_{n,z}\right]^\mathsf{T}$. The reconfigurability of the $n$-th \gls{OIRS} is achieved through a mechanical orientation system that allows independent adjustment of its tilt angles. Specifically, $\beta_n$ controls the vertical tilt (pitch), adjusting the inclination of the surface relative to the horizontal plane, while 
$\alpha_n$ governs the horizontal tilt (yaw), rotating the surface around its vertical axis, as shown in Fig.~\ref{fig:sysmodel_tilt}. 
Leveraging this reconfigurability, we define two distinct operating modes, controlled by a binary variable $a_n \in \{0,1\}$ with $n = 1, \dots, N$. In the first mode, referred to as \emph{deactivated} ($a_n = 0$), the \gls{OIRS} is tilted toward the ceiling to prevent illumination of the area of interest (as is the case of the \gls{OIRS} placed on the rightmost wall in Fig.~\ref{fig:sysmodel_full}). In the second mode, called \emph{active reflection} ($a_n = 1$), the \gls{OIRS} is oriented in a specific direction for controlled reflection. 

The \gls{PD} is located at an unknown position $\bm{u} = \left[ u_x \ u_y \ 0\right]^\mathsf{T}$.\footnote{\ed We assumed the PD to lie on the ground, so as to simplify some of the subsequent derivations. However, other system entities (LED and OIRS positions), as well as the LoS and NLoS distances and channel gains are defined for generic 3D geometries. This assumption is also consistent with practical scenarios in which the PD height is fixed and known (e.g., localization of mobile robots), a setting often referred to as 2.5D positioning.} 
Accordingly, $d = \|\bm{q} - \bm{u}\|$ denotes the \gls{LoS} distance between the \gls{LED} and the \gls{PD}, with corresponding irradiance and incidence angles $\theta$ and $\varphi$. 
The actual reflection point on the $n$-th \gls{OIRS} is denoted by $\bm{r}_n = \left[ r_{n,x} \ r_{n,y} \ r_{n,z}\right]^\mathsf{T}$. All vectors are expressed in the global coordinate reference system $O(xyz)$. 
{\ed To characterize the \gls{NLoS} channel, $\bm{r}_n$ is computed via a transformation between the global and local reference system of the $n$-th \gls{OIRS}; the corresponding derivation is provided in Appendix \ref{sec:app_b}}.
The distance between $\bm{r}_n$ and the \gls{PD} is given by $d_n = \| \bm{r}_n - \bm{u} \|$,  
while the distance between the \gls{LED} and $\bm{r}_n$  is denoted as  $s_n = \| \bm{q} - \bm{r}_n \|$. Additionally, $\theta_n$ represents the irradiance angle between the \gls{LED} and $\bm{r}_n$, whereas $\varphi_n$ is the incidence angle between $\bm{r}_n$ and the \gls{PD}.

\subsection{Visible Light Channel Gain Model} \label{subsec:ch_gain}
The channel gain over the \gls{LoS} path (\gls{LED}-\gls{PD}) in a \gls{VLC} system  is characterized as \cite{aboagye2023ris}
\begin{equation}
\label{eq:H_LoS}    
h = \frac{ A{\ed \cdot}T{\ed \cdot}G(m+1)}{2\pi d^2} \cos^m(\theta) \cos(\varphi),
\end{equation}
with $A$ the light detector area {\ed and} $T$ the optical filter gain.
{\ed The Lambertian index $m$ is computed as
\begin{equation}
\label{eq:m_lamb}    
m = - \frac{1}{\log_{2}{\cos({\Theta}_{1/2})}},
\end{equation}
with ${\Theta}_{1/2}$ denoting the half-intensity radiation angle.
Moreover, the} term $G$ represents the gain of the non-imaging concentrator and is expressed as
\begin{equation}
G = 
\begin{cases}
\frac{f^2}{\sin^2{\psi}}, &\text{$0 \leq \varphi \leq \psi$},\\
0, & \text{otherwise,}
\end{cases}
\end{equation}
with $\psi$ the \gls{PD} \gls{FoV} and $f$ the refractive index.

{\ed In indoor \gls{VLC} environments, \gls{NLoS} propagation is effectively limited to first-order specular reflections. Because of the nanometer-scale wavelength and the near-field operating regime, higher-order optical reflections suffer extreme attenuation, making penetration, diffraction, and diffuse scattering negligible. Therefore, multi-hop \gls{NLoS} paths, such as those arising from successive reflections across multiple OIRSs, do not provide any useful contribution \cite{sun2022joint, obeed2019optimizing}.
The first-order \gls{NLoS} contribution is modeled using \eqref{eq:H_LoS}, while accounting for both the \gls{LED}-\gls{OIRS} and \gls{OIRS}-\gls{PD} paths as \cite{sun2022joint}}
\begin{equation}
\label{eq:H_NLoS}    
h_n = \rho\frac{A{\ed \cdot}T{\ed \cdot}G(m+1)}{2\pi (s_n +d_n)^2} \cos^m(\theta_n) \cos(\varphi_n),
\end{equation}
with $\rho$ the \glspl{OIRS} reflection factor\footnote{\ed Notice that the reflection factor is real-valued since the OIRSs perform mechanical steering, rather than applying electronically controlled phase shifts. Therefore, the reflection does not introduce an additional phase rotation beyond the geometric phase already accounted for in the propagation model.}. 
As it can be noted, the path loss in \eqref{eq:H_NLoS} follows an additive model \cite{sun2022joint, Abdelhady2021Visible, ibne_mushfique2022mirrorvlc}. This is justified by the fact that, in \gls{VLC} systems, the wavelength is on the nanometer scale and, consequently,  typical indoor distances let signals undergo near-field propagation conditions.

\subsection{Noise Model} \label{subsec:noise}
The electric current induced by the noise at the PD follows a Gaussian distribution \cite{zhuang2018survey,komine2004fundamental} 
\begin{equation}
\label{eq:noise}
    \eta \sim \mathcal{N}(0,\underbrace{\sigma_\text{\tiny{S}}^2 +\sigma_\text{\tiny{T}}^2}_{\sigma^{2}}),
    \vspace{-0.1cm}
\end{equation}
with $\sigma^2$ the noise power, accounting for the contribution of two distinct sources: shot noise ($\sigma_\text{\tiny{S}}^2$) and thermal noise ($\sigma_\text{\tiny{T}}^2$).

The shot noise term represents a random variation in signal intensity due to the discrete nature of particle interactions, such as photons striking the sensor, causing fluctuations, especially at low signal levels.
This phenomenon can be expressed by means of its variance as \cite{zhuang2018survey}
\begin{equation}
\label{eq:s_noise}
    \sigma_\text{\tiny{S}}^2 = 2 q{\ed \cdot}R{\ed \cdot}P{\ed \cdot}B + 2 q{\ed \cdot}I_1 B,
\end{equation}
where $q$ is the electron charge, $R$ is the  \gls{PD} responsivity, $B$ is the bandwidth, and $I_1$ is the current induced by the background light. The total received power $P = p \left( h + \sum_{n=1}^{N} a_n h_n \right)$ includes contributions from the \gls{LoS} component $h$ and the \gls{NLoS} paths $h_n$ reflected by the active \glspl{OIRS} (i.e., with $a_n = 1$), with $p$ being the radiometric transmitted power.

The second term represents the thermal noise, resulting from the thermal agitation of electrons within the \gls{PD} circuit \cite{zhuang2018survey}. Specifically, its variance can be expressed as \cite{Zhang2014Theoretical}
\begin{equation}
\label{eq:t_noise}
    \sigma_\text{\tiny{T}}^2 = \frac{8\pi \kappa \tau}{G_0} \nu A{\ed \cdot}I_2 B^2 + \frac{16 \pi^2 \kappa \tau \varrho}{g} \nu^2 A^2 I_3 B^3,
\end{equation}
where $\kappa$ is the Boltzmann's constant, $\tau$ is the absolute temperature, $G_0$ is the open-loop voltage gain, and $\nu$ is the fixed capacitance of \gls{PD}. Furthermore, $\varrho$ is the channel noise factor, $g$ the transconductance, $I_2$ and $I_3$ are the noise bandwidth factors \cite{komine2004fundamental}. It is worth noting that $\sigma^2  =  \sigma_\text{\tiny{S}}^2 + \sigma_\text{\tiny{T}}^2$ has an explicit dependency on both the \gls{LoS} distance $d$ and the \gls{NLoS} distances $\{d_n\}_{n=1}^N$ (related to the \gls{OIRS}-\gls{PD} paths), through the received signal power $P$ appearing in $\sigma_\text{\tiny{S}}^2$. This signal-dependent nature of the noise variance will be explicitly used in the derivation of our estimation algorithms.

\subsection{Received Signal} \label{subsec:current}
To ease the notation, we assume that the \gls{LED} and the \gls{PD} have the same orientation\footnote{The notation easily generalizes to arbitrary orientations, as shown in \cite{Zhang2014Theoretical}.} \cite{Bozanis2023Indoor, ibne_mushfique2022mirrorvlc}, i.e., $\cos{(\theta)} = \cos{(\varphi)} = q_z/d$. Accordingly, the \gls{LoS} gain in \eqref{eq:H_LoS} can be rewritten as
\begin{equation}
\label{eq:d_from_H_LoS}    
h = { \frac{A{\ed \cdot}T{\ed \cdot}G (m+1)(q_z)^{m+1}}{2\pi d^{m+3}}}.
\end{equation}
Similarly, the \gls{NLoS} contribution from the $n$-th \gls{OIRS} in \eqref{eq:H_NLoS} can be expressed as 
\begin{equation}
\label{eq:d_from_H_LoS3}    
h_n =\rho \frac{A{\ed \cdot}T{\ed \cdot}G (m+1) \left(q_z - r_{n,z}\right)^m}{2\pi {s_n}^m(s_n + d_n)^2 d_n} r_{n,z},
\end{equation}
with $\cos^m(\theta_n) = \left((q_z - r_{n,z})/s_n\right)^m$ and $\cos(\varphi_n) = r_{n,z}/d_n$. 

Leveraging \eqref{eq:d_from_H_LoS} and \eqref{eq:d_from_H_LoS3}, the current induced at the \gls{PD} by the received optical signal can be expressed as
\begin{equation}\label{eq:gencur}
    \mu =  R{\ed \cdot}p \Big(h + \sum_{n=1}^{N} a_n h_n\Big) + \eta.
\end{equation}
    Accordingly, we define the \gls{SNR}  incorporating both the \gls{LoS} and \gls{NLoS} contributions (originating from active \glspl{OIRS}) as
\begin{equation}\label{eq:snr}
    \text{SNR} = \frac{\left[R{\ed \cdot}p (h + \sum_{n=1}^{N} a_n h_n)\right]^2}{\sigma^2}.
\end{equation}

\subsection{Problem Definition}
We tackle the problem of estimating the  \gls{PD} position $\bm{u}$, using information conveyed by both \gls{LoS} and \gls{NLoS} paths. To this aim, an indirect (two-step) adaptive localization paradigm is pursued: in a first step, we estimate the \gls{LoS} distance $d$ between the \gls{LED} and the \gls{PD}, along with the set of \gls{NLoS} distances $\{d_n\}_{n=1}^N$, each corresponding to a signal path via the $n$-th \gls{OIRS}. Then, in the second step, these estimated distances are used to initialize a \emph{low-complexity} iterative localization algorithm. The resulting \gls{PD} position estimate is then fed back into an adaptive beam-steering strategy to dynamically adjust the  control angles $\alpha_n$ and $\beta_n$ of each \gls{OIRS}. This process aims to progressively reorient all \glspl{OIRS} toward the \gls{PD}, so enhancing the accuracy of the localization process.

\section{LoS (LED-PD) Distance Estimation} \label{sec:los_est}
To estimate the \gls{LoS} distance $d$, we initially consider a configuration where all the $N$ \glspl{OIRS} are set in \emph{deactivated} mode (i.e., $a_n = 0 \, \, \forall n=1,\ldots,N$). The \gls{PD} collects  $K \in \mathbb{N}$ independent and identically distributed (i.i.d.) samples of noncoherent light emitted by the \gls{LED}, denoted as $\mu_{0,k}$, with $k = 1, \dots, K$, and expressed following \eqref{eq:gencur} as
\begin{equation}
\label{eq:mu_zero}
    \mu_{0,k} = R{\ed \cdot}p{\ed \cdot}h + \eta_{0,k},
\end{equation}
with $\eta_{0,k}\sim \mathcal{N}(0,\sigma^{2}_0)$ and 
$\sigma^{2}_0 = \sigma_\text{\tiny{T}}^2 + 2 q{\ed \cdot}R{\ed \cdot}p{\ed \cdot}h{\ed \cdot}B + 2{\ed \cdot}q I_1 B$. Accordingly, we have  $\mu_{0,k}\sim \mathcal{N}(\bar{\mu}_0,\sigma^{2}_0)$, with $\bar{\mu}_0 = R{\ed \cdot}p{\ed \cdot}h$.

\subsection{Maximum Likelihood LoS Distance Estimation}\label{ML_LOS_subsect}
In this section, we derive the \gls{ML} estimator of the \gls{LoS} distance $d$, leveraging the dependency appearing in both the channel gain \eqref{eq:d_from_H_LoS} and  the shot noise variance term \eqref{eq:s_noise}.
To this aim, using \eqref{eq:s_noise}-\eqref{eq:d_from_H_LoS} we rewrite
\begin{equation}
     \sigma^{2}_0(d) = a + b \bar{\mu}_0(d),
     \label{eq:noiseLoS}
\end{equation}
with $a = \sigma_\text{\tiny{T}}^2 +  b{\ed \cdot}I_1$ and $b =2{\ed \cdot}q{\ed \cdot}B$. Furthermore, using \eqref{eq:d_from_H_LoS} we make explicit the dependency of  $\bar{\mu}_0(d)$ on the unknown  $d$ as
\begin{equation}
\bar{\mu}_0(d) =  \frac{\xi}{d^{m+3}},
\end{equation}
{\ed where we collect all the known coefficients as 
\begin{equation} 
\xi = {\frac{R{\ed \cdot}p{\ed \cdot}A{\ed \cdot}T{\ed \cdot}G(m+1)(q_z)^{m+1}}{2\pi}}.
     \label{eq:xi}
\end{equation}
}

The likelihood function for the \gls{LoS} distance estimation problem is given by
\begin{equation}
     \label{eq:lh_los}
     L^\LoS(\bm{\mu}_0; d) = \prod^{K}_{k = 1} f(\mu_{0,k} ; d) = \displaystyle \frac{\ed\exp\!\left[{-\frac{\sum _{k=1}^K \left(\mu_{0,k} - \frac{\xi}{d^{m+3}}\right)^2}{2\left(a + b \frac{\xi}{d^{m+3}}\right)}}\right]}{{\left[2 \pi(a + b \frac{\xi}{d^{m+3}})\right]^{\frac{K}{2}}}},
\end{equation}
with {\ed$\exp[\cdot]$ the  exponential function,} $\bm{\mu}_0 = [\mu_{0,1} \ \cdots \ \mu_{0,K}]^\mathsf{T}$ a vector stacking the $K$ samples and $f(\mu_{0,k} ; d)$ denotes the probability density function (pdf) of $\mu_{0,k}$ given $d$.
The \gls{MLE} problem can be then formulated as
\begin{equation}
     \label{eq:ml_los_problem}
     \hat{d}^\ML =  \argmin_{d} \mathcal{L}^\LoS(\bm{\mu}_0; d), 
\end{equation}
where, for convenience, we transformed the original maximization problem through the negative log-likelihood function
\begin{align}
\label{eq:ll_los_full}
&\mathcal{L}^\LoS(\bm{\mu}_0; d) \eqdef -\ln L^\LoS(\bm{\mu}_0; d) \nonumber \\
&= \frac{K}{2}\ln{\left[2 \pi \left(a + b \frac{\xi}{d^{m+3}}\right)\right]} 
+ \frac{\sum\limits_{k=1}^K \left(\mu_{0,k} - \frac{\xi}{d^{m+3}}\right)^2}{2(a + b \frac{\xi}{d^{m+3}})}. 
\end{align}
To solve the optimization problem in  \eqref{eq:ml_los_problem}, we seek the value of $d$ that satisfies $\ed
\partial \mathcal{L}^\LoS(\bm{\mu}_0; d)/\partial d = 0$.
Given that $a$, $b$, $m$, $d$, $\xi$, and $K$ are strictly positive real values, after some algebraic manipulations, we obtain the following closed-form solution
\begin{equation}\label{eq:ml_los}
\!\!\!\hat{d}^\ML \!=\! \left[\frac{\xi \left(\!\! \sqrt{\!(a \!+\! bS_1)^2  \!+\! b^2(S_2 \!-\! S_1^2) \!+\! \frac{b^4}{4}} \!+\! a \!+\! \frac{b^2}{2} \!\!\right)}{2aS_1 + bS_2 - ab} \right]^{\frac{1}{m+3}}\!\!\!\!, 
\end{equation} 
where we introduced the summation terms
\begin{equation}\label{eq:sums_obs}
S_1 = \frac{1}{K}\sum_{k=1}^{K} \mu_{0,k}, \quad S_2 = \frac{1}{K}\sum_{k=1}^{K} \mu_{0,k}^2.
\end{equation}  
By closely examining \eqref{eq:ml_los} and leveraging \eqref{eq:sums_obs}, we note that the sample mean of the observables, $S_1$, and the sample mean of their squared values, $S_2$, serve as \emph{sufficient statistics} for the \gls{ML} estimator of $d$. More specifically, \eqref{eq:ml_los} depends on both $S_1$ and $S_2$, not only as separate terms, but also through the term $(S_2 - S_1^2)$ capturing the sample variance of the observations.

\subsection{Relaxed ML LoS Distance Estimation}\label{sec::RML_LOS}
Although the ML estimator in \eqref{eq:ml_los} has a closed-form solution, eliminating the need for suboptimal alternatives, we investigate, for comparison purposes, a different \gls{LoS} distance estimation algorithm that takes advantage of an unstructured transformation of the noise variance $\sigma^{2}_{0}(d)$: it neglects the dependence of the latter from the unknown $d$ and, more in general, its signal-dependent structure in \eqref{eq:noiseLoS}.
Under this relaxed model, the dependency on the sought $d$ remains only in $\bar{\mu}_0(d)$ (mean of the observables), while the variance $\sigma^2_0$ is treated as an additional unknown nuisance parameter to be estimated. The corresponding relaxed \gls{ML} (RML) estimation problem, again formulated as an equivalent minimization problem based on the negative log-likelihood function, can be expressed as
\begin{equation}
     \label{eq:rmll_los_problem}
     \hat{d}^\RML =  \argmin_d \left[ \argmin_{\sigma^2_0} \tilde{\mathcal{L}}^\LoS(\bm{\mu}_0; d,\sigma^2_0) \right], 
\end{equation}
with
\begin{equation}
\!\!\!\!\tilde{\mathcal{L}}^\LoS(\bm{\mu}_0; d,\sigma^2_0)  \!=\! \frac{K}{2}\ln{(2 \pi \sigma_0^2)} + \frac{1}{2 \sigma_0^2} \sum\limits_{k=1}^K \!\left(\!\mu_{0,k} \!-\! \frac{\xi}{d^{m+3}}\!\right)^{\!2}\!\!.
 \label{eq:ll_los_rel}
\end{equation}
It is simple to observe that the inner minimization in \eqref{eq:rmll_los_problem} with respect to $\sigma^2_0$ is solved by
\begin{equation}
    \hat{\sigma}_0^2 = \frac{1}{K} \sum\limits_{k=1}^{K} \left(\mu_{0,k}-\frac{\xi}{d^{m+3}}\right)^2.
\end{equation}
Substituting this value back in  \eqref{eq:ll_los_rel}, neglecting unnecessary constant terms, and considering a monotonic transformation of the
log-likelihood function leads to a compressed
cost function which, with a slight abuse of notation, is denoted by
\begin{equation}
\tilde{\mathcal{L}}^\LoS(\bm{\mu}_0; d)  = \frac{1}{K}\sum\limits_{k=1}^K \left(\!\mu_{0,k} - \frac{\xi}{d^{m+3}}\!\right)^2,
\end{equation}
and the RML estimation problem reduces to
\begin{equation}
     \label{eq:rmll_los_problem_final}
     \hat{d}^\RML =  \argmin_d \tilde{\mathcal{L}}^\LoS(\bm{\mu}_0; d).
\end{equation}
The solution of the RML estimation problem can be then obtained in closed form as
\begin{equation} \label{eq:rml_los}
     \hat{d}^\RML =   \sqrt[m+3]{\frac{\xi}{S_1}}.
\end{equation}
By directly comparing \eqref{eq:rml_los} and \eqref{eq:ml_los}, it is noteworthy that, in the case of the RML estimator, only the sample mean  $S_1$ serves as a sufficient statistic for its computation.
\section{NLoS (OIRS-PD) Distance Estimation} \label{sec:nlos_est}
This section deals with the estimation of the distances $\{d_n\}_{n=1}^N$, each corresponding to a \gls{NLoS} path received by the \gls{PD} via the $n$-th \gls{OIRS}.
To prevent interference among \glspl{OIRS}, we employ a structured activation protocol in which only one \gls{OIRS} operates in \emph{active reflection} mode at a time (which is tantamount to having $\sum_{n=1}^{N} a_n = 1$). The activation follows a predefined sequence known to the \gls{PD}, that collects $K_n \in \mathbb{N}$ samples of the incoming signals from the $n$-th \gls{OIRS}, denoted by $\mu_{n,k}$ with $k = 1, \dots, K_n$, including \gls{LoS} and \gls{NLoS} as
\begin{equation}
    \mu_{n,k}  =  R{\ed \cdot}p(h + h_n)+\eta_{n,k},
    \label{eq:mnk}
\end{equation}
where $\eta_{n,k}\sim \mathcal{N}(0,\sigma^{2}_n)$, with $\sigma^{2}_n = a +  b{\ed \cdot}R{\ed \cdot}p (h+h_n)$, denotes the additive noise. {\ed Since the LoS path typically dominates the received power, we first compensate for it by subtracting its  contribution, thereby isolating the weaker NLoS reflection. We can thus take advantage of the LoS estimation step in Sec. \ref{sec:los_est} and remove the estimated LoS term from \eqref{eq:mnk} as}
\begin{equation}
    \chi_{n,k} = \mu_{n,k} - \bar{\mu}_0(\hat{d}) \approx R{\ed \cdot}p{\ed \cdot}h_n\!+\eta_{n,k},\!\!
    \label{eq:NLoS_contrib}
\end{equation}
with $\hat{d}$ being either $\hat{d}^\ML$ or $\hat{d}^\RML$. After compensating for the \gls{LoS} component in the mean of $\mu_{n,k}$, the new observations approximately follow $\chi_{n,k}\sim \mathcal{N}(\bar{\chi}_n,\sigma^{2}_n)$, with $\bar{\chi}_n = R{\ed \cdot}p{\ed \cdot}h_n$.
We now observe that $h_n$ in \eqref{eq:d_from_H_LoS3} explicitly depends on the reflection point $\bm{r}_n$, which, in turn, is a function of the unknown \gls{PD} position $\bm{u}$. To streamline the derivation and avoid overloading the notation, the following development of the \gls{NLoS} distance estimators is conducted with respect to $\bm{r}_n$. Later in Section \ref{sec:algorithm}, we will then present the complete iterative localization algorithm that starts with an initial misalignment --- using $\bm{w}_n$ in place of $\bm{r}_n$ in the \gls{NLoS} distance estimators --- and progressively refines the orientation of the \gls{OIRS} network toward the \gls{PD} through an adaptive beam steering strategy.

\subsection{Maximum Likelihood NLoS Distance Estimation}\label{sec:NLOS_ML}
The \gls{ML} estimator of the \gls{NLoS} distance $d_n$ is derived following the same lead of Sec. \ref{ML_LOS_subsect}. More specifically, we first make explicit the dependency of $\sigma^{2}_n$ on the desired $d_n$ as
\begin{equation}
      \sigma^{2}_n(d_n) = a + b [\bar{\chi}_n(d_n) + \bar{\mu}_0(\hat{d})],
\end{equation}
where we keep using the estimate $\hat{d}$ (either $\hat{d}^\ML$ or $\hat{d}^\RML$) in place of $d$.
Also in this case, we recast $\bar{\chi}_n(d_n)$  to expose its dependence on $d_n$ as
\begin{equation}
 \bar{\chi}_n(d_n) = \frac{\omega}{(s_n + d_n)^2 d_n},
\end{equation}
{\ed with 
\begin{equation}
\omega= \frac{R{\ed \cdot}p{\ed \cdot}\rho{\ed \cdot}A{\ed \cdot}T{\ed \cdot}G (m+1) \left(q_z - r_{n,z}\right)^m r_{n,z}}{2\pi {s_n}^m}.
\end{equation}}
The likelihood function for the \gls{NLoS} distance estimation problem is given by
\begin{align}
\label{eq:nlos_like}
\!L^\NLoS(\bm{\chi}_n&;d_n)\!  = \prod^{K_n}_{k = 1} f(\chi_{n,k} ; d_n) =\nonumber \\
     & =\!\frac{\ed\exp\left[{-\frac{\sum _{k=1}^{K_n} \left(\chi_{n,k} - \frac{\omega}{(s_n+d_n)^2 d_n}\right)^2}{2\left[a + b \left(\frac{\xi}{\hat{d}^{m+3}} + \frac{\omega }{(s_n+d_n)^2 d_n}\right)\right]}}\right]}{\left\{ 2\pi \left[ a + b \left( \frac{\xi}{\hat{d}^{m+3}} + \frac{\omega}{(s_n + d_n)^2 d_n} \right) \right] \right\}^{\frac{K_n}{2}}},
\end{align}
where $\bm{\chi}_n = [\chi_{n,1} \ \cdots \ \chi_{n,K_n}]^\mathsf{T}$ and $f(\chi_{n,k} ; d_n)$ denotes the pdf of $\chi_{n,k}$ given $d_n$. By resorting to the negative log-likelihood function, the \gls{MLE} problem is formulated as
\begin{equation}
     \label{eq:mll_nlos_problem}
     \hat{d}^\ML_n =  \argmin_{d_n}
 \mathcal{L}^\NLoS(\bm{\chi}_n; d_n), 
\end{equation}
with
\begin{align} \label{eq:ll_nlos_full}
& \mathcal{L}^\NLoS(\bm{\chi}_n; d_n)  = -\ln L^\NLoS(\bm{\chi}_n; d_n)  = \nonumber \\ & =  \frac{K_n}{2}\ln 2\pi + \frac{K_n}{2}\ln\!{\left[\!a\! +\! b\!  \left(\frac{\omega }{(s_n+d_n)^2 {d_n}}+ \frac{\xi}{\hat{d}^{m+3}} \!\right)\!\right]}  \nonumber \\ & + \frac{\sum_{k=1}^{K_n} (\chi_{n,k} - \frac{\omega }{(s_n+d_n)^2 d_n})^2}{2 \left[a + b  \left(\frac{\omega }{(s_n+d_n)^2 {d_n}}+ \frac{\xi}{\hat{d}^{m+3}} \right)\right]}.
\end{align}
Unlike the \gls{LoS} scenario, a closed-form solution for $d_n$ that minimizes \eqref{eq:mll_nlos_problem} is not available in the \gls{NLoS} case. Since $\mathcal{L}^\NLoS(\bm{\chi}_n; d_n)$ is highly non-linear, but depends only on the unknown $d_n$, the \gls{ML} estimate $\hat{d}^\ML_n$ can be obtained by performing a grid search over the space of $d_n$. To avoid the need for numerical optimization, in the following we propose a relaxed formulation of the \gls{ML} estimator tailored to the \gls{NLoS} scenario.

\subsection{Relaxed ML NLoS Distance Estimation}\label{sec:NLOS_RML}
Similarly to the \gls{LoS} case, we adopt an unstructured transformation of the noise variance $\sigma^{2}_{n}(d_n)$, relaxing its dependency on the unknown $d_n$. 
As a result, $\sigma^2_n$ becomes an additional nuisance parameter to be estimated jointly with the desired $d_n$. The corresponding RML estimation problem is expressed, in terms of the negative log-likelihood function, as
\begin{equation}
     \label{eq:rmll_nlos_problem}
     \hat{d}_n^\RML =  \argmin_{d_n} \left[ \operatorname*{min}_{\sigma^{2}_{n}} \tilde{\mathcal{L}}^\NLoS(\bm{\chi}_n; d_n,\sigma^{2}_{n}) \right],
\end{equation}
with
\begin{align}
& \tilde{\mathcal{L}}^\NLoS(\bm{\chi}_n; d_n,\sigma^{2}_{n})  = -\ln L^\NLoS(\bm{\chi}_n; d_n,\sigma^{2}_{n})   =\nonumber \\ &= 
\frac{K_n}{2}\ln{(2 \pi \sigma_n^2)} + \frac{1}{2 \sigma_n^2} \sum\limits_{k=1}^{K_n} \left(\chi_{n,k} - \frac{\omega }{(s_n+d_n)^2 d_n}\right)^2. 
\label{eq:ll_nlos_full_r}
\end{align}
The inner minimization in \eqref{eq:rmll_nlos_problem} yields a closed-form expression for the estimate of $\sigma_n^2$ as
\begin{equation}
    \!\hat{\sigma}_n^2 \!= \!\frac{1}{K_n} \sum\limits_{k=1}^{K_n} \left(\chi_{n,k} - \frac{\omega }{(s_n+d_n)^2 d_n}\right)^2.
\end{equation}
Substituting the above estimate into  \eqref{eq:ll_nlos_full_r}, omitting constant terms, and applying a monotonic transformation, we obtain the  following compressed cost function 
\begin{equation}\label{eq::RML_NLOS_compressed}
\tilde{\mathcal{L}}^\NLoS(\bm{\chi}_n; d_n)  = \frac{1}{K_n}
\sum\limits_{k=1}^{K_n} \left(\chi_{n,k} - \frac{\omega }{(s_n+d_n)^2 d_n}\right)^2,
\end{equation}
and the RML estimation problem in \eqref{eq:rmll_nlos_problem} reduces to 
\begin{equation}
     \label{eq:rmll_nlos_problem_final}
     \hat{d}_n^\RML =  \argmin_{d_n} \tilde{\mathcal{L}}^\NLoS(\bm{\chi}_n; d_n).
\end{equation}
Minimization in \eqref{eq:rmll_nlos_problem_final} is conducted by differentiating $\tilde{\mathcal{L}}^\NLoS(\bm{\chi}_n; d_n)$ with respect to $d_n$, manipulating the resulting expression to obtain a more tractable depressed cubic equation in $d_n$, leading to the following closed-form solution
\begin{equation}\label{eq:rml_nlos}
     \hat{d}_n^\RML =   \frac{\gamma  T_1 s^2_n}{3 \upsilon }+\frac{\upsilon }{3 \gamma T_1}-\frac{2 {s_n}}{3},
\end{equation}
with $T_1 = \frac{1}{K_n} \sum _{k=1}^{K_n} \chi_{n,k}$, $\gamma = \sqrt[3]{2}$, and
\begin{equation}\label{eq::upsilon}
\upsilon \!= \!\left(\!3 \sqrt{3\omega  T_1^4 \!\left(27 \omega \!+\!4 T_1 s^3_n\right)}\!+\!2 T_1^3 s^3_n\!+\!27 \omega  T_1^2\right)^\frac{1}{3}\!\!\!.
\end{equation}
From \eqref{eq:rml_nlos}--\eqref{eq::upsilon}, it is evident that, similarly to the RML distance estimator in the \gls{LoS} case, the sample mean $T_1$ of the new observations $\chi_{n,k}$ represents a sufficient statistic also for the RML estimator in the \gls{NLoS} scenario.

\section{Iterative Indirect Localization Algorithm with Adaptive Beam Steering}\label{sec:algorithm}
In this section, we present a low-complexity iterative localization algorithm based on an indirect estimation paradigm, where the position of the \gls{PD} is inferred through a two-step process: an initial \emph{distance estimation} phase, followed by a subsequent \emph{position estimation} step that leverages the distance estimates obtained in the first stage. The algorithm operates without requiring any prior information on $\bm{u}$,  and employs a successive adaptive beam-steering strategy to iteratively reorient the \gls{OIRS} network toward the \gls{PD} position. {\ed The processing steps involved in the proposed algorithm are detailed in Sec. \ref{subsec::AlgDescri}, while Sec. \ref{subsec::Complexity} presents a theoretical analysis of the associated computational complexity.}

\subsection{\ed Algorithm Description}\label{subsec::AlgDescri}

\textit{\gls{LoS} \& \gls{NLoS} Distance Estimation:} 
Following the indirect localization paradigm, the first stage involves the estimation of the \gls{LoS} distance $d$, as well as of the \gls{NLoS} distances $\{d_n\}_{n=1}^N$. The former is carried out by setting all $N$ \glspl{OIRS} in deactivated mode ($a_n = 0, \forall n$), and using either the ML estimator ($\hat{d}^\ML$ in \eqref{eq:ml_los}) or the RML estimator ($\hat{d}^\RML$ in \eqref{eq:rml_los}). Once an estimate of the \gls{LoS} distance is obtained, the algorithm proceeds with the \gls{NLoS} distance estimation, using either the \gls{ML} estimator derived in Sec. \ref{sec:NLOS_ML}, or the RML estimator in Sec. \ref{sec:NLOS_RML}. At this point, we observe that both the ML cost function in~\eqref{eq:ll_nlos_full} --- whose numerical minimization with respect to $d_n$ for each $n$-th \gls{OIRS}-\gls{PD} link yields the \gls{ML} estimate --- and the closed-form RML estimator in~\eqref{eq:rml_nlos}-\eqref{eq::upsilon} depend on the reflection point $\bm{r}_n$, which in turn is a function of the unknown \gls{PD} position $\bm{u}$. Consequently, a direct evaluation of the \gls{NLoS} estimators would require some form of a priori information about $\bm{u}$. This limitation is particularly critical at the beginning of the localization procedure, when no  estimate of $\bm{u}$ is yet available. 

To overcome this need, we propose to initially evaluate the \gls{NLoS} distance estimators using $\bm{w}_n$ in place of $\bm{r}_n$ --- assuming that all \glspl{OIRS} are already perfectly steered towards the \gls{PD} position --- which leads to having $\tilde{d}_n = \| \bm{w}_n - \bm{u} \|$ and $\tilde{s}_n = \| \bm{q} - \bm{w}_n \|$ in place of $d_n$ and $s_n$ in \eqref{eq:ll_nlos_full} and \eqref{eq:rml_nlos}-\eqref{eq::upsilon}, respectively.
This is tantamount to paying the price of an initial misalignment as a trade-off to enable the estimation of the \gls{NLoS} distances as either $\{\hat{d}^\ML_n\}_{n=1}^N$ or $\{\hat{d}^\RML_n\}_{n=1}^N$, without any prior information on $\bm{u}$. This initial approximation is progressively refined over the iterations of the localization algorithm through a suitable adaptive beam steering strategy, which will be presented later in this section.

\textit{Position Estimation:}  
Once the \gls{PD} has obtained the estimates of the \gls{LoS} and \gls{NLoS} distances --- denoted, for notational convenience, by $\hat{d}$ and $\{\hat{d}_n\}_{n=1}^N$, respectively (with their exact values depending on the chosen estimation method, either \gls{ML} or RML) --- it proceeds to estimate its own position. To this aim, we stack the estimated distances into a vector $\hat{\bm{d}} = [\hat{d} \ \hat{d}_1 \ \cdots \ \hat{d}_N]^T \in \mathbb{R}^{N+1}$ and set up a system of $N+1$ (with $N\geq2$ to make the problem well-posed) nonlinear equations linking $\hat{\bm{d}}$ with the sought \gls{PD} position $\bm{u}$ as
\begin{equation}\label{eq:d_sqr}
    \hat{\bm{d}}^2 = \bm{f}(\bm{u}) + \bm{\zeta},
\end{equation}
with $\bm{\zeta} = [\zeta \ \zeta_1 \ \cdots \ \zeta_N]^T$
denoting the vector of the generic errors associated to the \gls{LoS} and \gls{NLoS} distance estimation process, respectively, $\hat{\bm{d}}^2$ the element-wise square of $\hat{\bm{d}}$, and the nonlinear vector function $\bm{f}(\bm{u}) \in \mathbb{R}^{N+1}$ is given by
\begin{equation}\label{eq:f_u}
    \bm{f}(\bm{u}) = \left[
        \|\bm{u} - \bm{q}\|^2,\;
        \|\bm{u} - \bm{w}_{1}\|^2, \;
        \cdots,\;
        \|\bm{u} - \bm{w}_{N}\|^2
    \right]^T.
\end{equation} 
{\ed It is worth noting that, in this step, we rely solely on the known positions of the OIRSs and the LED, as these are system parameters that we can control, whereas no prior knowledge of the PD position is required.}
The \gls{PD} position is then estimated by solving a Weighted Least Squares (WLS) optimization problem formulated as
\begin{equation}\label{eq:WLS}
    \hat{\bm{u}} = \min_{\bm{u}} \bm{\epsilon}(\bm{u})^\mathsf{T} \bm{W}\bm{\epsilon}(\bm{u}),
\end{equation}
where $\bm{\epsilon}(\bm{u}) \eqdef \hat{\bm{d}}^2 - \bm{f}(\bm{u})$, and the diagonal weighting matrix $\bm{W} = \mathrm{diag}(\text{w}_0, \text{w}_1, \ldots, \text{w}_N)$. We anticipate that the weights in $\bm{W}$ will be set according to the inverse of the \gls{CRLB} associated with the distance estimation problems, whose derivation is proposed in Sec. \ref{sec:bounds}.
Considering the limited computational capabilities available on a low-cost \gls{PD}, we solve \eqref{eq:WLS} using a low-complexity \gls{IWLS} approach. More specifically, denoting with  $\hat{\bm{u}}^{(i)}$ the \gls{PD} position estimate at current iteration $i$, we employ the following iterative update rule \cite[Ch. 8]{kay1993fundamentals}
\begin{equation}\label{eq::IWLS_update}
    \hat{\bm{u}}^{[i+1]} = \hat{\bm{u}}^{[i]} + \left( (\bm{H}^{[i]})^T \bm{W} \bm{H}^{[i]} \right)^{-1} (\bm{H}^{[i]})^T \bm{W} \bm{\epsilon}^{[i]},
\end{equation}
with $\bm{H}^{[i]} = \left. \frac{\partial \bm{f}(\bm{u})}{\partial \bm{u}} \right|_{\bm{u} = \hat{\bm{u}}^{[i]}}$ the Jacobian of $\bm{f}(\bm{u})$, $\bm{\epsilon}^{[i]} = \hat{\bm{d}}^2 - \bm{f}(\hat{\bm{u}}^{[i]})$, and $\bm{f}(\bm{u}) \approx \bm{f}(\hat{\bm{u}}^{[i]}) + \bm{H}^{[i]} (\bm{u} - \hat{\bm{u}}^{(i)})$. The IWLS algorithm
stops when the difference between the current and the updated estimate $\norm{\hat{\bm{u}}^{[i+1]} - \hat{\bm{u}}^{[i]}} < \varepsilon$ for a predefined threshold $\varepsilon$ or if a maximum number of
iterations is reached.

\begin{algorithm}[!t]
\caption{Iterative Localization with Adaptive Beam Steering}
\label{algo1}
\begin{algorithmic}[1]
\State \textbf{Input:} LED position $\bm{q}$, OIRS positions $\{\bm{w}_n\}_{n=1}^N$, threshold $\varepsilon$, maximum iterations $I_{\max}$
\State \textbf{Output:} Estimated PD position $\hat{\bm{u}}$

\Statex \textit{\textbf{Step 1: Distance Estimation}}
\State Deactivate all OIRSs: $a_n \gets 0, \ \forall n$
\State Estimate LoS distance $\hat{d}$ with \eqref{eq:ml_los} (ML) or \eqref{eq:rml_los} (RML) using $\bm{\mu}_0$
\For{$n = 1,\dots,N$}
    \State Activate the $n$-th \gls{OIRS}: $a_n \gets 1$
    \State Estimate NLoS distance $\hat{d}_n$ with \eqref{eq:mll_nlos_problem} (ML) or \eqref{eq:rml_nlos}
    \Statex \quad\, (RML) using $\bm{r}_n \gets \bm{w}_n$ and $\bm{\chi}_n$
    \State Deactivate the $n$-th \gls{OIRS}: $a_n \gets 0$
\EndFor
\State Assemble distance vector: $\hat{\bm{d}} \gets [\hat{d}, \hat{d}_1, \dots, \hat{d}_N]^T$

\Statex \textit{\textbf{Step 2: Position Estimation via IWLS}}
\State Initialize $\hat{\bm{u}}^{[0]}$ (e.g., random or centroid of OIRSs)
\For{$i = 0$ to $I_{\max} - 1$}
    \State Compute $\bm{\epsilon}^{[i]}$ and the Jacobian $\bm{H}^{[i]}$
    \State Set weights $\bm{W}$ according to inverse of CRLB
    \Statex \quad\, evaluated in $\hat{d}$ and $\{\hat{d}_n\}_{n=1}^N$ 
    \State IWLS update of position estimate using \eqref{eq::IWLS_update}
    \If{$\|\hat{\bm{u}}^{[i+1]} - \hat{\bm{u}}^{[i]}\| < \varepsilon$}
        \State \textbf{break}
    \EndIf
\EndFor
\State Final estimate: $\hat{\bm{u}} \gets \hat{\bm{u}}^{[i+1]}$

\Statex \textit{\textbf{Step 3: OIRS Adaptive Beam Steering}}
\For{$n = 1,\dots,N$}
    \State Employ \eqref{eq:oirsnorm} to compute normal vector of $n$-th \gls{OIRS}
    \State Update \gls{OIRS} tilt angles according to \eqref{eq:a_b_angles}
\EndFor

\State Return to Step 1 with updated \glspl{OIRS} angles for position estimate refinement
\end{algorithmic}
\end{algorithm}

\textit{\gls{OIRS} Adaptive Beam Steering:} The \gls{PD} position estimate $\hat{\bm{u}}$ obtained via the IWLS procedure is subsequently exploited to adaptively steer the beams of all $N$ \glspl{OIRS} toward the estimated \gls{PD} direction. In particular, the tilt angles $\alpha_n$ and $\beta_n$ of each $n$-th \gls{OIRS} are reconfigured by aligning its surface normal vector to an optimal bisector direction between the \gls{LED} and the estimated \gls{PD} position. This is achieved by setting the unit normal vector $\bm{o}_n$ of the $n$-th \gls{OIRS} as 
\begin{equation}\label{eq:oirsnorm}
    \bm{o}_n = \frac{\frac{\bm{q}-\bm{w}_n}{\|\bm{q}-\bm{w}_n\|} + \frac{\hat{\bm{u}}-\bm{w}_n}{\|\hat{\bm{u}}-\bm{w}_n\|}}%
    {\sqrt{2 + 2\left(\frac{\bm{q}-\bm{w}_n}{\|\bm{q}-\bm{w}_n\|}\right)^\mathsf{T} \frac{\hat{\bm{u}}-\bm{w}_n}{\|\hat{\bm{u}}-\bm{w}_n\|}}}.
\end{equation}
Accordingly, the tilt angles are updated based on the orientation of the unit normal vector $\bm{o}_n$ as follows
\begin{equation}\label{eq:a_b_angles}
\beta_n = \sin^{-1} (\bm{o}_n^T \bm{e}_3), \ \ \
\alpha_n = \sin^{-1} (\bm{o}_n^T \bm{e}_1/\cos{(\beta_n)}),
\end{equation}
with $\mathbf{e}_j$ (with either $j=1$ or $j=3$) denoting the $j$-th canonical basis vector in $\mathbb{R}^{3\times1}$. This adaptive reconfiguration steers each \gls{OIRS} with the aim of compensating for the initial mismatch between the actual reflection point $\bm{r}_n$ and the geometric center of the $n$-th \gls{OIRS}, $\bm{w}_n$. Notably, when the estimated position $\hat{\bm{u}}$ closely approximates the true \gls{PD} position $\bm{u}$, the beam steering strategy defined in \eqref{eq:oirsnorm} and \eqref{eq:a_b_angles} naturally drives the reflection alignment towards $\bm{r}_n \approxeq \bm{w}_n, \: \forall n$.

\textit{Check and Iterate:}
After executing all the preceding steps, the \gls{PD} uses a simple criterion to determine whether a further refinement of its position estimate is required. Among various possible stopping conditions, we adopt a threshold on the Euclidean distance 
between the current and previous \gls{PD} position estimates, that is, the localization procedure stops when no significant changes in $\hat{\bm{u}}$ occur, meaning that an additional reconfiguration of the \glspl{OIRS} is no more beneficial. We anticipate that, remarkably, the proposed algorithm needs a very few iterations to improve the quality of the \gls{PD} position estimate; we denote by $M$ the total number of such iterations.
The main steps of the proposed localization algorithm  with adaptive beam steering are summarized in Alg.~\ref{algo1}.

\subsection{\ed Computational Complexity Analysis}\label{subsec::Complexity}
{\ed In this section, we analyze the computational complexity of the iterative localization algorithm with adaptive beam steering proposed in Sec. \ref{subsec::AlgDescri}. To conduct the analysis, we separately inspect each individual step of the  approach, as follows.
\\
\indent \emph{Complexity of Distance Estimation: }
As the first step of our algorithm, we obtain the ML  estimate of the \gls{LoS} distance $d$ using the expressions in \eqref{eq:ml_los_problem}-\eqref{eq:ll_los_full}. Owing to its closed-form nature, the \gls{ML} \gls{LoS} estimator involves only simple arithmetic operations, resulting in an asymptotic complexity of $O(K)$, associated to the cost required to compute its sufficient statistics $S_1 = \frac{1}{K}\sum_{k=1}^K \mu_{0,k}$ and $S_2 = \frac{1}{K}\sum_{k=1}^K \mu^2_{0,k}$.

After estimating the \gls{LoS} distance, the next step involves computing the estimates of the $N$ \gls{NLoS} distances $d_1, \ldots, d_N$. 
For each NLoS component $n$ (originating from the $n$-th active OIRS), the computational cost of the ML NLoS estimator derived in Sec. \ref{ML_LOS_subsect} scales asymptotically as
$O(Q\cdot K_n)$, where $O(K_n)$ accounts for the processing of the $K_n$ \gls{NLoS} samples in $\bm{\chi}_n$ according to \eqref{eq:nlos_like}, while $Q$ denotes the number of grid evaluation points used to perform the minimization in \eqref{eq:mll_nlos_problem}. 
Consequently, when applied to all $N$ \gls{NLoS} paths, the total computational cost of the \gls{ML} \gls{NLoS} estimator is
$O\left(Q\sum_{n=1}^{N} K_n\right) = O(Q\cdot K_\Sigma)$. On the other hand,  the proposed RML \gls{NLoS} estimator, provided in \eqref{eq:rml_nlos}-\eqref{eq::upsilon} of Sec. \ref{sec:NLOS_RML}, eliminates the need for a grid search and, remarkably, reduces the overall asymptotic complexity to $O(K_\Sigma)$.
\\
\indent \emph{Complexity of Position Estimation via IWLS:} Each iteration of the IWLS position estimation algorithm requires the evaluation of the update rule in \eqref{eq::IWLS_update}, which entails matrix multiplications between the $N \times 2$ Jacobian matrix $\boldsymbol{H}^{[i]}$ 
and the $N \times N$ weighting matrix $\bm{W}$. 
This results in a per-iteration computational cost on the order of $O(N^2)$. Considering a number of iterations $I$ (upper bounded by $I_\text{max}$), the overall cost of the IWLS algorithm is $O(I\cdot N^2)$.
\\
\indent \emph{Complexity of \gls{OIRS} Adaptive Beam Steering:} The final \gls{OIRS} re-alignment step introduces only an additional $O(N)$ term, which arises from the computation of the updated unit normal vectors $\boldsymbol{o}_n$, according to \eqref{eq:oirsnorm}, and the corresponding \gls{OIRS} tilt angles $\alpha_n$ and $\beta_n$ using \eqref{eq:a_b_angles}.
\\
\indent \emph{Complexity of Iterative Position Estimate Refinement:} The PD position estimate obtained through the IWLS procedure can be further refined by iteratively reapplying the localization algorithm after all $N$ OIRSs have been adaptively steered toward the estimated PD direction. 
As discussed in Sec.~\ref{subsec::AlgDescri}, the iterative process terminates when an additional reconfiguration of the OIRS network 
no longer yields a significant improvement in the position estimate. The overall asymptotic computational complexity of the proposed iterative localization algorithm with adaptive beam steering is thus given by
\begin{equation}
O\big(M(K + K_\Sigma + I\cdot N^2)\big),
\end{equation}
with $M$ the total number of iterations. 
It is worth noting that the above expression highlights the low-complexity nature of the proposed approach: as it will be confirmed by the numerical results in Sec. \ref{sec:results}, the number of iterations $M$ required to refine the PD position estimate is very low (as small as two or three). As to the number of collected samples $K$ and $K_\Sigma$, they are also limited to a few hundreds, and the number of OIRSs $N$ --- with respect to which the complexity scales quadratically --- is clearly very modest. 
Conversely, if the ML NLoS estimator is employed, 
the expression of the overall asymptotic complexity becomes $O\big(M(K + Q\cdot K_\Sigma + I\cdot N^2)\big)$ and the number of grid points $Q$ becomes a dominant contributor to the overall computational burden, markedly increasing the total complexity.
}

\section{Cramér-Rao Lower Bound} \label{sec:bounds}
In this section, we derive the \gls{FIM} and the corresponding \gls{CRLB} for both i) \gls{LoS} and \gls{NLoS} distance estimation, and ii) \gls{PD} position estimation. These results lead to the definition of the \gls{DEB} and, subsequently, the \gls{PEB}. Both the \gls{DEB} and \gls{PEB} will be used in Sec. \ref{sec:results} as  performance benchmarks for the proposed  algorithms.

\subsection{Distance Error Bounds}
We start by deriving the DEB on the estimation of the \gls{LoS} distance $d$, knowing that any unbiased estimator $\hat{d}$ of $d$ has a \gls{MSE} lower bounded by \cite{kay1993fundamentals}
\begin{equation}
    \mathbb{E}[(\hat{d}-d)^2] \geq J_d^{-1},
\end{equation}
where $\mathbb{E}[\cdot]$ denotes the expectation operator. The Fisher information $J_d$ for the \gls{LoS} model is
\begin{align}\label{eq:fim_los}
 J_d &= \!-\mathbb{E}\!\left[\!\frac{\partial^2 \ln L^\LoS(\boldsymbol{\mu}_0 ; d)}{\partial d^2}\!\right]\!  \nonumber \\ &= \! \frac{K (m\!+\!3)^2\mathrm{SNR}_0(d) (\sigma^2_0(d) \!+\! b^2/2)}{d^2 \sigma^2_0(d)},
\end{align}
with $\mathrm{SNR}_0(d) \eqdef \bar{\mu}^2_0(d)/\sigma^2_0(d)$ the \gls{SNR} in the \gls{LoS}  case (whose expression easily follows from \eqref{eq:snr}, after setting $a_n =0, \forall n$). Accordingly, the DEB on $d$ is given by
\begin{equation}\label{DEB_d}
    \mathrm{DEB}(d) = \sqrt{J_d^{-1}}.
\end{equation}
From \eqref{eq:fim_los} and \eqref{DEB_d}, we observe that the DEB exhibits intuitive scaling behavior: it decreases as the number of observations $K$ increases or under higher SNR conditions, while it grows with increasing $d$ (considering a fixed transmit power $p$). 

Similarly, any unbiased estimator $\hat{d}_n$ of the \gls{NLoS} distance $d_n$ can achieve \gls{MSE} performance lower bounded by $\mathbb{E}[(\hat{d}_n-d_n)^2] \geq J_{d_n}^{-1}$, with the Fisher information for the \gls{NLoS} model
\begin{align}\label{eq:fim_nlos}
J_{d_n} &= -\mathbb{E}\left[\frac{\partial^2 \ln L^\NLoS(\bm{\chi}_n ; d_n) }{\partial {d_n}^2}\right]  \nonumber \\ 
&=  \frac{K_n \mathrm{SNR}_n(d_n)(s_n + 3d_n)^2(\sigma^2_n(d_n) + b^2/2)}{d^2_n(s_n + d_n)^2\sigma^2_n(d_n)},
\end{align}
where $\mathrm{SNR}_n(d_n) \eqdef \bar{\chi}^2_n(d_n)/\sigma^2_n(d_n)$ captures the SNR of the \gls{NLoS} reflection generated by the $n$-th \gls{OIRS} (obtained from  \eqref{eq:snr} by setting only the $n$-th coefficient $a_n = 1$, with all others set to zero, and after \gls{LoS} compensation, as per \eqref{eq:NLoS_contrib}). Accordingly, the DEB on $d_n$ is given by
\begin{equation}\label{DEB_dn}
    \mathrm{DEB}(d_n) = \sqrt{J_{d_n}^{-1}}.
\end{equation}
Despite its slightly more involved structure, \eqref{eq:fim_nlos} shares a similar scaling behavior with \eqref{eq:fim_los}, with the accuracy in the estimation of $d_n$ that tends to improve as $\mathrm{SNR}_n(d_n)$ increases or as more observations $K_n$ are available, whereas it degrades at a faster rate than \eqref{eq:fim_los} as $d_n$ increases, owing to the increased path loss over the reflected path. 

As anticipated, the DEB for both \gls{LoS} and \gls{NLoS} distance estimation is used to set the entries of the weighting matrix $\bm{W}$ in the \gls{IWLS} update rule \eqref{eq::IWLS_update}, specifically as $\text{w}_0 = 1/\mathrm{DEB}(\hat{d})$ and $\text{w}_n = 1/\mathrm{DEB}(\hat{d}_n)$, where $\hat{d}$ and $\{\hat{d}_n\}_{n=1}^N$ are the distances estimated via either the ML or RML approach.

\subsection{Position Error Bound}
Let $\boldsymbol{d} =  [d \ d_1 \ \cdots \ d_N]^T \in \mathbb{R}^{N+1\times1}$ denote the vector stacking the unknown \gls{LoS} and \gls{NLoS} distances. The joint likelihood function, accounting for all the information associated to the \gls{LoS} and \gls{NLoS} paths, can be expressed as 
\begin{equation}\label{eq:full_likelihood}
 L(\boldsymbol{\mu} ;\boldsymbol{d}) =  L^\LoS(\bm{\mu}_0; d) \times \prod _{n=1}^{N} L^\NLoS(\boldsymbol{\mu}_n ; d,d_n),
\end{equation}
where $\bm{\mu}_n = [\mu_{n,1} \ \cdots \ \mu_{n,K_n}]^\mathsf{T}$ is the vector stacking the $K_n$ samples of the signals received by the \gls{PD} via the $n$-th \gls{OIRS}, in the form of \eqref{eq:mnk}. 
The complete \gls{FIM} $\bm{J}_{\bm{d}} \in \mathbb{R}^{(N+1)\times (N+1)}$ in the distance domain is defined as
\begin{align}
 \label{eq:dist_fim}
 \!\!\!\!\bm{J}_{\boldsymbol{d}} &= -\mathbb{E}\left[\frac{\partial^2 \ln  L(\boldsymbol{\mu} ;\boldsymbol{d})}{\partial \boldsymbol{d} \partial \boldsymbol{d}^T}\right] 
\nonumber \\ & = -\mathbb{E}{\begin{bmatrix}
\frac{\partial^2 \ln  L(\boldsymbol{\mu} ;\boldsymbol{d})}{\partial d \partial d} & \frac{\partial^2 \ln  L(\boldsymbol{\mu} ;\boldsymbol{d})}{\partial d \partial d_1} & \cdots & \frac{\partial^2 \ln  L(\boldsymbol{\mu} ;\boldsymbol{d})}{\partial d \partial d_N}  \\
\frac{\partial^2 \ln  L(\boldsymbol{\mu} ;\boldsymbol{d})}{\partial d_1 \partial d}  & \frac{\partial^2 \ln  L(\boldsymbol{\mu} ;\boldsymbol{d})}{\partial d_1 \partial d_1}  & \cdots & \frac{\partial^2 \ln  L(\boldsymbol{\mu} ;\boldsymbol{d})}{\partial d_1 \partial d_N}  \\
\vdots  & \vdots  &  \ddots & \vdots  \\
\frac{\partial^2 \ln  L(\boldsymbol{\mu} ;\boldsymbol{d})}{\partial d_N \partial d}  & \frac{\partial^2 \ln  L(\boldsymbol{\mu} ;\boldsymbol{d})}{\partial d_N \partial d_1}  & \cdots  & \frac{\partial^2 \ln  L(\boldsymbol{\mu} ;\boldsymbol{d})}{\partial d_N \partial d_N} 
\end{bmatrix}}.
\end{align}
The expression for each entry of $\bm{J}_{\boldsymbol{d}}$ is given in Appendix \ref{sec:app_a}.

We now derive the FIM in the position domain
by means of a transformation from the vector of distances $\bm{d}$ to the vector of location parameters $\bm{\upsilon}= [u_x \ u_y]^T$. The \gls{FIM} of $\bm{\upsilon}$ follows by applying a $2 \times (N+1)$ transformation matrix $\bm{T}$ as
\begin{equation} \label{eq:peb}
 \bm{J}_{\bm{\upsilon}} = \bm{T}\bm{J}_{\boldsymbol{d}}\bm{T}^\mathsf{T},
\end{equation}
where
\begin{equation}
\!\!\boldsymbol{T}\! =\! \frac{\partial \boldsymbol{d}^T}{\partial \boldsymbol{\upsilon}}\! =\! {\begin{bmatrix}
{\partial d}/{\partial u_x} & {\partial d_1}/{\partial u_x} & \cdots & {\partial d_N}/{\partial u_x}  \\
{\partial d}/{\partial u_y} & {\partial d_1}/{\partial u_y} & \cdots & {\partial d_N}/{\partial u_y}
\end{bmatrix}}, 
\end{equation}
with
\begin{align}
& \frac{\partial d}{\partial u_x}= \frac{u_x-q_x}{\|\bm{u}-\bm{q}\|}, \qquad \frac{\partial d}{\partial u_y}= \frac{u_y-q_y}{\|\bm{u}-\bm{q}\|},\\
& \frac{\partial d_n}{\partial u_x}= \frac{u_x-r_{n,x}}{\|\bm{u}-\bm{r}_n\|}, \qquad \!\!\!\frac{\partial d_n}{\partial u_y}= \frac{u_y-r_{n,y}}{\|\bm{u}-\bm{r}_n\|}. 
\end{align}
Finally, the \gls{PEB} is given by 
\begin{equation}\label{eq::PEB}
\mathrm{PEB} = \sqrt{\mathrm{tr}\left(\bm{J}^{-1}_{\bm{\upsilon}}\right)},
\end{equation}
with $\mathrm{tr}(\cdot)$ the trace operator.

{\color{red}
}

\section{\ed{Simulation Results}}\label{sec:results}
In this section, we present simulation results to evaluate the performance of the proposed \gls{LoS} and \gls{NLoS} distance estimators --- both \gls{ML} and RML --- and of the complete iterative localization algorithm with adaptive beam steering,  in comparison with the theoretical bounds
derived from the Fisher information analysis conducted in Sec. \ref{sec:bounds}, {\ed as well as with the state-of-the-art direct ML localization approach.}
We consider an indoor localization scenario representative of a room with dimensions $5\times5\times3$\,m. A single \gls{LED}  is mounted on the ceiling at position $\bm{q} = \left[ 2.5 \ 2.5 \ 3 \right]^\mathsf{T}$\,m. Initially, the environment includes only one ($N = 1$) \gls{OIRS} installed on a wall at position $\bm{w}_1 = \left[ 2.5 \ 0 \ 1.5 \right]^\mathsf{T}$\,m. The \gls{OIRS} has dimensions of $1\times1$\,m and is initially oriented such that its normal vector $\bm{o}_1$, as defined in \eqref{eq:oirsnorm}, is aligned with the normal to the $x$–$z$ wall plane, i.e., $\alpha_1 = 0$ and $\beta_1 = 0$.
Unless otherwise stated, all the results in this section are averaged over $10^4$ independent Monte Carlo trials, using the following setting of system parameters: $\psi=70^\circ$, $\Theta_{1/2}=70^\circ$, $f = 1.5$, $\rho = 0.95$, $A = \SI{0.2}{cm^2}$, $R = \SI{0.54}{A/W}$, $T = 1$, $I_1 = \SI{5}{pA}$, $I_2 = 0.562$, $I_3 = 0.0868$, $\tau = \SI{295}{K}$, $G_0=10$, $\varrho=1.5$, $g = \SI{30}{mS}$, $\nu = \SI{112}{pF/cm^2}$ and $B=\SI{5}{MHz}$~\cite{Zhang2014Theoretical, komine2004fundamental}.

\subsection{Distance Estimation Performance}\label{results:distance}
The analysis begins by evaluating the performance of the \gls{LoS}/\gls{NLoS} distance estimators proposed in Sec. \ref{sec:los_est} and \ref{sec:nlos_est}.
\subsubsection{LoS Distance Estimation Analysis} 
\begin{figure}
    \setlength{\plotWidth}{0.85\linewidth}
    \setlength{\plotHeight}{0.48\linewidth}
%
%
\makeatletter
\newcommand\notsotiny{\@setfontsize\notsotiny\@vipt\@viipt}
\makeatother

\pgfplotsset{every axis/.append style={
  label style={font=\footnotesize},
  legend style={font=\notsotiny},
  tick label style={font=\scriptsize},
}}

\definecolor{mycolor1}{rgb}{0.00000,0.44700,0.74100}%
\definecolor{mycolor2}{rgb}{0.63529,0.07843,0.18431}%
\definecolor{mycolor3}{rgb}{0.46667,0.67451,0.18824}%
\definecolor{mycolor4}{rgb}{0.00000,0.44706,0.74118}%
\hspace{0cm}\begin{tikzpicture}

\begin{axis}[%
width=\plotWidth,
height=\plotHeight,
at={(0\plotWidth,0\plotWidth)},
scale only axis,
axis line style= thick,
xlabel shift={-4pt},
ylabel shift={-4pt}, 
xmin=10,
xmax=25,
xlabel style={font=\color{white!15!black}},
xlabel={\footnotesize SNR [dB]},
ymode=log,
ymin=0.01,
ymax=0.4,
yminorticks=true,
scaled y ticks=false,  
ylabel style={font=\color{white!15!black}},
ylabel={\footnotesize RMSE [m]},
yticklabel style = {
           /pgf/number format/fixed,
           /pgf/number format/precision = 2},
axis background/.style={fill=white},
xmajorgrids,
ymajorgrids,
yminorgrids,
grid style={dashdotted,white!85!black},
legend style={at={(0,0)}, anchor=south west, legend cell align=left, align=left, legend columns=3, draw=white!15!black}
]
\addplot [color=mycolor1, dashed, line width=0.75pt, mark size=1.8pt, mark=square, mark options={solid, mycolor1}]
  table[row sep=crcr]{%
10	0.375951205806346\\
11	0.310130412493494\\
12	0.268768031388489\\
13	0.205135701526903\\
14	0.181252471136308\\
15	0.156567914273116\\
16	0.139467241236522\\
17	0.121000005612209\\
18	0.107630131883646\\
19	0.094900885193423\\
20	0.084642671793312\\
21	0.0743191285581238\\
22	0.0663817768247938\\
23	0.0583998443549598\\
24	0.052452967633487\\
25	0.0470548638815054\\
};
\addlegendentry{$\hat{d}^\RML$}

\addplot [color=mycolor4, line width=0.75pt, mark size=3.8pt, mark=x, mark options={solid, mycolor4}]
  table[row sep=crcr]{%
10	0.376259692621479\\
11	0.310593212290416\\
12	0.269019717304868\\
13	0.205177824367905\\
14	0.181284528749541\\
15	0.156592543567466\\
16	0.13948705456615\\
17	0.12101497695481\\
18	0.107641700434785\\
19	0.0949092639330622\\
20	0.0846497147027478\\
21	0.0743245031227331\\
22	0.0663867417582645\\
23	0.0584034783998087\\
24	0.0524556212699858\\
25	0.0470571697685753\\
};
\addlegendentry{$\hat{d}^\ML$}

\addplot [color=mycolor4, dashdotted, line width=0.75pt]
  table[row sep=crcr]{%
10	0.260194138136051\\
11	0.231898269725306\\
12	0.206679550458915\\
13	0.184203343254206\\
14	0.164171402497315\\
15	0.146317916505089\\
16	0.130405980365551\\
17	0.116224452353654\\
18	0.103585152208871\\
19	0.0923203640974677\\
20	0.0822806111252884\\
21	0.0733326718701878\\
22	0.0653578126146993\\
23	0.0582502118217421\\
24	0.0519155559471183\\
25	0.0462697879560822\\
};
\addlegendentry{$\text{DEB}(d)$ $\color{mycolor1}\bm{[K = 1]}$}

\addplot [color=mycolor2, dashed, line width=0.75pt, mark size=1.8pt, mark=square, mark options={solid, mycolor2}]
  table[row sep=crcr]{%
10	0.158236573281176\\
11	0.141827745383551\\
12	0.125769946643091\\
13	0.111181007210848\\
14	0.0991258345128163\\
15	0.0860404904285613\\
16	0.07659555319985\\
17	0.0674401935489964\\
18	0.0606981285585946\\
19	0.0539478872138353\\
20	0.0480459263522904\\
21	0.0422619983483415\\
22	0.037961633866923\\
23	0.0338137412845864\\
24	0.0300661672811264\\
25	0.0268210495581979\\
};
\addlegendentry{$\hat{d}^\RML$}

\addplot [color=mycolor2, line width=0.75pt, mark size=3.8pt, mark=x, mark options={solid, mycolor2}]
  table[row sep=crcr]{%
10	0.158253513558159\\
11	0.141840051238638\\
12	0.125778264684921\\
13	0.111189753503744\\
14	0.099131484438783\\
15	0.0860441236127296\\
16	0.076598301684381\\
17	0.0674428075951833\\
18	0.0606998587511235\\
19	0.0539495583742894\\
20	0.0480469397312037\\
21	0.0422631088030093\\
22	0.0379625395225816\\
23	0.0338141420886187\\
24	0.0300663012336757\\
25	0.026821619152145\\
};
\addlegendentry{$\hat{d}^\ML$}

\addplot [color=mycolor2, dashdotted, line width=0.75pt]
  table[row sep=crcr]{%
10	0.150223155694412\\
11	0.133886528450514\\
12	0.119326494093445\\
13	0.106349849813445\\
14	0.0947844034250633\\
15	0.0844766884814781\\
16	0.0752899278679879\\
17	0.0671022188527992\\
18	0.0598049155118402\\
19	0.0533011870633573\\
20	0.0475047329822722\\
21	0.0423386378446474\\
22	0.0377343507067417\\
23	0.0336307754756355\\
24	0.0299734602011978\\
25	0.026713874531791\\
};
\addlegendentry{$\text{DEB}(d)$ $\color{mycolor2}\bm{[K = 3]}$}

\addplot [color=mycolor3, dashed, line width=0.75pt, mark size=1.8pt, mark=square, mark options={solid, mycolor3}]
  table[row sep=crcr]{%
10	0.121336230426662\\
11	0.108385640104378\\
12	0.0954835338346263\\
13	0.0824081833627948\\
14	0.0751908885048061\\
15	0.0661385623249695\\
16	0.0594862233578683\\
17	0.0519502155131387\\
18	0.0467643730999038\\
19	0.0410440332053606\\
20	0.0369050222971287\\
21	0.0330490567159729\\
22	0.0294256512528574\\
23	0.0259979582077475\\
24	0.0234580649178703\\
25	0.020800710635354\\
};
\addlegendentry{$\hat{d}^\RML$}

\addplot [color=mycolor3, line width=0.75pt, mark size=3.8pt, mark=x, mark options={solid, mycolor3}]
  table[row sep=crcr]{%
10	0.121344048207885\\
11	0.10839126518058\\
12	0.0954875205618111\\
13	0.0824115963332383\\
14	0.0751937390856895\\
15	0.0661404513577146\\
16	0.0594880093247671\\
17	0.0519520909469228\\
18	0.0467657722526877\\
19	0.0410446333029145\\
20	0.0369055157862614\\
21	0.0330492374690897\\
22	0.0294259994610044\\
23	0.0259983221890157\\
24	0.0234584017442721\\
25	0.0208008372908683\\
};
\addlegendentry{$\hat{d}^\ML$}

\addplot [color=mycolor3, dashdotted, line width=0.75pt]
  table[row sep=crcr]{%
10	0.116362356043836\\
11	0.103708058994073\\
12	0.0924299048770463\\
13	0.0823782394398266\\
14	0.0734196831890951\\
15	0.0654353615263034\\
16	0.058319327353975\\
17	0.0519771552220914\\
18	0.0463246883597398\\
19	0.0412869219658938\\
20	0.0367970079412741\\
21	0.0327953678546853\\
22	0.0292289023734322\\
23	0.0260502866674354\\
24	0.02321734243749\\
25	0.0206924782348602\\
};
\addlegendentry{$\text{DEB}(d)$ $\color{mycolor3}\bm{[K = 5]}$}

\end{axis}

\end{tikzpicture}%
    \caption{RMSE of the ML and RML \gls{LoS} distance estimators, along with the corresponding DEB, as a function of the SNR, for different values of $K$.}
    \label{fig:fig2_tikz}
\end{figure}
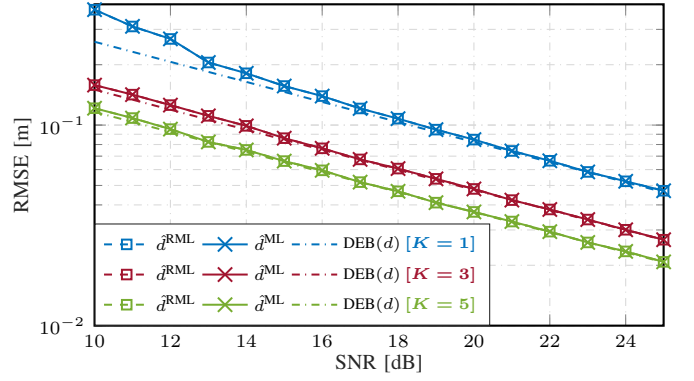

Fig. \ref{fig:fig2_tikz} illustrates the performance of the \gls{ML} and RML \gls{LoS} estimators, given  in \eqref{eq:ml_los} and \eqref{eq:rml_los}, respectively, for a \gls{PD} located at $\bm{u} = [3 \ 3 \ 0]^\mathsf{T}$. The results are presented in terms of the \gls{RMSE} as a function of the \gls{SNR}, considering three different sample sizes, namely $K \in \{1, 2, 5\}$, and are compared against the \gls{DEB} derived in \eqref{DEB_d}. The topmost (blue) curves in Fig. \ref{fig:fig2_tikz}  reveal that both \gls{ML} and RML achieve satisfactory estimation accuracy over all the considered span of SNR and, notably, attain the theoretical bound already at $\text{SNR} \approx \SI{15}{dB}$, despite relying on only a single observation (i.e., $K=1$). As the number of samples $K$ increases, both estimators exhibit enhanced accuracy and approach the \gls{DEB} at progressively lower \gls{SNR} values. Specifically, for $K=5$, the estimators attain the \gls{DEB} at $\text{SNR} = \SI{10}{dB}$  and provide a RMSE close to cm-level for high SNR values. Overall, the RMSE is reduced by approximately 58\% for $K=3$ and by 67.7\% for $K=5$, compared to the single-sample case ($K=1$). On the one hand, these results show that the proposed estimators  provide accurate \gls{LoS} distance estimation even with a very small number of samples $K$.
On the other hand, it is worth noting that the RML estimator achieves performance comparable to that of the ML estimator, despite relying on a sub-optimal processing --- resulting from the relaxation introduced in the signal model, as discussed in Sec. \ref{sec::RML_LOS} ---  based only on the sample mean $S_1$ as a sufficient statistic.
\subsubsection{NLoS Distance Estimation Analysis} 
\begin{figure}
    \centering
    \setlength{\plotWidth}{0.85\linewidth}
    \setlength{\plotHeight}{0.48\linewidth}
\input{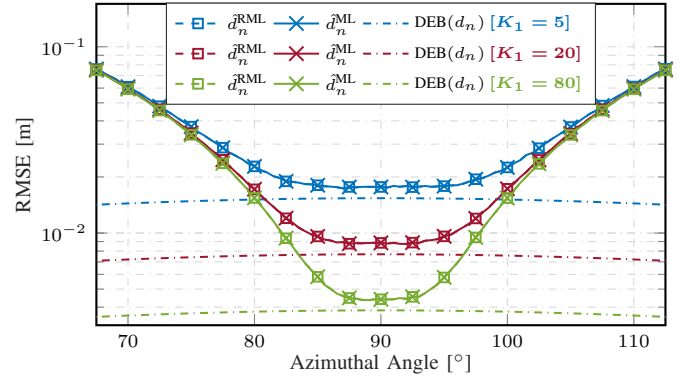}
    \caption{RMSE of the ML and RML NLoS distance estimators, along with the corresponding DEB, for different values of $K_1$, as a function of the azimuthal angle representing the change in the \gls{PD} position along a circular arc.}
    \label{fig:fig3_tikz}
\end{figure}

We now analyze the performance of the ML and RML \gls{NLoS} estimators given in \eqref{eq:mll_nlos_problem} and \eqref{eq:rml_nlos}. As discussed in Sec. \ref{sec:algorithm}, both estimators are evaluated using $\bm{w}_1$ in place of $\bm{r}_1$ to overcome the need of any prior information on $\bm{u}$. To analyze the impact of this misalignment, we let the \gls{PD} having a fixed distance of \SI{2.5}{m} from the \gls{OIRS} center --- so that any impact on the performance can be mainly attributed to the pointing mismatch --- and then vary its position along a circular arc. Fig. \ref{fig:fig3_tikz} shows the \gls{RMSE} as a function of the azimuth angle (capturing the variation in the \gls{PD} position along the circular arc), for different values of collected samples $K_1 \in \{5,20,80\}$. The results are obtained for a fixed transmit power $p_{[\text{lm}]} = \SI{1000}{lm}$ (converted to photometric power in lumens), keeping the initial orientation of the \gls{OIRS}, namely $\alpha_1 = 0$ and $\beta_1 = 0$. 

The RMSE curves in Fig. \ref{fig:fig3_tikz} indicate that both the ML and RML \gls{NLoS} distance estimators closely approach the \gls{DEB} for an azimuthal angle equal to $\SI{90}{\degree}$, which corresponds to the configuration where the \gls{OIRS} is perfectly aligned with the \gls{PD} located at $\bm{u}=[2.5,2.5,0]$ (i.e., center of the room), when  $\bm{r}_1 = \bm{w}_1$ and no misalignment is present. Remarkably, this behavior is observed even for the most challenging case among the three, namely when only $K_1 = 5$ samples are available. Although the DEB increases due to the reduced sample size (as evident from \eqref{eq:fim_nlos}-\eqref{DEB_dn}), both the algorithms still attain it, highlighting their good performance even for a very small sample size.
As the azimuthal angle deviates from the optimal alignment, the mismatch in the \gls{OIRS} orientation leads to a gradual increase in the gap between the RMSE and the \gls{DEB}. Nonetheless, it is notable to observe that the estimators maintain RMSE values close to the DEB within a certain angular interval and, overall, exhibit an estimation error that remains limited to the order of a few centimeters even for \gls{OIRS} orientation errors as large as \SI{20}{\degree}.
From these results, we can outline that the estimators exhibit a good robustness against \gls{OIRS} misalignment conditions;  consequently, the resulting \gls{NLoS} distance estimates can be effectively used (together with an estimate of the \gls{LoS} distance) to compute an estimate of the \gls{PD} position via the IWLS algorithm, which is subsequently refined through the iterative localization process proposed in  Sec. \ref{sec:algorithm}.
Fig. \ref{fig:fig3_tikz} further highlights that, also in the \gls{NLoS} case, the proposed RML estimator guarantees an accuracy practically identical to that of the \gls{ML} estimator. This is a remarkable outcome, as the RML estimator admits a closed-form solution (given in \eqref{eq:rml_nlos}), in contrast to the \gls{ML} estimator, which requires a numerical grid search to be implemented (as per \eqref{eq:mll_nlos_problem}).

\begin{figure}
    \centering
    \setlength{\plotWidth}{0.85\linewidth}
    \setlength{\plotHeight}{0.48\linewidth}
    \input{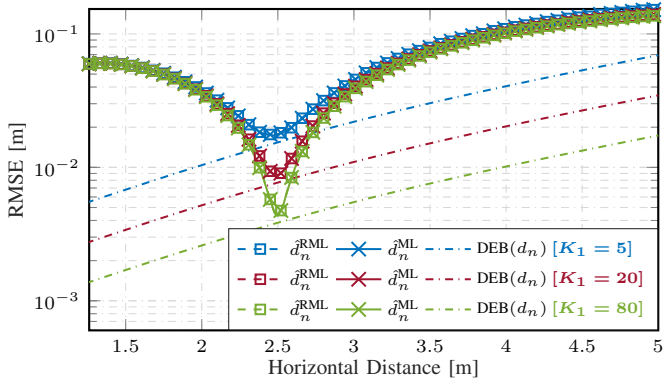}
    \caption{RMSE of the ML and RML NLoS distance estimators, along with the corresponding DEB, for different values of $K_1$, as a function of the horizontal distance between the \gls{PD} and the \gls{OIRS}.}
    \label{fig:fig4_tikz}
\end{figure}

To corroborate the analysis on the impact of the \gls{OIRS} misalignment, we now consider a different setup where the azimuthal angle is fixed at $\SI{90}{\degree}$, while the horizontal distance between the \gls{OIRS} center and the \gls{PD} is varied by progressively moving the position of the latter away along the $y$-axis. Fig. \ref{fig:fig4_tikz} reports the \gls{RMSE} curves as a function of the horizontal distance, under the same settings of $p$ and $K_1$ used in Fig. \ref{fig:fig3_tikz}. In line with the results presented in Fig. \ref{fig:fig3_tikz}, the RMSEs of both estimators closely approach the DEB at a horizontal distance of \SI{2.5}{m}, where the \gls{OIRS} is perfectly aligned with the \gls{PD}. A noteworthy trend emerges as the horizontal distance decreases below \SI{2.5}{m}: while the DEB continues to decrease --- owing to improved \gls{SNR} conditions resulting from reduced propagation loss, given the fixed transmit power $p$ --- the RMSEs of the estimators instead exhibit a progressive increase. This apparent discrepancy underscores the influence of the \gls{OIRS} orientation mismatch in the \gls{NLoS} estimators: although the received signal becomes stronger, the growing impact of the \gls{OIRS} misalignment leads to increased RMSEs, practically overriding the  benefits of a higher \gls{SNR}. On the other hand, for horizontal distances greater than \SI{2.5}{m}, the performance is affected by both reduced \gls{SNR} and increased \gls{OIRS} misalignment, resulting in a general upward trend in the RMSE curves.

\begin{figure}
    \setlength{\plotWidth}{0.85\linewidth}
    \setlength{\plotHeight}{0.495\linewidth}
\input{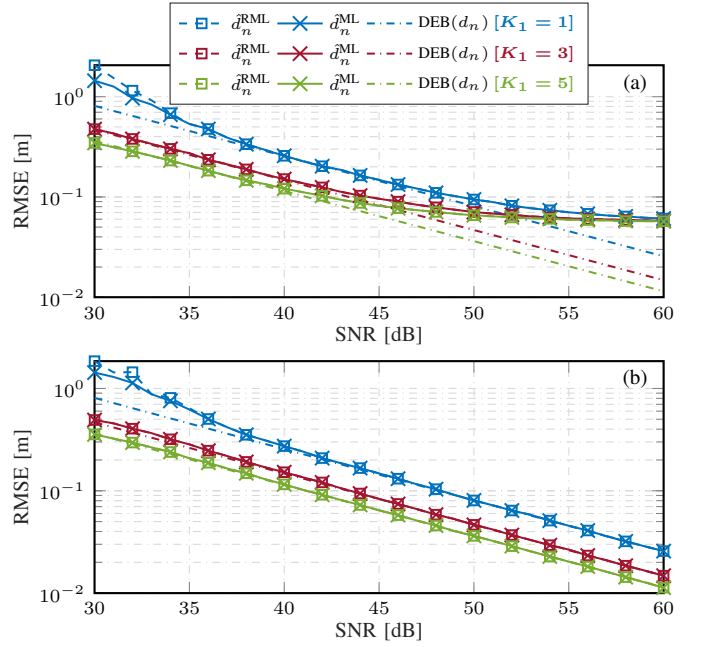}
    \caption{RMSE of ML and RML NLoS distance estimators, compared against the DEB, for different values of $K_1$, as a function of the SNR with (a) initial orientation of \gls{OIRS} and (b) \gls{OIRS} perfectly steered toward the \gls{PD}.}
    \label{fig:fig5_tikz}
\end{figure}

To conclude the analysis, in Fig. \ref{fig:fig5_tikz} we illustrate the trend of the \gls{RMSE} as a function of the SNR, for a fixed \gls{PD} position $\bm{u}=[3\ 3\ 0]^\mathsf{T}$, and considering the same lower number of samples as for the \gls{LoS} case, i.e., $K_1 \in \{1,3,5\}$. 
More specifically, the results in Fig. \ref{fig:fig5_tikz}(a) are obtained by keeping fixed the initial orientation of the \gls{OIRS} (namely $\alpha_1 = 0$ and $\beta_1 = 0$), which thus gives rise to the presence of a mismatch in the \gls{NLoS} estimators. As evident from the topmost (blue) curves, in the challenging scenario with only $K_1 = 1$ sample, both ML and RML estimators exhibit relatively higher RMSEs, with the former offering slightly better accuracy over the latter, though at the cost of a greater computational complexity. From $\text{SNR} \geq \SI{35}{dB}$, the RML estimator provides an accuracy close to that of the ML, and both estimators closely approach the DEB.  Notably, increasing the number of samples to just $K_1 = 3$ proves sufficient to close the initial gap between the RML and ML estimators. However, for SNR values beyond \SI{45}{dB}, the estimators incur a performance saturation, regardless of the number of processed samples $K_1$. This plateau arises from the inherent model mismatch in the \gls{NLoS} estimators, due to the use of $\bm{w}_1$ instead of the actual $\bm{r}_1$. As commonly observed in mismatched or misspecified estimators (e.g., \cite{MissEstim2},\cite{MismEstim1}), such a discrepancy imposes a fundamental limit on the achievable RMSE, independent of the available SNR or sample size. As further confirmation, in Fig. \ref{fig:fig5_tikz}(b) we report the RMSE curves when the $\alpha_1$ and $\beta_1$ angles of the \gls{OIRS} are adjusted to steer the beam towards the \gls{PD} position. As it can be noticed, the plateau disappears and the RMSEs keep decreasing as the SNR increases, with the estimators achieving cm-level accuracy in the high SNR regime.
This analysis motivates the need for an algorithm as the one proposed in Sec. \ref{sec:algorithm}, which aims to iteratively mitigate the misalignment via an adaptive beam steering strategy and whose  performance is analyzed in the next section.

\subsection{Position Estimation Performance}\label{subsec:sim_pos_est}
\begin{figure}
    \centering
    \setlength{\plotWidth}{0.85\linewidth}
    \setlength{\plotHeight}{0.37\linewidth}
%
%
\makeatletter
\newcommand\notsotiny{\@setfontsize\notsotiny\@vipt\@viipt}
\makeatother

\pgfplotsset{every axis/.append style={
  label style={font=\footnotesize},
  legend style={font=\notsotiny},
  tick label style={font=\scriptsize},
}}

\definecolor{mycolor1}{rgb}{0.00000,0.44700,0.74100}%
\definecolor{mycolor2}{rgb}{0.63529,0.07843,0.18431}%
\definecolor{mycolor3}{rgb}{0.46667,0.67451,0.18824}%
\definecolor{mycolor4}{rgb}{0.00000,0.44706,0.74118}%
\hspace{0cm}\begin{tikzpicture}

\begin{axis}[%
width=\plotWidth,
height=\plotHeight,
at={(0\plotWidth,0\plotWidth)},
scale only axis,
axis line style= thick,
xmin=1,
xmax=5,
xtick={1,2,3,4,5},
xlabel shift={-4pt},
ylabel shift={-4pt}, 
xlabel style={font=\color{white!15!black}},
xlabel={\footnotesize Iterations $M$},
ymode=log,
ymin=0.0008,
ymax=0.07,
yminorticks=true,
ylabel style={font=\color{white!15!black}},
ylabel={\footnotesize RMSE [m]},
yticklabel style = {
           /pgf/number format/fixed,
           /pgf/number format/precision = 2},
axis background/.style={fill=white},
xmajorgrids,
ymajorgrids,
yminorgrids,
grid style={dashdotted,white!85!black},
legend style={at={(1,1)}, anchor=north east, legend cell align=left, align=left, legend columns=3, draw=white!15!black}
]
\addplot [color=IEEEdarkorange, dashed, line width=1.2pt, mark size=1.8pt, mark=square, mark options={solid, IEEEdarkorange}]
  table[row sep=crcr]{%
1	0.0595180930244224\\
2	0.00670740749614944\\
3	0.00479375601492261\\
4	0.00483529840554022\\
5	0.00480206781680102\\
};
\addlegendentry{ILS}

\addplot [color=IEEEdarkorange, line width=1.2pt, mark size=1.8pt, mark=o, mark options={solid, IEEEdarkorange}]
  table[row sep=crcr]{%
1	0.00543825774174782\\
2	0.00254010433509858\\
3	0.00252162499269895\\
4	0.00251175471104411\\
5	0.00253289088565172\\
};
\addlegendentry{IWLS}

\addplot [color=IEEEdarkorange, dashdotted, line width=1.2pt]
  table[row sep=crcr]{%
1	0.00230799669360444\\
2	0.00230799669360444\\
3	0.00230799669360444\\
4	0.00230799669360444\\
5	0.00230799669360444\\
};
\addlegendentry{PEB \color{IEEEdarkorange}$\bm{[p_{[\text{lm}]} = 1000 \,\text{lm}]}$}

\addplot [color=IEEEslategray, dashed, line width=1.2pt, mark size=1.8pt, mark=square, mark options={solid, IEEEslategray}]
  table[row sep=crcr]{%
1	0.0593688719072035\\
2	0.00514897607245075\\
3	0.00211918332598089\\
4	0.00208362285142806\\
5	0.00208454002832099\\
};
\addlegendentry{ILS}

\addplot [color=IEEEslategray, line width=1.2pt, mark size=1.8pt, mark=o, mark options={solid, IEEEslategray}]
  table[row sep=crcr]{%
1	0.00483528952727081\\
2	0.00110548179717837\\
3	0.00109538879544199\\
4	0.00109592233737286\\
5	0.00108743179698267\\
};
\addlegendentry{IWLS}

\addplot [color=IEEEslategray, dashdotted, line width=1.2pt]
  table[row sep=crcr]{%
1	0.00100799879343269\\
2	0.00100799879343269\\
3	0.00100799879343269\\
4	0.00100799879343269\\
5	0.00100799879343269\\
};
\addlegendentry{PEB \color{IEEEslategray}$\bm{[p_{[\text{lm}]} = 3000 \,\text{lm}]}$}

\end{axis}
\end{tikzpicture}%
    \caption{RMSE of the proposed localization algorithm with either IWLS or ILS in comparison with the PEB, for two different levels of transmit power, as a function of the number of iterations $M$ for \gls{OIRS} adaptive beam steering.}
    \label{fig:fig7_tikz}
\end{figure}
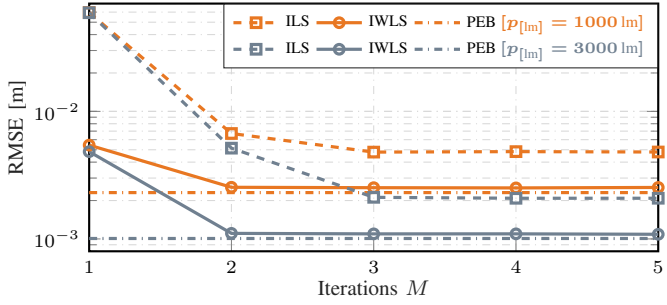

We now assess the performance of the proposed iterative localization algorithm with adaptive beam steering. Given its satisfactory performance, we use the RML approach to estimate the \gls{NLoS} distances, thus significantly reducing the complexity of the localization process thanks to its closed-form expression.
For the analyses, we consider an initial setup with $N=4$ \glspl{OIRS}, each mounted at the center of a wall, i.e., $\bm{w}_1 = \left[2.5 \ 0 \ 1.5 \right]^\mathsf{T}$, $\bm{w}_2 = \left[2.5 \ 5 \ 1.5 \right]^\mathsf{T}$, $\bm{w}_3 = \left[0 \ 2.5 \ 1.5 \right]^\mathsf{T}$, and $\bm{w}_4 = \left[0 \ 5 \ 1.5 \right]^\mathsf{T}$.
The number of collected samples is set to $K=50$ for the \gls{LoS} distance estimation and to $K_n = 100, \forall n=1,\ldots,N$, for the \gls{NLoS} distance estimation, respectively. For the sake of comparison, alongside the theoretical lower bound given by the PEB (derived in \eqref{eq::PEB}), we also include in the analyses a variant of the proposed algorithm that employs classical \gls{ILS} instead of the \gls{IWLS} approach, readily obtained by using $\bm{W} = \bm{I}$ in \eqref{eq:WLS} and \eqref{eq::IWLS_update}. Both ILS and IWLS algorithms are executed with a maximum number of iterations set to $I_{\max} = 100$ and a convergence threshold of $\epsilon = 10^{-6}$ as stopping criteria. {\ed For a thorough assessment, we add to the analysis also the direct ML localization approach, which represents the state-of-the-art for the considered problem \cite{TARHAN2025104799}. To adapt such an approach to our scenario, we consider the observation vectors collected by the PD under LoS and NLoS conditions, $\bm{\mu}_0, \bm{\mu}_1,\ldots,\bm{\mu}_N$, and denote by $f(\mu_{0,k};\bm{\upsilon})$ and $f(\mu_{n,k};\bm{\upsilon})$ the pdfs of the corresponding LoS and NLoS samples. These pdfs have the same functional form as $f(\mu_{0,k}; d)$ and $f(\mu_{n,k}; d)$, but their means and variances are parameterized through the LoS and NLoS distances, $d(\bm{\upsilon})$ and $d_n(\bm{\upsilon})$, expressed as functions of the unknown PD position $\bm{\upsilon} = [u_x\ u_y]^\mathsf{T}$. The joint likelihood function for the direct ML estimator can be thus written as
\begin{equation}
  L^{\text{dir}}(\bm{\mu}_{0},\bm{\mu}_{1},\dots,\bm{\mu}_{N}; \bm{\upsilon})
  = \prod^{K}_{k = 1} f(\mu_{0,k} ; \bm{\upsilon}) \prod_{n=1}^N \prod^{K_{n}}_{k_n = 1} f(\mu_{n,k_n} ; \bm{\upsilon}),
  \label{eq:lh_direct_compact}
\end{equation}
and, accordingly, the direct ML  estimate is obtained as
\begin{equation}
  \hat{\bm{\upsilon}}^{\text{\tiny ML}} 
  = \argmin_{\bm{\upsilon}}\, \mathcal{L}^{\text{dir}}(\bm{\mu}_{0},\bm{\mu}_{1},\dots,\bm{\mu}_{N}; \bm{\upsilon}),
  \label{eq:ml_direct_compact}
\end{equation}
with $\mathcal{L}^{\text{dir}}(\cdot) \triangleq -\ln L^{\text{dir}}(\cdot)$.  
Due to the highly non-convex nature of $\mathcal{L}^{\text{dir}}(\cdot)$, $\hat{\bm{\upsilon}}^{\text{\tiny ML}}$ must be computed via a 2D exhaustive grid search over the localization area.}

\begin{figure}
    \setlength{\plotWidth}{0.88\linewidth}
    \setlength{\plotHeight}{0.37\linewidth}
\input{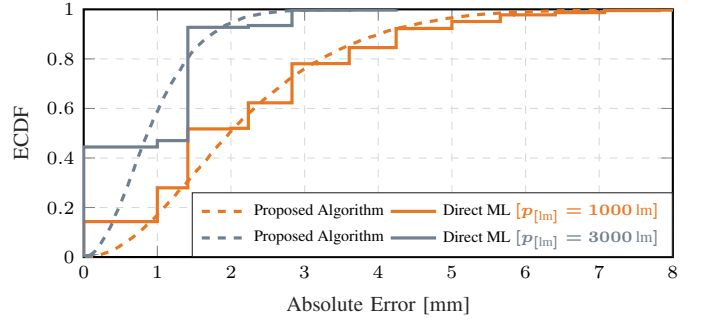}
    \caption{\ed ECDFs of the absolute position estimation error obtained with the proposed iterative algorithm and the direct ML estimator,  for two different levels of transmit power.}
    \label{fig:ECDF_comparison}
\end{figure}

Fig.~\ref{fig:fig7_tikz} illustrates the \gls{RMSE} of the proposed localization algorithm as a function of the number of iterations $M$ used to reconfigure the \gls{OIRS} network via the proposed adaptive beam steering strategy. The results are shown for two different transmit power levels, $p_{[\text{lm}]} \in \{1000, 3000\}\si{lm}$, and are compared against the \gls{PEB}. In agreement with the distance estimation results in Sec. \ref{results:distance}, both ILS and IWLS suffer from the initial \gls{OIRS} misalignment at the first iteration, yielding comparable RMSEs that deviate from the PEB. Notably, by the second iteration, the IWLS rapidly converges toward the PEB, significantly outperforming the ILS, which instead stabilizes only from the third iteration, but at RMSE levels approximately 48\% higher. These results highlight the satisfactory performance of the proposed IWLS, not only in terms of improved \gls{PD} position estimation --- reaching mm-level accuracy for either lower or higher values of $p_{[\text{lm}]}$ --- but also in terms of a more effective re-orientation of the \gls{OIRS} network toward the \gls{PD} within just $M = 2$ iterations, ensuring near-optimal localization performance with minimal overhead. 

{\ed We now conduct a comparison between the proposed iterative localization algorithm with adaptive beam steering and the direct ML estimator in \eqref{eq:lh_direct_compact}-\eqref{eq:ml_direct_compact}, which represents the state-of-the-art algorithm \cite{TARHAN2025104799}. In Fig. \ref{fig:ECDF_comparison}, we report the complete error distribution through the ECDF curves of the estimation errors for the PD location $\bm{\upsilon}$, achieved by the proposed} {\ed algorithm --- using the ML LoS and RML NLoS distance estimators, respectively, together with the IWLS procedure for the subsequent position estimation and $M = 2$ iterations for refinement --- and by the direct ML estimator, for two different transmit power levels. 
We can observe that both the proposed algorithm and the direct ML estimator consistently provide mm-level accuracy, with the ECDFs of the direct ML presenting a stepped shape due to the discrete nature of its search space. Remarkably, the ECDFs of the proposed algorithm are practically overlapped with those of the direct ML, demonstrating that the proposed iterative localization framework can attain nearly identical performance but at an extremely reduced cost, as it will be shown in Sec. \ref{results:runtimes}.}

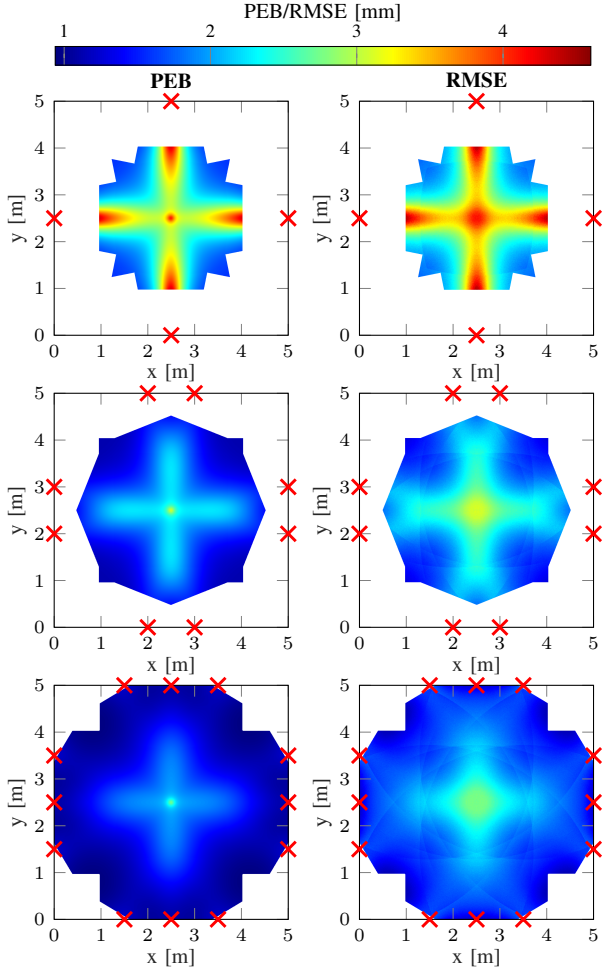
\begin{figure}
    \centering
    \setlength{\plotWidth}{0.76\linewidth}
    \setlength{\plotHeight}{0.36\linewidth}
    \makeatletter
\newcommand\notsotiny{\@setfontsize\notsotiny\@vipt\@viipt}
\makeatother

\pgfplotsset{every axis/.append style={
  label style={font=\footnotesize},
  legend style={font=\notsotiny},
  tick label style={font=\scriptsize},
}}
\begin{tikzpicture}

\begin{axis}[%
width=0.460\plotWidth,
height=0.460\plotWidth,
at={(0\plotWidth,1.148\plotWidth)},
scale only axis,
point meta min=0,
point meta max=0.993,
xmin=0,
xmax=5,
xlabel shift={-4pt},
ylabel shift={-4pt}, 
xlabel style={font=\color{white!15!black}},
xlabel={\footnotesize x [m]},
ymin=0,
ymax=5,
ylabel style={font=\color{white!15!black}},
ylabel={\footnotesize y [m]},
xtick={0,1,2,3,4,5}, 
ytick={0,1,2,3,4,5}, 
axis on top=true,
title={\footnotesize \textbf{PEB}},
title style={at={(0.5,0.95)}, anchor=south}
]
\addplot [forget plot] graphics [xmin=-0.03125, xmax=5.03125, ymin=-0.03125, ymax=5.03125] {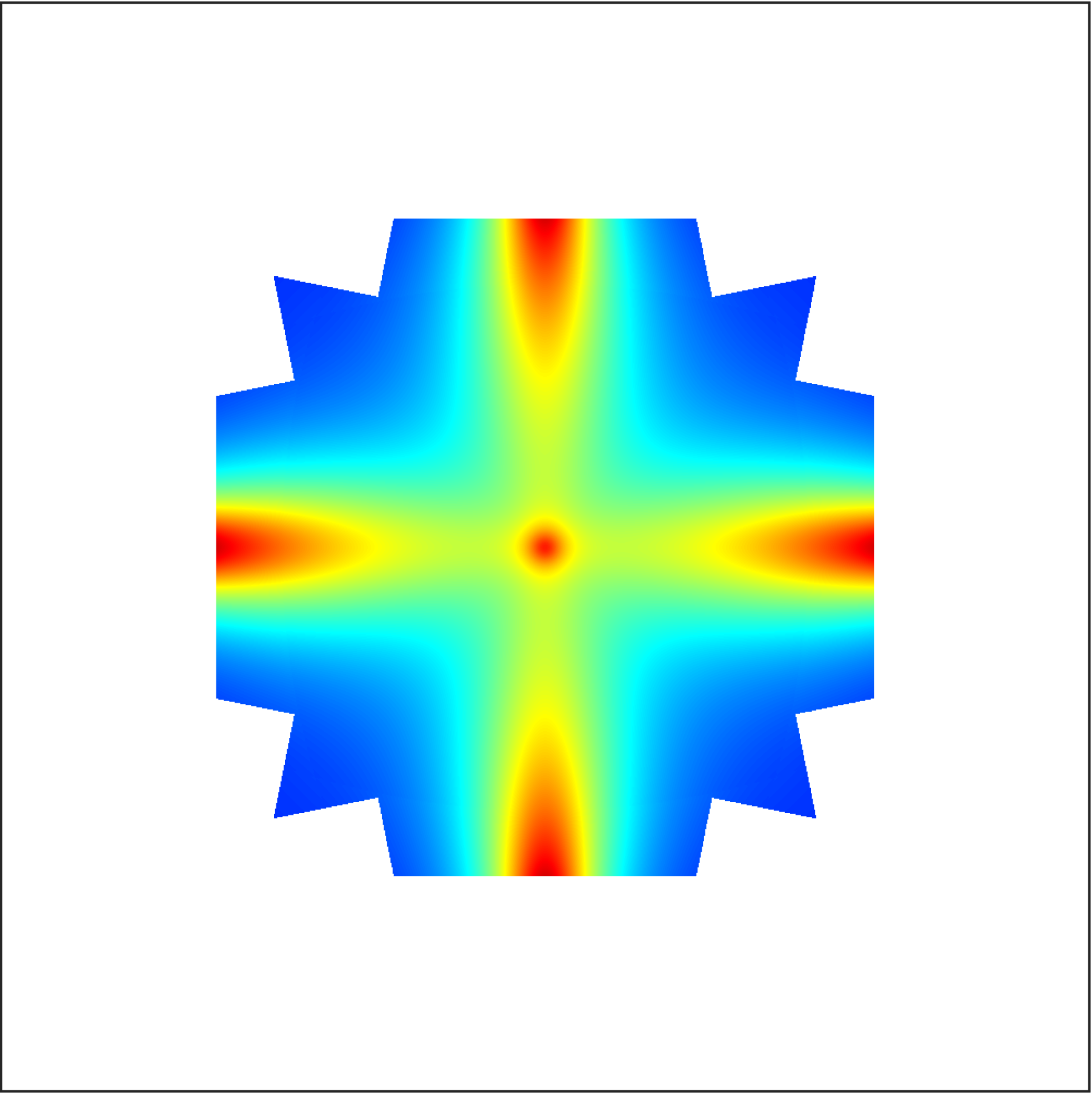};
\addplot [red, mark=x, mark size=4pt, mark options={line width=1.1pt}] coordinates {(0,2.5)};
\addplot [red, mark=x, mark size=4pt, mark options={line width=1.1pt}] coordinates {(5,2.5)};
\addplot [red, mark=x, mark size=4pt, mark options={line width=1.1pt}] coordinates {(2.5,0)};
\addplot [red, mark=x, mark size=4pt, mark options={line width=1.1pt}] coordinates {(2.5,5)};  
\end{axis}

\begin{axis}[%
width=0.460\plotWidth,
height=0.460\plotWidth,
at={(0.60\plotWidth,1.148\plotWidth)},
scale only axis,
point meta min=0,
point meta max=0.993,
xmin=0,
xmax=5,
xlabel shift={-4pt},
ylabel shift={-4pt}, 
xlabel style={font=\color{white!15!black}},
xlabel={\footnotesize x [m]},
ymin=0,
ymax=5,
ylabel style={font=\color{white!15!black}},
ylabel={\footnotesize y [m]},
    xtick={0,1,2,3,4,5}, 
    ytick={0,1,2,3,4,5}, 
    axis on top=true,
title={\footnotesize \textbf{\ed RMSE}},
title style={at={(0.5,0.95)}, anchor=south}
]
\addplot [forget plot] graphics [xmin=-0.03125, xmax=5.03125, ymin=-0.03125, ymax=5.03125] {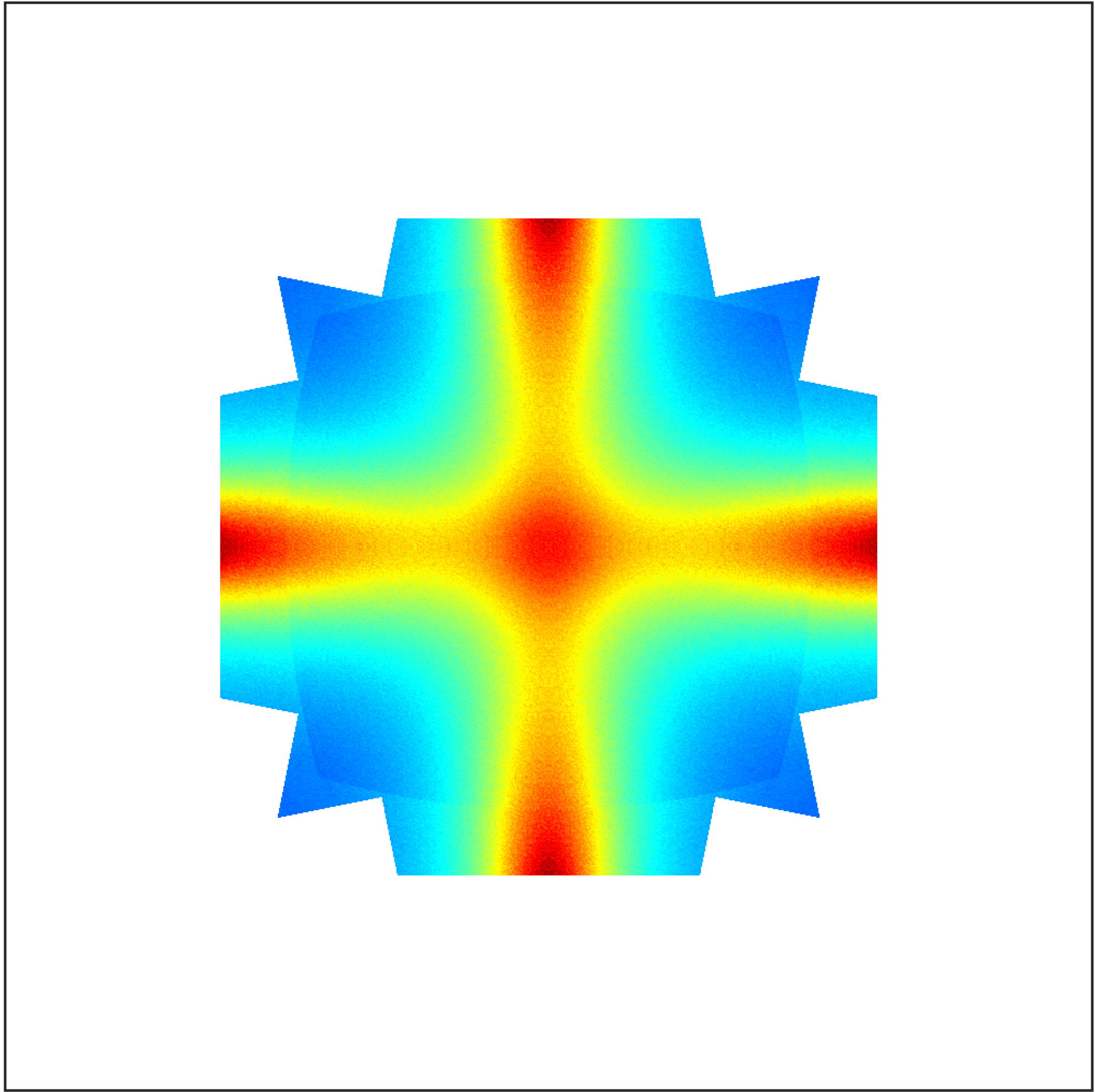};
\addplot [red, mark=x, mark size=4pt, mark options={line width=1.1pt}] coordinates {(0,2.5)};
\addplot [red, mark=x, mark size=4pt, mark options={line width=1.1pt}] coordinates {(5,2.5)};
\addplot [red, mark=x, mark size=4pt, mark options={line width=1.1pt}] coordinates {(2.5,0)};
\addplot [red, mark=x, mark size=4pt, mark options={line width=1.1pt}] coordinates {(2.5,5)};  
\end{axis}

\begin{axis}[%
width=0.460\plotWidth,
height=0.460\plotWidth,
at={(0\plotWidth,0.574\plotWidth)},
scale only axis,
point meta min=0,
point meta max=0.993,
xmin=0,
xmax=5,
xlabel shift={-4pt},
ylabel shift={-4pt}, 
xlabel style={font=\color{white!15!black}},
xlabel={\footnotesize x [m]},
ymin=0,
ymax=5,
ylabel style={font=\color{white!15!black}},
ylabel={\footnotesize y [m]},
    xtick={0,1,2,3,4,5}, 
    ytick={0,1,2,3,4,5}, 
    axis on top=true
]
\addplot [forget plot] graphics [xmin=-0.03125, xmax=5.03125, ymin=-0.03125, ymax=5.03125] {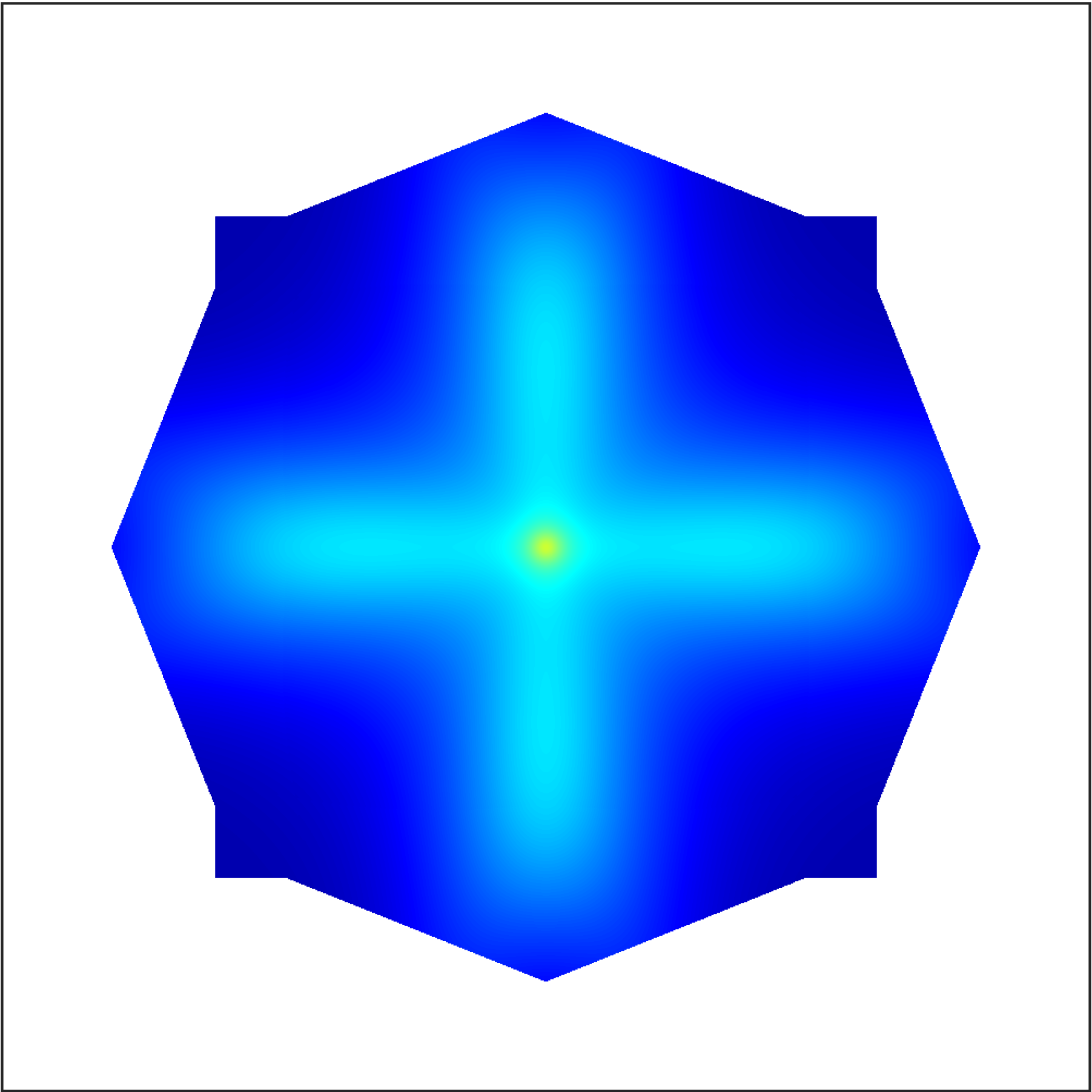};
\addplot [red, mark=x, mark size=4pt, mark options={line width=1.1pt}] coordinates {(0,3)};
\addplot [red, mark=x, mark size=4pt, mark options={line width=1.1pt}] coordinates {(0,2)};
\addplot [red, mark=x, mark size=4pt, mark options={line width=1.1pt}] coordinates {(5,3)};
\addplot [red, mark=x, mark size=4pt, mark options={line width=1.1pt}] coordinates {(5,2)};
\addplot [red, mark=x, mark size=4pt, mark options={line width=1.1pt}] coordinates {(3,0)};
\addplot [red, mark=x, mark size=4pt, mark options={line width=1.1pt}] coordinates {(2,0)};
\addplot [red, mark=x, mark size=4pt, mark options={line width=1.1pt}] coordinates {(3,5)};
\addplot [red, mark=x, mark size=4pt, mark options={line width=1.1pt}] coordinates {(2,5)};   
\end{axis}

\begin{axis}[%
width=0.460\plotWidth,
height=0.460\plotWidth,
at={(0.60\plotWidth,0.574\plotWidth)},
scale only axis,
point meta min=0,
point meta max=0.993,
xmin=0,
xmax=5,
xlabel shift={-4pt},
ylabel shift={-4pt}, 
xlabel style={font=\color{white!15!black}},
xlabel={\footnotesize x [m]},
ymin=0,
ymax=5,
ylabel style={font=\color{white!15!black}},
ylabel={\footnotesize y [m]},
    xtick={0,1,2,3,4,5}, 
    ytick={0,1,2,3,4,5}, 
    axis on top=true
]
\addplot [forget plot] graphics [xmin=-0.03125, xmax=5.03125, ymin=-0.03125, ymax=5.03125] {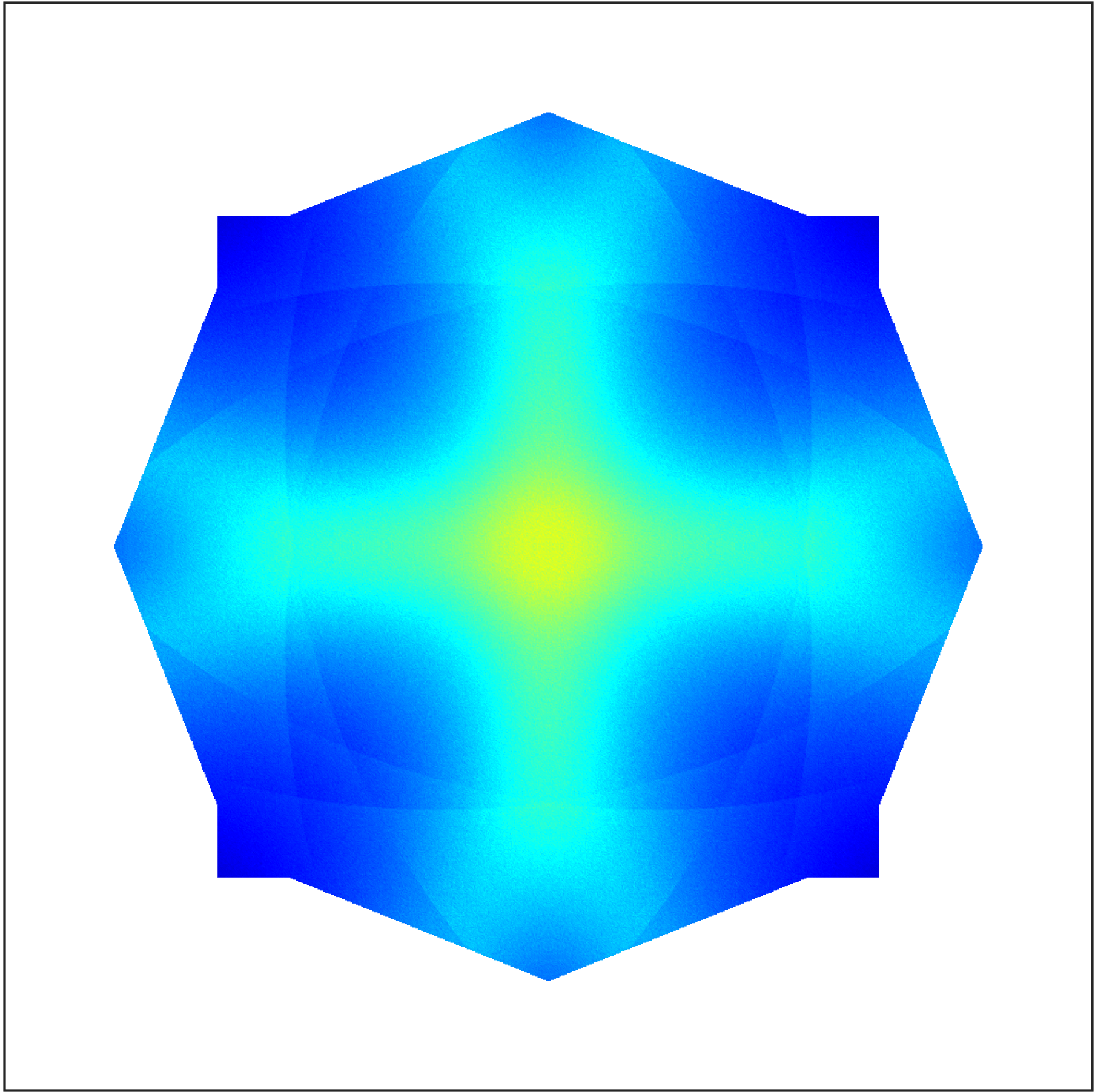};
\addplot [red, mark=x, mark size=4pt, mark options={line width=1.1pt}] coordinates {(0,3)};
\addplot [red, mark=x, mark size=4pt, mark options={line width=1.1pt}] coordinates {(0,2)};
\addplot [red, mark=x, mark size=4pt, mark options={line width=1.1pt}] coordinates {(5,3)};
\addplot [red, mark=x, mark size=4pt, mark options={line width=1.1pt}] coordinates {(5,2)};
\addplot [red, mark=x, mark size=4pt, mark options={line width=1.1pt}] coordinates {(3,0)};
\addplot [red, mark=x, mark size=4pt, mark options={line width=1.1pt}] coordinates {(2,0)};
\addplot [red, mark=x, mark size=4pt, mark options={line width=1.1pt}] coordinates {(3,5)};
\addplot [red, mark=x, mark size=4pt, mark options={line width=1.1pt}] coordinates {(2,5)};   
    
\end{axis}

\begin{axis}[%
width=0.460\plotWidth,
height=0.460\plotWidth,
at={(0\plotWidth,0\plotWidth)},
scale only axis,
point meta min=0,
point meta max=0.993,
xmin=0,
xmax=5,
xlabel shift={-4pt},
ylabel shift={-4pt}, 
xlabel style={font=\color{white!15!black}},
xlabel={\footnotesize x [m]},
ymin=0,
ymax=5,
ylabel style={font=\color{white!15!black}},
ylabel={\footnotesize y [m]},
xtick={0,1,2,3,4,5}, 
ytick={0,1,2,3,4,5}, 
    axis on top=true
]
\addplot [forget plot] graphics [xmin=-0.03125, xmax=5.03125, ymin=-0.03125, ymax=5.03125] {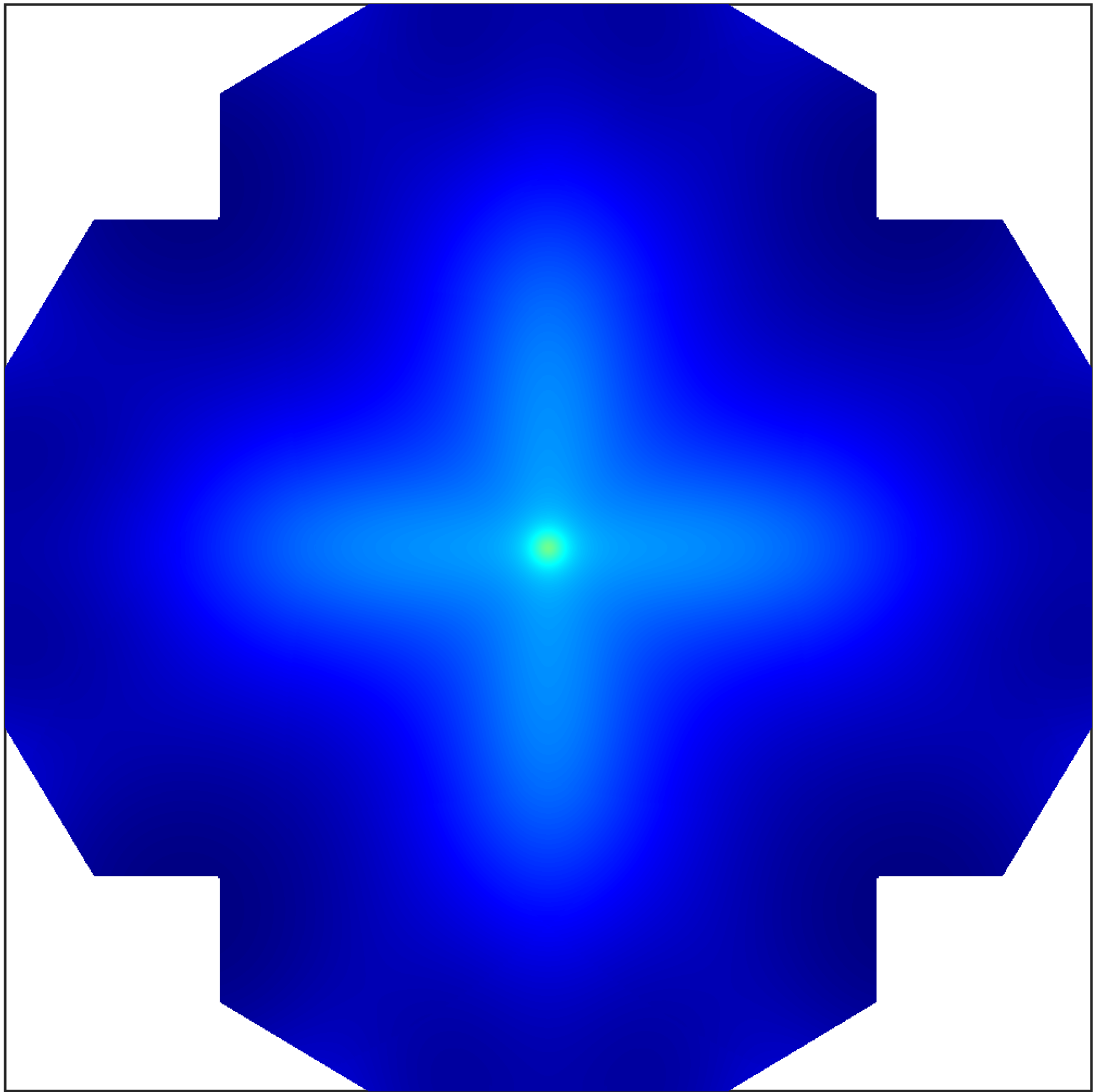};
\addplot [red, mark=x, mark size=4pt, mark options={line width=1.1pt}] coordinates {(0,1.5)};
\addplot [red, mark=x, mark size=4pt, mark options={line width=1.1pt}] coordinates {(0,2.5)};
\addplot [red, mark=x, mark size=4pt, mark options={line width=1.1pt}] coordinates {(0,3.5)};
\addplot [red, mark=x, mark size=4pt, mark options={line width=1.1pt}] coordinates {(5,1.5)};
\addplot [red, mark=x, mark size=4pt, mark options={line width=1.1pt}] coordinates {(5,2.5)};
\addplot [red, mark=x, mark size=4pt, mark options={line width=1.1pt}] coordinates {(5,3.5)};
\addplot [red, mark=x, mark size=4pt, mark options={line width=1.1pt}] coordinates {(1.5,0)};
\addplot [red, mark=x, mark size=4pt, mark options={line width=1.1pt}] coordinates {(2.5,0)};
\addplot [red, mark=x, mark size=4pt, mark options={line width=1.1pt}] coordinates {(3.5,0)};
\addplot [red, mark=x, mark size=4pt, mark options={line width=1.1pt}] coordinates {(1.5,5)};
\addplot [red, mark=x, mark size=4pt, mark options={line width=1.1pt}] coordinates {(2.5,5)};
\addplot [red, mark=x, mark size=4pt, mark options={line width=1.1pt}] coordinates {(3.5,5)};
\end{axis}

\begin{axis}[%
width=0.460\plotWidth,
height=0.460\plotWidth,
at={(0.60\plotWidth,0\plotWidth)},
scale only axis,
xmin=0,
xmax=5,
xlabel shift={-4pt},
ylabel shift={-4pt}, 
xlabel style={font=\color{white!15!black}},
xlabel={\footnotesize x [m]},
ymin=0,
ymax=5,
ylabel style={font=\color{white!15!black}},
ylabel={\footnotesize y [m]},
xtick={0,1,2,3,4,5},
ytick={0,1,2,3,4,5},
axis on top=true,
colormap/jet,
point meta min=0.9388586017867116, 
point meta max= 4.6000,  
colorbar horizontal,
colorbar style={anchor=south west, at={(-1.30,3.65)}, height=7, width= 0.8\linewidth, xtick={0, 1, 2, 3, 4}, xticklabel pos=upper, xticklabels={0, 1, 2, 3, 4},title={\footnotesize PEB/RMSE [mm]} }
]
\addplot [forget plot] graphics [xmin=-0.03125, xmax=5.03125, ymin=-0.03125, ymax=5.03125] {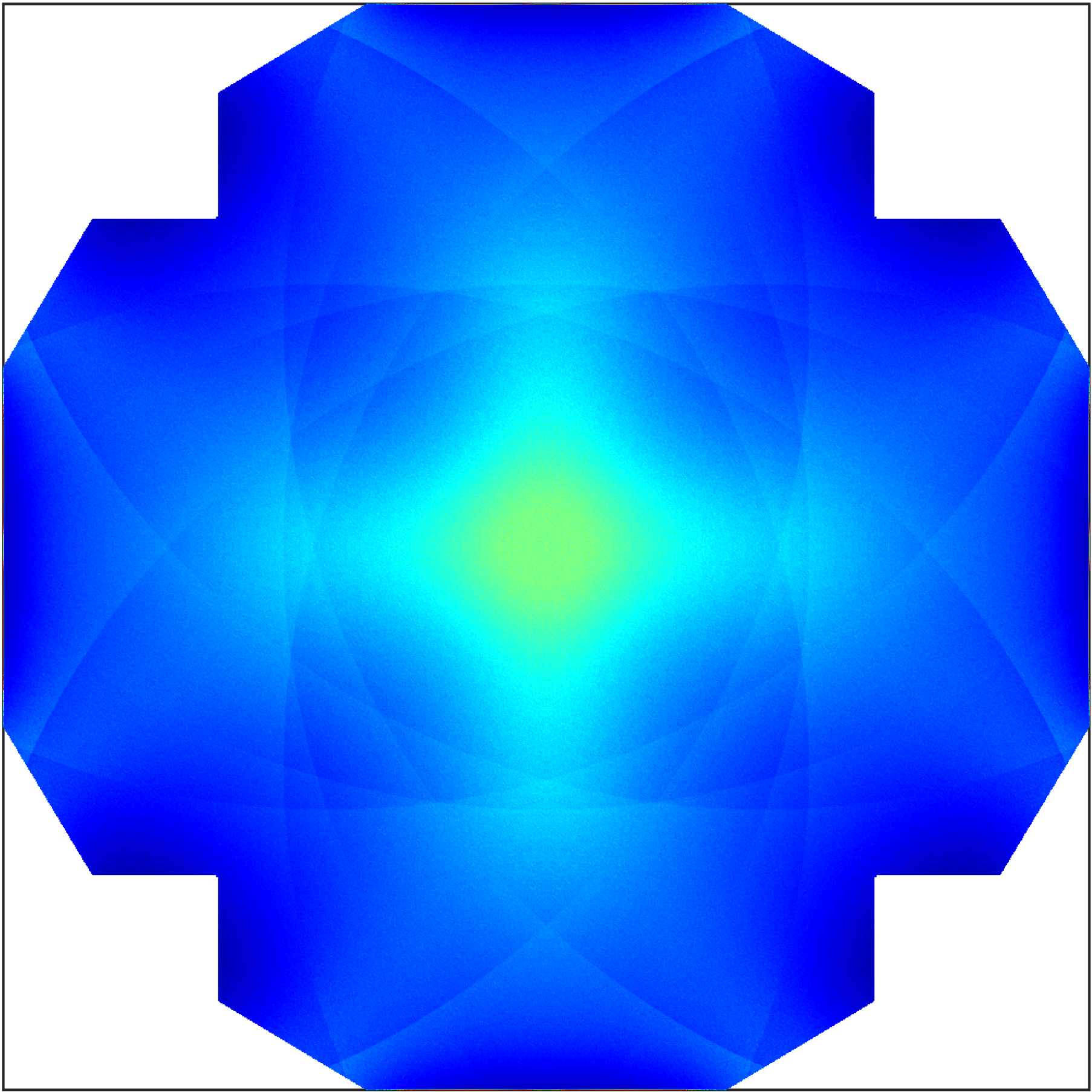};
\addplot [red, mark=x, mark size=4pt, mark options={line width=1.1pt}] coordinates {(0,1.5)};
\addplot [red, mark=x, mark size=4pt, mark options={line width=1.1pt}] coordinates {(0,2.5)};
\addplot [red, mark=x, mark size=4pt, mark options={line width=1.1pt}] coordinates {(0,3.5)};
\addplot [red, mark=x, mark size=4pt, mark options={line width=1.1pt}] coordinates {(5,1.5)};
\addplot [red, mark=x, mark size=4pt, mark options={line width=1.1pt}] coordinates {(5,2.5)};
\addplot [red, mark=x, mark size=4pt, mark options={line width=1.1pt}] coordinates {(5,3.5)};
\addplot [red, mark=x, mark size=4pt, mark options={line width=1.1pt}] coordinates {(1.5,0)};
\addplot [red, mark=x, mark size=4pt, mark options={line width=1.1pt}] coordinates {(2.5,0)};
\addplot [red, mark=x, mark size=4pt, mark options={line width=1.1pt}] coordinates {(3.5,0)};
\addplot [red, mark=x, mark size=4pt, mark options={line width=1.1pt}] coordinates {(1.5,5)};
\addplot [red, mark=x, mark size=4pt, mark options={line width=1.1pt}] coordinates {(2.5,5)};
\addplot [red, mark=x, mark size=4pt, mark options={line width=1.1pt}] coordinates {(3.5,5)};
\end{axis}

\end{tikzpicture}%


    \caption{Localization performance across the entire $xy$-plane of the room as a function of the number of available \glspl{OIRS} $N$. Red crosses denote the $(x,y)$ coordinates of the \glspl{OIRS} positioned along the walls of the room.}
    \label{fig:fig8_tikz}
\end{figure}

To provide a more comprehensive analysis, we now evaluate the performance in localizing the \gls{PD} across the entire $xy$-plane, considering an increasing number of \glspl{OIRS}, $N \in \{4, 8, 12\}$, evenly distributed along the walls. The $5 \times 5\ \si{m^2}$ area is discretized with a resolution of $1\ \si{mm}$, and the RMSE at each  position is computed over $10^3$ independent trials. Fig.~\ref{fig:fig8_tikz} compares the \gls{PEB} (on the left) and the \gls{RMSE} (on the right) of the proposed localization algorithm, obtained for a fixed $p_{[\text{lm}]} = 1000 \;\si{lm}$ and considering only two iterations of the adaptive beam steering strategy. The blank regions indicate areas where localization is not feasible due to an insufficient number of visible \glspl{OIRS}. As expected, increasing the number of \glspl{OIRS} enhances both the spatial coverage within the room and the localization accuracy, as reflected by lower values of \gls{PEB} and \gls{RMSE}. Notably, the spatial pattern of the \gls{RMSE} is similar to that of the \gls{PEB}, both exhibiting a characteristic cross-shaped region of slightly higher RMSE values. This behavior can be attributed to two main factors: first, the increased distance from the majority of the \glspl{OIRS} leads to weaker signal reception (e.g., \gls{PD} at the center of the room) and, thus, less accurate distance estimation; second, configurations where the \gls{PD} and one (or multiple) \glspl{OIRS} are nearly collinear result in poor geometric diversity. Overall, the proposed localization algorithm achieves very good performance, with RMSE values close to the PEB and a spatial coverage increasing from approximately 30\% for $N = 4$ to around 46\% for $N = 8$, and reaching nearly 80\% when $N = 12$ \glspl{OIRS} are available.

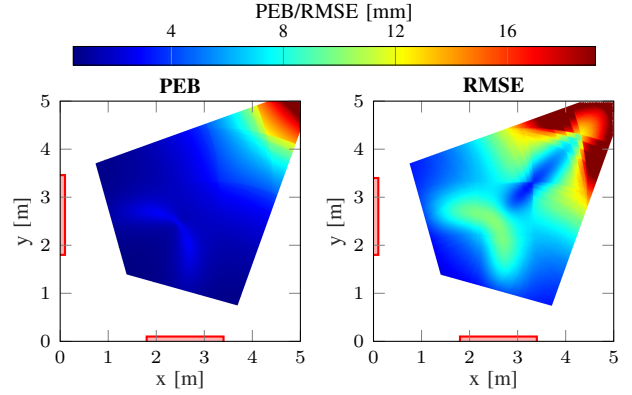
\begin{figure}
    \setlength{\plotWidth}{0.78\linewidth}
    \setlength{\plotHeight}{0.38\linewidth}
    \makeatletter
\newcommand\notsotiny{\@setfontsize\notsotiny\@vipt\@viipt}
\makeatother

\pgfplotsset{every axis/.append style={
  label style={font=\footnotesize},
  legend style={font=\notsotiny},
  tick label style={font=\scriptsize},
}}
\begin{tikzpicture}

\begin{axis}[%
width=0.460\plotWidth,
height=0.460\plotWidth,
at={(0\plotWidth,0\plotWidth)},
scale only axis,
point meta min=0,
point meta max=0.993,
xmin=0,
xmax=5,
xlabel shift={-4pt},
ylabel shift={-4pt}, 
xlabel style={font=\color{white!15!black}},
xlabel={\footnotesize x [m]},
ymin=0,
ymax=5,
ylabel style={font=\color{white!15!black}},
ylabel={\footnotesize y [m]},
xtick={0,1,2,3,4,5}, 
ytick={0,1,2,3,4,5}, 
axis on top=true,
title={\footnotesize \textbf{PEB}},
title style={at={(0.5,0.93)}, anchor=south}
]
\addplot [forget plot] graphics [xmin=-0.03125, xmax=5.03125, ymin=-0.03125, ymax=5.03125] {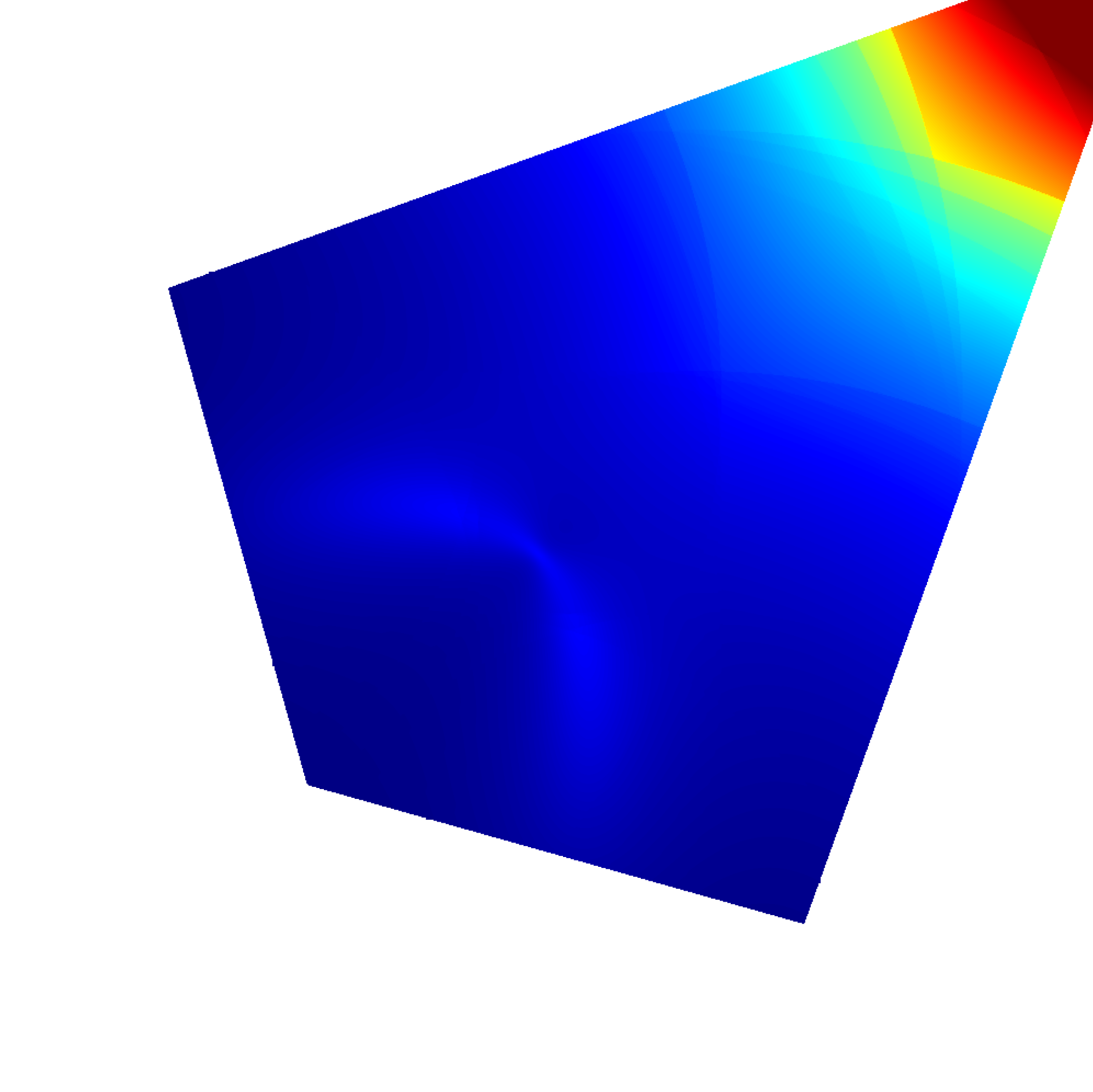};

\draw[fill=red, fill opacity=0.3, draw=red, thick] 
        (axis cs:-0.1, 1.8) rectangle (axis cs:0.1, 3.46);
        
        \draw[fill=red, fill opacity=0.3, draw=red, thick] 
        (axis cs:1.8, -0.1) rectangle (axis cs:3.4, 0.1);
\end{axis}

\begin{axis}[%
width=0.460\plotWidth,
height=0.460\plotWidth,
at={(0.60\plotWidth,0\plotWidth)},
scale only axis,
xmin=0,
xmax=5,
xlabel shift={-4pt},
ylabel shift={-4pt}, 
xlabel style={font=\color{white!15!black}},
xlabel={\footnotesize x [m]},
ymin=0,
ymax=5,
ylabel style={font=\color{white!15!black}},
ylabel={\footnotesize y [m]},
xtick={0,1,2,3,4,5},
ytick={0,1,2,3,4,5},
title={\footnotesize \textbf{RMSE}},
title style={at={(0.5,0.93)}, anchor=south},
axis on top=true,
colormap/jet,
point meta min=0.493623431293656,
point meta max=19.1420508677134,
colorbar horizontal,
colorbar style={anchor=south west, at={(-1.25,1.15)}, height=7, width= \plotWidth, xtick={0, 4, 8, 12, 16}, xticklabel pos=upper, xticklabels={0, 4, 8, 12, 16},title={\footnotesize PEB/RMSE [mm]} }
]
\addplot [forget plot] graphics [xmin=-0.03125, xmax=5.03125, ymin=-0.03125, ymax=5.03125] {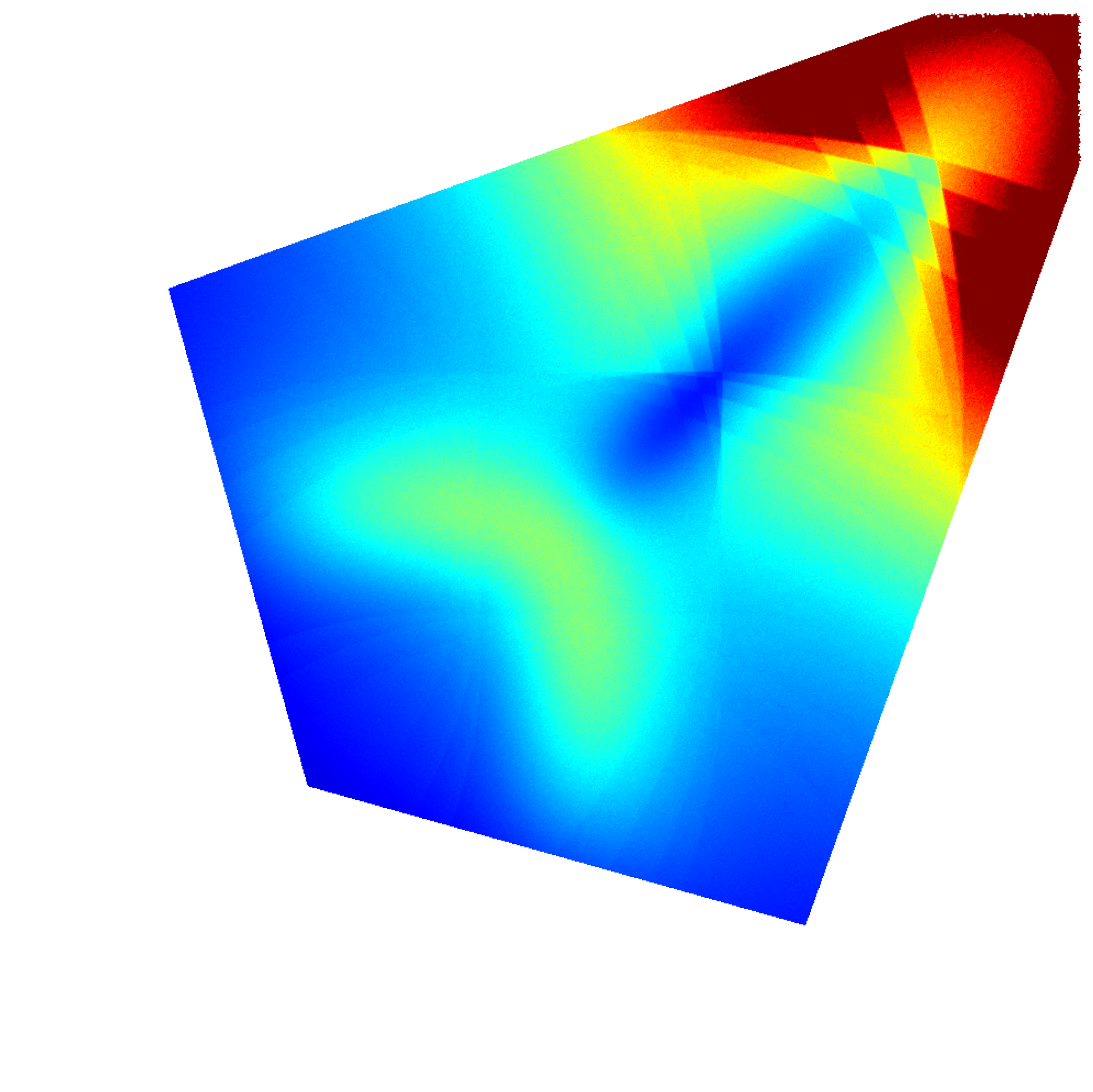};

\draw[fill=red, fill opacity=0.3, draw=red, thick] 
        (axis cs:-0.1, 1.8) rectangle (axis cs:0.1, 3.4);

        \draw[fill=red, fill opacity=0.3, draw=red, thick] 
        (axis cs:1.8, -0.1) rectangle (axis cs:3.4, 0.1);
\end{axis}

\end{tikzpicture}%


    \caption{\ed Localization performance across the entire $xy$-plane of the room for the multi-element OIRS configuration. Red boxes denote the locations of the two $4 \times 4$ element OIRS.}
    \label{fig:fig9_tikz}
\end{figure}

{\ed To conclude the analysis, we investigate the performance of the proposed algorithm when the localization process is supported by two multi-element OIRSs available in the considered area. Specifically, we consider a $4 \times 4$ element configuration} {\ed for each OIRS,  resulting in two multi-element surfaces with physical dimensions of $1.6  \times 1.6$ m. In Fig. \ref{fig:fig9_tikz}, we report the performance of the proposed iterative localization algorithm in terms of RMSE, in comparison with the theoretical PEB. 
It can be noted that the proposed algorithm achieves an accuracy very close to the PEB, differing by only $\SI{15}{mm}$ in the most unfavorable case.  As already observed in Fig. \ref{fig:fig8_tikz}, slightly higher RMSEs appear in the regions where the LED and the OIRSs intersect geometrically, a condition that leads to reduced diversity due to the PD becoming nearly collinear with one OIRS and the LED.
Moreover, although the overall reflecting aperture provided by the two multi-element OIRSs is smaller than in the single-element OIRSs scenario in Fig. \ref{fig:fig8_tikz} (i.e., $\SI{3.2}{m^2}$ versus $\SI{4.0}{m^2}$), the percentage of the area where positioning is feasible increases to $\SI{44.3}{\%}$, underscoring the importance of OIRSs placement over their  dimensions.}

\subsection{\ed Complexity and Average Runtime Analysis}\label{results:runtimes}

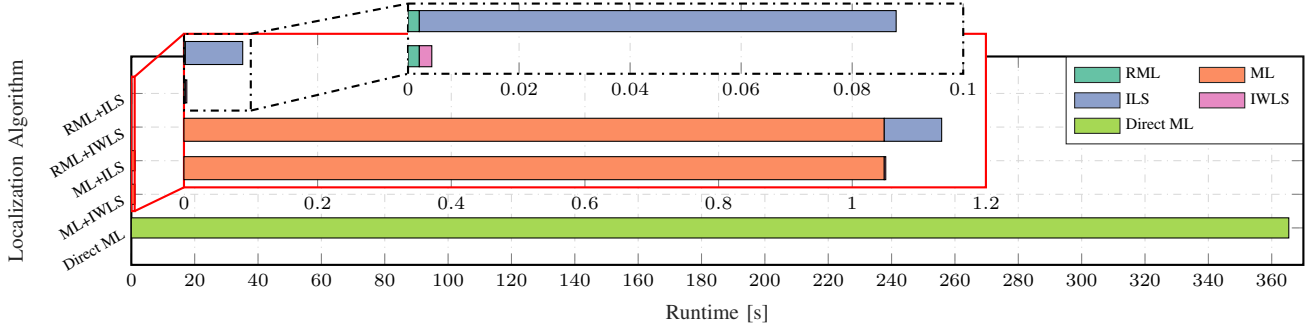
\begin{figure*}
    \centering
    \setlength{\plotWidth}{0.9\linewidth}
    \setlength{\plotHeight}{0.16\linewidth}
%

 \makeatletter
\newcommand\notsotiny{\@setfontsize\notsotiny\@vipt\@viipt}
\makeatother

\pgfplotsset{every axis/.append style={
  label style={font=\footnotesize},
  legend style={font=\notsotiny},
  tick label style={font=\scriptsize},
}}

\definecolor{mycolor1}{rgb}{0.4, 0.760, 0.647}   
\definecolor{mycolor2}{rgb}{0.988, 0.552, 0.384} 
\definecolor{mycolor3}{rgb}{0.552, 0.627, 0.796} 
\definecolor{mycolor4}{rgb}{0.905, 0.541, 0.765} 
\definecolor{mycolor5}{rgb}{0.651, 0.847, 0.329} 
\begin{tikzpicture}

\begin{axis}[%
name=mainaxis,
width=0.95\plotWidth,
height=0.95\plotHeight,
at={(0\plotWidth,0\plotWidth)},
scale only axis,
axis line style= thick,
bar width=0.6,
xmin=0,
xmax=370,
xlabel style={font=\color{white!15!black}},
xlabel={\footnotesize Runtime [s]},
y dir=reverse,
ymin=-0.1,
ymax=6.1,
ytick={1,2,3,4,5},
yticklabels={{RML+ILS},{RML+IWLS},{ML+ILS},{ML+IWLS},{Direct ML}},
ylabel style={font=\color{white!15!black}},
ylabel={\footnotesize Localization Algorithm},
axis background/.style={fill=white},
xmajorgrids,
ymajorgrids,
yticklabel style={
        font=\notsotiny,
        rotate=30,
        anchor=east
    },
grid style={dashdotted,white!85!black},
legend style={at={(1,1)}, legend cell align=left, anchor=north east,legend columns=2, draw=white!15!black}
]
\addplot[xbar stacked, fill=mycolor1, area legend] table[row sep=crcr] {%
0.002035700000000	1\\
0.002035700000000	2\\
0	3\\
0	4\\
0	5\\
};
\addlegendentry{RML}

\addplot[xbar stacked, fill=mycolor2, area legend] table[row sep=crcr] {%
0	1\\
0	2\\
1.047901370000001	3\\
1.047901370000001	4\\
0	5\\
};
\addlegendentry{ML}

\addplot[xbar stacked, fill=mycolor3, area legend] table[row sep=crcr] {%
0.085881905000000	1\\
0	2\\
0.085881905000000	3\\
0	4\\
0	5\\
};
\addlegendentry{ILS}

\addplot[xbar stacked, fill=mycolor4, area legend] table[row sep=crcr] {%
0	1\\
0.002245730000000	2\\
0	3\\
0.002245730000000	4\\
0	5\\
};
\addlegendentry{IWLS}

\addplot[xbar stacked, fill=mycolor5, area legend] table[row sep=crcr] {%
0	1\\
0	2\\
0	3\\
0	4\\
365.39	5\\
};
\addlegendentry{Direct ML}

 \draw[red, thick] 
        (axis cs:0, 0.5) rectangle (axis cs:1.055, 4.5);

\coordinate (redTL) at (axis cs:1.055, 0.5);
\coordinate (redBL) at (axis cs:1.055, 4.5);

\end{axis}

\begin{axis}[
    at=(mainaxis.north east),
    anchor=north east,
    xshift=-4.2cm, 
    yshift=0.3cm, 
    width=0.65\plotWidth,
    height=0.7\plotHeight,
    xmin=0, xmax=1.2,       
    ymin=0.5, ymax=4.5,     
    y dir=reverse,
    scale only axis,
    bar width=0.6,
    axis background/.style={fill=white},
    xmajorgrids,
    ymajorgrids,
    xtick={0,0.2,0.4,0.6,0.8,1.0,1.2},
    ytick=\empty,
    grid style={dashdotted,white!85!black},
    axis line style={red, thick}
  ]
\addplot[xbar stacked, fill=mycolor1, area legend] table[row sep=crcr] {%
0.002035700000000	1\\
0.002035700000000	2\\
0	3\\
0	4\\
0	5\\
};

\addplot[xbar stacked, fill=mycolor2, area legend] table[row sep=crcr] {%
0	1\\
0	2\\
1.047901370000001	3\\
1.047901370000001	4\\
0	5\\
};

\addplot[xbar stacked, fill=mycolor3, area legend] table[row sep=crcr] {%
0.085881905000000	1\\
0	2\\
0.085881905000000	3\\
0	4\\
0	5\\
};

\addplot[xbar stacked, fill=mycolor4, area legend] table[row sep=crcr] {%
0	1\\
0.002245730000000	2\\
0	3\\
0.002245730000000	4\\
0	5\\
};

\addplot[xbar stacked, fill=mycolor5, area legend] table[row sep=crcr] {%
0	1\\
0	2\\
0	3\\
0	4\\
5	5\\ 
};
 \draw[black, thick, dashdotted] 
        (axis cs:0, 0.5) rectangle (axis cs:0.1, 2.5);

        \coordinate (inset1TL) at (axis cs:0, 0.5);
\coordinate (inset1BL) at (axis cs:0, 4.5);
\coordinate (blueTL) at (axis cs:0.1, 0.5);
\coordinate (blueBL) at (axis cs:0.1, 2.5);
  \end{axis}

  \begin{axis}[
    at=(mainaxis.north east),
    anchor=north east,
    xshift=-4.5cm, 
    yshift=0.7cm, 
    width=0.45\plotWidth,
    height=0.32\plotHeight,
    xmin=0, xmax=0.1,       
    ymin=0.5, ymax=2.5,     
    y dir=reverse,
    scale only axis,
    bar width=0.6,
    axis background/.style={fill=white},
    xmajorgrids,
    ymajorgrids,
    xtick={0,0.02,0.04,0.06, 0.08,0.1},
    xticklabel style={/pgf/number format/fixed},
    ytick=\empty,
    grid style={dashdotted,white!85!black},
    axis line style={black, thick, dashdotted}
  ]
\addplot[xbar stacked, fill=mycolor1, area legend] table[row sep=crcr] {%
0.002035700000000	1\\
0.002035700000000	2\\
0	3\\
0	4\\
0	5\\
};

\addplot[xbar stacked, fill=mycolor2, area legend] table[row sep=crcr] {%
0	1\\
0	2\\
1.047901370000001	3\\
1.047901370000001	4\\
0	5\\
};

\addplot[xbar stacked, fill=mycolor3, area legend] table[row sep=crcr] {%
0.085881905000000	1\\
0	2\\
0.085881905000000	3\\
0	4\\
0	5\\
};

\addplot[xbar stacked, fill=mycolor4, area legend] table[row sep=crcr] {%
0	1\\
0.002245730000000	2\\
0	3\\
0.002245730000000	4\\
0	5\\
};

\addplot[xbar stacked, fill=mycolor5, area legend] table[row sep=crcr] {%
0	1\\
0	2\\
0	3\\
0	4\\
5	5\\ 
};

\coordinate (inset2TL) at (axis cs:0, 0.5);
\coordinate (inset2BL) at (axis cs:0, 2.5);
        
  \end{axis}

  \draw[red, thick] (redTL) -- (inset1TL);
\draw[red, thick] (redBL) -- (inset1BL);

\draw[black, thick, dashdotted] (blueTL) -- (inset2TL);
\draw[black, thick, dashdotted] (blueBL) -- (inset2BL);
\end{tikzpicture}%
    \caption{\ed Average runtime comparison between the proposed iterative localization algorithm with adaptive beam steering, using either the ML or RML estimator in the distance estimation step combined with either the ILS or IWLS in the position estimation step, and the direct ML position estimation algorithm.}
    \label{fig:runtime}
\end{figure*}

{\ed To complement the theoretical complexity analysis of Sec. \ref{subsec::Complexity}, we evaluate the average execution time of the proposed iterative localization algorithm, also in comparison with the direct ML estimator. 
The results, presented in Fig. \ref{fig:runtime}, are obtained averaging over 100 independent Monte Carlo trials. To provide a fair comparison, both distance and position grid searches --- required to implement the ML NLoS distance estimator and direct ML position estimator, respectively --- are performed over the same number of evaluation points per each dimension, i.e., $Q=5000$. For our proposed approaches, which follow an indirect (two-step) localization paradigm and  iteratively apply adaptive beam steering, we report the aggregate average runtimes over all the $M$ iterations, decomposed in terms of the time associated with the distance estimation step (performed employing the ML LoS estimator and either the ML or RML NLoS estimator), and the time required for the subsequent position estimation step (through either ILS or IWLS).  
As it can be observed, the direct ML estimator incurs an extremely high computation time owing to the need to discretize the entire area and conduct a 2D exhaustive search over the $Q^2$ candidate positions. On the other hand, the proposed  iterative algorithms execute in less than 0.5\% of the time required by the direct ML, thereby confirming their markedly lower cost. In particular, during the} {\ed distance estimation step, the proposed RML NLoS estimator yields a remarkable reduction in complexity compared to the ML NLoS estimator (which we recall requires a 1D grid search), achieving a solution in approximately 0.35\% of the time required by the ML NLoS estimator, thanks to its closed-form expression. Furthermore, in the position estimation step, the adoption of the proposed IWLS procedure, employing a weighting matrix derived from the CRLB, enables convergence in only about 3.2\% of the time required by the ILS. Overall, the position estimation results and complexity analysis confirm the effectiveness of the proposed iterative localization framework with adaptive beam steering.}

\section{Conclusion} \label{sec:conclusion}
This work presented a novel approach for visible light-based indoor localization that leverages only a single \gls{LED} and multiple distributed single-element \glspl{OIRS}. More specifically, we proposed an indirect estimation paradigm built upon computationally efficient, closed-form estimators: the optimal ML estimator for the \gls{LoS} distance, and a tailored RML estimator for the \gls{NLoS} distances, the latter circumventing the numerical optimization required by its optimal ML counterpart. Building upon these estimators, we designed a low-complexity \gls{IWLS} localization algorithm, whose weights are derived from the \gls{CRLB} to account for path-dependent uncertainty on distance estimation. Notably, the algorithm integrates an adaptive beam-steering mechanism to iteratively refine the orientation of the OIRSs and face initial pointing mismatches, so enabling the localization process without any prior knowledge on the \gls{PD} position. Fundamental Fisher information analyses were conducted to derive the CRLBs for distance and position estimation under the proposed setup. Extensive numerical evaluations demonstrated the effectiveness of the proposed distance and position estimation algorithms across various SNR regimes and OIRS misalignment conditions. Specifically, the ML \gls{LoS} estimator unveiled excellent estimation performance even with a very small sample size, attaining the DEB already for low values of the SNR. On the other hand, the RML \gls{NLoS} estimator provided performance close to the optimal ML, but at a significantly reduced cost. Moreover, both ML and RML NLoS estimators exhibited a good robustness to \gls{OIRS} orientation errors of up to \SI{20}{\degree}, guaranteeing accuracy in the order of a few centimeters. Overall, the proposed IWLS algorithm combined with the adaptive beam steering strategy required only a couple of iterations to re-orient the \gls{OIRS} network toward the \gls{PD}, achieving RMSEs close to the PEB and offering a spatial coverage of about 80\% of the area when an average of three \glspl{OIRS} are deployed on each wall. {\ed Remarkably, when compared with the state-of-the-art direct ML estimator, the proposed iterative localization algorithm achieved nearly identical estimation accuracy while requiring less than 
0.5\% of the execution time of the direct ML.}

Possible directions of future work include the joint optimization of \gls{LED} and \glspl{OIRS} placement, as well as the extension of the proposed approach to multi-\gls{PD} localization scenarios. Another promising direction involves investigating the use of metasurface-based \glspl{OIRS}, capable of of absorbing or phase-shifting selectively different wavelengths, to further enhance localization performance.

\appendices


{\ed
\section{Reflection Point Geometric Model}\label{sec:app_b}
We derive the geometric model of the reflection point $\bm{r}_n$, relating it to the \gls{LED}, \gls{PD}, and the $n$-th \gls{OIRS}.
To compute $\bm{r}_n$, we translate the reference frame so that it is centered at $\bm{w}_n$, with the $z$-axis oriented downward, the $x$-axis aligned parallel to the ground, and the $y$-axis pointing outward from the \gls{OIRS}, as shown in Fig. \ref{fig:scenario_sx_sz}.
\begin{figure}
    \centering
\includegraphics[width=0.5\linewidth]{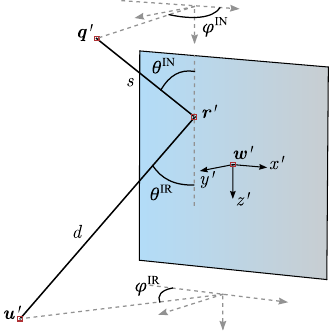}%
    \caption{\ed Geometric representation of the reflection point on the \gls{OIRS} surface.}
   \label{fig:scenario_sx_sz}
\end{figure}
The positions of the LED, the reflection point, the center of the OIRS, and the PD in the new reference frame are denoted as $\bm{q}' = \left[ q'_x \ q'_y \ q'_z \right]^\mathsf{T}$, $\bm{r}' = \left[ r'_x \ 0 \ r'_z \right]^\mathsf{T}$, $\bm{w}' = \left[ 0 \ 0 \ 0 \right]^\mathsf{T}$, and $\bm{u}' = \left[ u'_x \ u'_y \ u'_z \right]^\mathsf{T}$, respectively, where from now on we drop the dependency on $n$ for brevity.
Accordingly, we define $s = \| \bm{q}' - \bm{r}' \|$ as the distance from the LED to the reflection point, where $\theta^{\text{\tiny IN}}$ and $\varphi^{\text{\tiny IN}}$ denote the corresponding azimuth and elevation angles of incidence. Similarly, $d = \| \bm{r}' - \bm{u}' \|$ represents the distance from the reflection point to the PD, with $\theta^{\text{\tiny IR}}$ and $\varphi^{\text{\tiny IR}}$  the azimuth and elevation angles of irradiance, as  in Fig. \ref{fig:scenario_sx_sz}.

Using polar coordinates, $\bm{u}'$ can be expressed as
\begin{align} \label{eq:cart_pol_conv}
    \begin{cases}
        u'_x = r'_x - d \sin(\theta^{\text{\tiny IR}})\cos(\varphi^{\text{\tiny IR}}),\\
        u'_y = d\sin(\theta^{\text{\tiny IR}})\sin(\varphi^{\text{\tiny IR}}),\\
        u'_z = r'_z + d\cos(\theta^{\text{\tiny IR}}),\\
    \end{cases} 
\end{align}
with $d=|u'_z - r'_z|/\cos(\theta^{\text{\tiny IR}})$
and, without loss of generality, $r'_x<0$. By plugging these values back in (\ref{eq:cart_pol_conv}) and imposing Snell's law (i.e., $\theta^{\text{\tiny IR}} = \theta^{\text{\tiny IN}}$ and $\varphi^{\text{\tiny IR}} = \varphi^{\text{\tiny IN}}$), we obtain
\begin{align} \label{eq:cart_pol_conv_2}
    \begin{cases}
        u'_x = r'_x - \frac{\sqrt{(q'_x-r'_x)^2} \sqrt{(u'_z-r'_z)^2}}{\sqrt{(q'_z-r'_z)^2}},\\
        u'_y = \frac{\sqrt{{q'_y}^2} \sqrt{(u'_z-r'_z)^2}}{\sqrt{(q'_z-r'_z)^2}},\\
        u'_z = r'_z + \sqrt{(u'_z-r'_z)^2},
    \end{cases}
\end{align}

with 
\begin{align}
    & \!\cos(\theta^{\text{\tiny IN}})\!=\!\frac{\sqrt{(q'_z\!-\!r'_z)^2}}{\norm{\bm{q}' \!- \!\bm{r}'}},\ \ \ \ \ \ \ \ 
      \sin(\theta^{\text{\tiny IN}})\!=\!\frac{\sqrt{(q'_x\!-\!r'_x)^2\!+\!{q'_y}^2}}{\norm{\bm{q}' \!-\! \bm{r}'}} \nonumber \\&
      \!\cos(\varphi^{\text{\tiny IN}})\!=\!\frac{\sqrt{(q'_x\!-\!r'_x)^2}}{\sqrt{(q'_x\!-\!r'_x)^2\!+\!{q'_y}^2}},
      \sin(\varphi^{\text{\tiny IN}})\!=\!\frac{\sqrt{{q'_y}^2}}{\sqrt{(q'_x\!-\!r'_x)^2\!+\!{q'_y}^2}}. \nonumber
\end{align}
The final expression of $\bm{r}'$ follows by solving (\ref{eq:cart_pol_conv_2}), namely
\begin{equation}\label{eq::r_u_relation}
    \begin{split}
        &r'_x= \frac{q'_x u'_y+q'_y u'_x}{q'_y+u'_y},\ \ r'_z= \frac{q'_y u'_z+q'_z u'_y}{q'_y+u'_y}.
    \end{split}
\end{equation}
The transformation initially applied can be then applied in the opposite way to map $\bm{r}'$ back to the global reference system.
}

\section{Entries of FIM in \eqref{eq:dist_fim}}\label{sec:app_a}
We consider the log-likelihood of \eqref{eq:full_likelihood}, which is
\begin{equation}\label{eq:full_loglike}
\ln  L(\boldsymbol{\mu} ;\boldsymbol{d})= \mathcal{L}^\LoS(\bm{\mu}_0; d) + \sum\limits_{n=1}^{N} \mathcal{L}^\NLoS(\bm{\mu}_n; d, d_n),
\end{equation}
where $\mathcal{L}^\LoS(\bm{\mu}_0; d)$ and $\mathcal{L}^\NLoS(\bm{\mu}_n; d, d_n)$ are the positive forms of \eqref{eq:ll_los_full} and \eqref{eq:ll_nlos_full}. 
Due to the additive structure of \eqref{eq:full_loglike}, the partial derivatives needed for each element of the \gls{FIM} in \eqref{eq:dist_fim} can be computed separately for the \gls{LoS} and \gls{NLoS} terms.
For compactness, we introduce the following summation terms 
\begin{align}\label{eq:sums}
& \!\mathcal{S}_0 \!=\!\! \sum\limits_{k=1}^K \!\!\left(\!\mu_{0,k} \!-\!\! \frac{\xi}{d^{m+3}}\!\!\right)\!, \!\mathcal{S}_n \!=\!\! \sum _{k=1}^{K_n} \!\!\left(\!\mu_{n,k} \!-\!\frac{\xi}{d^{m+3}}\!-\!\frac{\omega }{d_n (d_n\!+\!s_n)^2}\!\!\right)\!, \nonumber \\ &
\!\mathcal{T}_0\!=\!\! \sum\limits_{k=1}^K \!\!{\left(\!\mu_{0,k} \!-\!\! \frac{\xi}{d^{m+3}}\!\!\right)\!\!}^2\!,
\!\mathcal{T}_n\!=\!\! \sum _{k=1}^{K_n} \!\!{\left(\!\mu_{n,k} \!-\!\frac{\xi}{d^{m+3}}\!-\!\frac{\omega }{d_n (d_n\!+\!s_n)^2}\!\!\right)\!\!}^2\!.
\end{align}

As for the $(0,0)$-th element of $\bm{J}_{\boldsymbol{d}}$, the derivatives of the \gls{LoS} and the $n$-th \gls{NLoS} parts, $\forall n = 1, \dots,N$, are given by
\begin{align}\label{eq:partial_dd}
& \!\!\frac{\partial^2   \mathcal{L}^\LoS(\bm{\mu}_0; d)}{\partial d \partial d} \!= \!\frac{\bar{\mu_0}  (m^2\!+\!7m\!+\!12)\! \left(\frac{b }{\sigma_0^2} \mathcal{T}_0 \!+\! 2 \mathcal{S}_0 \!-\! bK\right)}{2 d^2 \sigma_0^2} \nonumber \\ 
& \!\!- \frac{\bar{\mu_0}^2 (m\!+\!3)^2 \!\left(\frac{b^2 }{\sigma_0^2} \mathcal{T}_0\!+\!2 b\mathcal{S}_0 \!+\! K \!\left(\sigma_0^2\!-\!\frac{b^2}{2}\right)\!\right)}{ d^2 \sigma_0^4}, \end{align}
\vspace{-0.2cm}
\begin{align}\label{eq:partial_ddn}
& \!\!\frac{\partial^2   \mathcal{L}^\NLoS(\bm{\mu}_n;d, d_n)}{\partial d \partial d}\!= \!\frac{\bar{\mu_0}  (m^2\!+\!7m\!+\!12)\! \left(\frac{b}{\sigma_0^2}  \mathcal{T}_n\!+\!2  \mathcal{S}_n\!-\!bK_n \right)}{2 d^2 \sigma_0^2} \nonumber \\ & 
\!\!-\frac{\bar{\mu_0} ^2 (m\!+\!3)^2 \!\left(\frac{b^2}{\sigma_0^2} \mathcal{T}_n\!+\!2 b\mathcal{S}_n \!+\! K_n \left(\sigma_0^2\!-\!\frac{b^2}{2}\right)\right)}{d^2 \sigma_0^4}.
\end{align}
For the $(n,n)$-th diagonal entries of $\bm{J}_{\boldsymbol{d}}$, only the derivative of the log-likelihood term associated with the $n$-th \gls{NLoS} component is non-zero, and is expressed as
\begin{align}\label{eq:partial_d_nd_n}
&\!\! \frac{\partial^2 \mathcal{L}^\NLoS(\bm{\mu}_n;d, d_n)}{\partial d_n \partial d_n}\!= \!\frac{\bar{\chi}_n  \!\left(6 d_n^2\!+\!4 d_n s_n\!+\!s_n^2\right)\! \left(\frac{b}{\sigma_n^2} \mathcal{T}_n\!+\!2 \mathcal{S}_n \!-\!bK_n \right)}{d_n^2 \sigma_n^2 (d_n\!+\!s_n)^2} \nonumber \\ &
\!\!-\frac{\bar{\chi}_n^2 (3 d_n\!+\!s_n)^2 \!\left( \frac{b^2}{\sigma_n^2} \mathcal{T}_n\!+\!2 b\mathcal{S}_n\!+\! K_n (\sigma_n^2\!-\!\frac{b^2}{2}) \right)}{d_n^2 \sigma_n^4 (d_n\!+\!s_n)^2}. 
\end{align}

A similar consideration holds for the $(0,n)$-th element, which is equal to the symmetric $(n,0)$-th entry, resulting in
\begin{align}\label{eq:partial_dd_n}
& \!\!\frac{\partial^2 \mathcal{L}^\NLoS(\bm{\mu}_n;d, d_n)}{\partial d \partial d_n}= \frac{\partial^2   \mathcal{L}^\NLoS(\bm{\mu}_n;d, d_n)}{\partial d_n \partial d}=  \\ & \!\!-\frac{\bar{\mu}_n \bar{\chi}_n (m\!+\!3)  (3 d_n\!+\!s_n) \!\left(\! \frac{b^2}{\sigma_n^2} \mathcal{T}_n\!+\!2 b\mathcal{S}_n \! + \! K_n  ( \sigma_n^2 \!- \! \frac{b^2}{2} ) \right)}{d{\ed \cdot}d_n \sigma_n^4 (d_n\!+\!s_n)}. \nonumber
\end{align}
All other elements are zero.
We can now take the (negative) expectation of each element which, when applied to the summations (\ref{eq:sums}) appearing in (\ref{eq:partial_dd}), (\ref{eq:partial_d_nd_n}), and (\ref{eq:partial_dd_n}), yields
\begin{align}\label{eq:expect_sums}
& \mathbb{E}\left[\mathcal{S}_0\right]\!=\!0, \ \mathbb{E}\left[\mathcal{S}_n\right]\!=\!0, \
\mathbb{E}\left[\mathcal{T}_0\right]\!=\!K\!\left(\!a\!+\!b\!\left(\!\frac{\xi}{d^{m+3}}\!\right)\!\!\right)\!=\!K\sigma_0^2, \nonumber \\ &   
\mathbb{E}\left[\mathcal{T}_n\right]\!= \!K_n\!\left(\!a\!+\!b\!\left(\!\frac{\xi}{d^{m+3}}\!+\!\frac{\omega}{d_n (d_n+s_n)^2}\!\right)\!\!\right) \!= \!K_n\sigma_n^2. 
\end{align}
Therefore, we finally obtain
\begin{align}\label{eq:fim_00}
& \!\![\bm{J}_{\boldsymbol{d}}]_{0,0} =
J_d +
\sum\limits_{n=1}^{N} \left(\frac{K_n \bar{\mu}_0^2 (m\!+\!3)^2 \!\left(\sigma_n^2\!+\!b^2/2\right)}{d^2 \sigma_n^4}\right), \\
& \!\![\bm{J}_{\boldsymbol{d}}]_{0,n} \!=\![\bm{J}_{\boldsymbol{d}}]_{n,0} = 
\frac{K_n \bar{\mu}_0(m\!+\!3) \bar{\chi}_n \left(b^2\!+\!2 \sigma ^2\right) (3 d_n\!+\!s_n)}{2 d{\ed \cdot}d_n \sigma_n^4 (d_n\!+\!s_n)}, \nonumber
\end{align}
$[\bm{J}_{\boldsymbol{d}}]_{n,n} \!=\! J_{d_n} $, and $[\bm{J}_{\boldsymbol{d}}]_{n,n'}\! =\! 0 $ for $n\! \ne\! n' \!\ne \!0$.
As it can be observed, $[\bm{J}_{\boldsymbol{d}}]_{n,n}$ has the same structure as the \gls{NLoS} Fisher information in \eqref{eq:fim_nlos}, while the first element of the summation in (\ref{eq:fim_00}) reflects the \gls{LoS} Fisher information in \eqref{eq:fim_los}.

 \bibliographystyle{IEEEtran}
 \bibliography{bibliography}

\begin{thebibliography}{10}
\providecommand{\url}[1]{#1}
\csname url@samestyle\endcsname
\providecommand{\newblock}{\relax}
\providecommand{\bibinfo}[2]{#2}
\providecommand{\BIBentrySTDinterwordspacing}{\spaceskip=0pt\relax}
\providecommand{\BIBentryALTinterwordstretchfactor}{4}
\providecommand{\BIBentryALTinterwordspacing}{\spaceskip=\fontdimen2\font plus
\BIBentryALTinterwordstretchfactor\fontdimen3\font minus \fontdimen4\font\relax}
\providecommand{\BIBforeignlanguage}[2]{{%
\expandafter\ifx\csname l@#1\endcsname\relax
\typeout{** WARNING: IEEEtran.bst: No hyphenation pattern has been}%
\typeout{** loaded for the language `#1'. Using the pattern for}%
\typeout{** the default language instead.}%
\else
\language=\csname l@#1\endcsname
\fi
#2}}
\providecommand{\BIBdecl}{\relax}
\BIBdecl

\bibitem{CommMagaz1}
A.~Jovicic, J.~Li, and T.~Richardson, ``Visible light communication: opportunities, challenges and the path to market,'' \emph{\ed{IEEE Commun. Mag.}}, vol.~51, no.~12, pp. 26--32, 2013.

\bibitem{80211bb}
IEEE, ``{IEEE Approved Draft Standard for Information Technology--Telecommunications and Information Exchange Between Systems Local and Metropolitan Area Networks--Specific Requirements - Part 11: Wireless LAN Medium Access Control (MAC) and Physical Layer (PHY) Specifications Amendment 7: Light Communications},'' \emph{IEEE P802.11bb/D7.0, March 2023}, pp. 1--29, 2023.

\bibitem{Barbarossa6G}
E.~{Calvanese Strinati} and S.~Barbarossa, ``{6G networks: Beyond Shannon towards semantic and goal-oriented communications},'' \emph{\ed{Comput. Networks}}, vol. 190, p. 107930, 2021.

\bibitem{petrosino2023light}
A.~Petrosino, D.~Striccoli, O.~Romanov, G.~Boggia, and L.~A. Grieco, ``{Light Fidelity for Internet of Things: A survey},'' \emph{\ed{Opt. Switching Networking}}, p. 100732, 2023.

\bibitem{sun2022joint}
S.~Sun, F.~Yang, J.~Song, and Z.~Han, ``{Joint Resource Management for Intelligent Reflecting Surface–Aided Visible Light Communications},'' \emph{\ed{IEEE Trans. Wireless Commun.}}, vol.~21, no.~8, pp. 6508--6522, 2022.

\bibitem{zhuang2018survey}
Y.~Zhuang, L.~Hua, L.~Qi, J.~Yang, P.~Cao, Y.~Cao, Y.~Wu, J.~Thompson, and H.~Haas, ``{A survey of positioning systems using visible LED lights},'' \emph{\ed{IEEE Commun. Surv. Tutorials}}, vol.~20, no.~3, pp. 1963--1988, 2018.

\bibitem{Bozanis2023Indoor}
D.~Bozanis, N.~G. Evgenidis, V.~K. Papanikolaou, P.~S. Bouzinis, S.~A. Tegos, A.~A. Dowhuszko, P.~D. Diamantoulakis, and G.~K. Karagiannidis, ``{Indoor 3D Visible Light Positioning Analysis with Channel Estimation Errors},'' in \emph{2023 30th International Conference on Systems, Signals and Image Processing (IWSSIP)}, 2023, pp. 1--4.

\bibitem{Ma2023Centimeter}
S.~Ma, B.~Li, G.~Zhang, H.~Li, C.~Qiu, C.~Yu, S.~Li, and C.~Shen, ``{Centimeter-Level 3D Mobile Online Visible Light Positioning System With Single LED-Lamp},'' \emph{\ed{IEEE Internet Things J.}}, vol.~11, no.~1, pp. 418--429, 2024.

\bibitem{Liu2022Machine}
R.~Liu, Z.~Liang, K.~Yang, and W.~Li, ``{Machine Learning Based Visible Light Indoor Positioning With Single-LED and Single Rotatable Photo Detector},'' \emph{\ed{IEEE Photonics J.}}, vol.~14, no.~3, pp. 1--11, 2022.

\bibitem{RSS2022}
X.~Sun, Y.~Zhuang, J.~Huai, L.~Hua, D.~Chen, Y.~Li, Y.~Cao, and R.~Chen, ``{RSS-Based Visible Light Positioning Using Nonlinear Optimization},'' \emph{\ed{IEEE Internet Things J.}}, vol.~9, no.~15, pp. 14\,137--14\,150, 2022.

\bibitem{RSSFing2021}
I.~M. Abou-Shehada, A.~F. AlMuallim, A.~K. AlFaqeh, A.~H. Muqaibel, K.-H. Park, and M.-S. Alouini, ``{Accurate Indoor Visible Light Positioning Using a Modified Pathloss Model With Sparse Fingerprints},'' \emph{\ed{J. Lightwave Technol.}}, vol.~39, no.~20, pp. 6487--6497, 2021.

\bibitem{RobustRSS_2020}
B.~Zhou, A.~Liu, and V.~Lau, ``{Visible Light-Based User Position, Orientation and Channel Estimation Using Self-Adaptive Location-Domain Grid Sampling},'' \emph{\ed{IEEE Trans. Wireless Commun.}}, vol.~19, no.~7, pp. 5025--5039, 2020.

\bibitem{TOA1}
T.~Akiyama, M.~Sugimoto, and H.~Hashizume, ``{Time-of-arrival-based smartphone localization using visible light communication},'' in \emph{International Conference on Indoor Positioning and Indoor Navigation (IPIN)}, 2017, pp. 1--7.

\bibitem{TDOA1}
S.~M. Sheikh, H.~M. Asif, K.~Raahemifar, and F.~Al-Turjman, ``{Time Difference of Arrival Based Indoor Positioning System Using Visible Light Communication},'' \emph{IEEE Access}, vol.~9, pp. 52\,113--52\,124, 2021.

\bibitem{TDOA2}
X.~Cao, Y.~Zhuang, G.~Chen, X.~Wang, X.~Yang, and B.~Zhou, ``{A Visible Light Positioning System Based on a Particle Filter and Deep Learning},'' \emph{\ed{IEEE Trans. Aerosp. Electron. Syst.}}, vol.~60, no.~3, pp. 2735--2748, 2024.

\bibitem{AOA1}
B.~Soner and S.~Coleri, ``{Visible Light Communication Based Vehicle Localization for Collision Avoidance and Platooning},'' \emph{\ed{IEEE Trans. Veh. Technol.}}, vol.~70, no.~3, pp. 2167--2180, 2021.

\bibitem{AOA2}
Z.~Li, G.~Qiu, L.~Zhao, and M.~Jiang, ``{Dual-Mode LED Aided Visible Light Positioning System Under Multi-Path Propagation: Design and Demonstration},'' \emph{\ed{IEEE Trans. Wireless Commun.}}, vol.~20, no.~9, pp. 5986--6003, 2021.

\bibitem{AOA3}
K.~Zhang, Z.~Zhang, and B.~Zhu, ``{Beacon LED Coordinates Estimator With Selected AOA Estimators for Visible Light Positioning Systems},'' \emph{\ed{IEEE Trans. Wireless Commun.}}, vol.~23, no.~3, pp. 1713--1727, 2024.

\bibitem{aboagye2023ris}
S.~Aboagye, A.~R. Ndjiongue, T.~M.~N. Ngatched, O.~A. Dobre, and H.~V. Poor, ``{RIS-Assisted Visible Light Communication Systems: A Tutorial},'' \emph{\ed{IEEE Commun. Surv. Tutorials}}, vol.~25, no.~1, pp. 251--288, 2023 \ed.

\bibitem{RIS_mio}
A.~Fascista, M.~F. Keskin, A.~Coluccia, H.~Wymeersch, and G.~Seco-Granados, ``{RIS-Aided Joint Localization and Synchronization With a Single-Antenna Receiver: Beamforming Design and Low-Complexity Estimation},'' \emph{IEEE J. Sel. Top. Signal Process.}, vol.~16, no.~5, pp. 1141--1156, 2022.

\bibitem{Rev3_1}
H.~Sun, L.~Zhu, W.~Mei, and R.~Zhang, ``{Power-Measurement-Based Channel Autocorrelation Estimation for IRS-Assisted Wideband Communications},'' \emph{IEEE Trans. Wireless Commun.}, vol.~24, no.~6, pp. 4647--4662, 2025.

\bibitem{Rev3_2}
------, ``{Power Measurement-Based Channel Estimation for IRS-Enhanced Wireless Coverage},'' \emph{IEEE Trans. Wireless Commun.}, vol.~23, no.~12, pp. 19\,183--19\,198, 2024 \color{black}.

\bibitem{10198213}
S.~Sun, N.~An, F.~Yang, J.~Song, and Z.~Han, ``{Capacity Characterization Analysis of Optical Intelligent Reflecting Surface Assisted MISO VLC},'' \emph{\ed{IEEE Internet Things J.}}, vol.~11, no.~3, pp. 4801--4814, 2024.

\bibitem{10190313}
S.~Sun, W.~Mei, F.~Yang, N.~An, J.~Song, and R.~Zhang, ``{Optical Intelligent Reflecting Surface Assisted MIMO VLC: Channel Modeling and Capacity Characterization},'' \emph{\ed{IEEE Trans. Wireless Commun.}}, vol.~23, no.~3, pp. 2125--2139, 2024.

\bibitem{9543660}
S.~Aboagye, T.~M.~N. Ngatched, O.~A. Dobre, and A.~R. Ndjiongue, ``{Intelligent Reflecting Surface-Aided Indoor Visible Light Communication Systems},'' \emph{\ed{IEEE Commun. Lett.}}, vol.~25, no.~12, pp. 3913--3917, 2021.

\bibitem{Guzman2025}
B.~G. Guzman, M.~M. Cespedes, V.~P. Gil~Jimenez, A.~G. Armada, and M.~Brandt-Pearce, ``{Resource allocation exploiting reflective surfaces to minimize the outage probability in VLC},'' \emph{\ed{IEEE Trans. Wireless Commun.}}, pp. 1--1, 2025.

\bibitem{10526227}
N.~An, F.~Yang, L.~Cheng, J.~Song, and Z.~Han, ``{IRS-Assisted Aggregated VLC-RF System: Resource Allocation for Energy Efficiency Maximization},'' \emph{\ed{IEEE Trans. Wireless Commun.}}, vol.~23, no.~10, pp. 12\,578--12\,593, 2024.

\bibitem{Abdelhady2021Visible}
A.~M. Abdelhady, A.~K.~S. Salem, O.~Amin, B.~Shihada, and M.-S. Alouini, ``{Visible Light Communications via Intelligent Reflecting Surfaces: Metasurfaces vs Mirror Arrays},'' \emph{\ed{IEEE Open J. Commun. Soc.}}, vol.~2, pp. 1--20, 2021.

\bibitem{9681888}
A.~M. Abdelhady, O.~Amin, A.~K.~S. Salem, M.-S. Alouini, and B.~Shihada, ``{Channel Characterization of IRS-Based Visible Light Communication Systems},'' \emph{\ed{IEEE Trans. Commun.}}, vol.~70, no.~3, pp. 1913--1926, 2022.

\bibitem{Zhang:24}
Q.~Zhang, F.~Yang, Z.~Liu, S.~Sun, H.~Zhang, and J.~Song, ``{Hardware implementation of a mirror array-based optical intelligent reflecting surface for VLC: prototype and experimental results},'' \emph{Opt. Express}, vol.~32, no.~11, pp. 19\,252--19\,264, May 2024.

\bibitem{Wu:25}
Q.~Wu, J.~Zhang, Y.~yu~Zhang, G.~Xin, and D.~Li, ``{Reconfigurable intelligent surface-aided positioning-centered multi-LED integrated visible light communication and positioning system},'' \emph{Opt. Express}, vol.~33, no.~3, pp. 5227--5252, Feb 2025 \ed.

\bibitem{11161078}
Q.~Zhang, F.~Yang, N.~An, J.~Song, and Z.~Han, ``{Optical Intelligent Reflecting Surface-Aided Visible Light Positioning Based on RSS Fingerprint},'' in \emph{ICC 2025 - IEEE International Conference on Communications}, 2025, pp. 868--873.

\bibitem{Wang23}
Y.~Wang, S.~Wu, L.~Yu, C.~Xu, Z.~Wang, and X.~Cai, ``{RIS-Assisted Indoor Visible Light Positioning Based on Sparse Bayesian Learning},'' in \emph{2023 3rd International Conference on Intelligent Communications and Computing (ICC)}, 2023, pp. 90--97.

\bibitem{11096556}
S.~Xu, Y.~Zheng, S.~Xing, and Y.~Wu, ``{Simultaneous Position and Orientation Estimation in Single Optical IRS-Assisted Visible Light Systems Using Single LED and Single PD},'' \emph{IEEE Internet Things J.}, vol.~12, no.~19, pp. 41\,181--41\,196, 2025 \color{black}.

\bibitem{Kokdogan24}
F.~Kokdogan and S.~Gezici, ``{Intelligent Reflecting Surfaces for Visible Light Positioning Based on Received Power Measurements},'' \emph{\ed{IEEE Trans. Veh. Technol.}}, vol.~73, no.~9, pp. 13\,108--13\,121, 2024.

\bibitem{IDDRISU2025109867}
I.~Iddrisu and S.~Gezici, ``{Visible light positioning with intelligent reflecting surfaces under mismatched orientations},'' \emph{\ed{Signal Process.}}, vol. 230, p. 109867, 2025 \ed.

\bibitem{TARHAN2025104799}
E.~Tarhan, F.~Kokdogan, and S.~Gezici, ``{IRS aided visible light positioning with a single LED transmitter},'' \emph{Digital Signal Process.}, vol. 156, p. 104799, 2025 \color{black}.

\bibitem{10669070}
S.~Xu, Y.~Wu, L.~Zhang, S.~Xing, and F.~Wei, ``{Simultaneous Position and Orientation Estimation for Reconfigurable Intelligent Surfaces-Assisted Visible Light Communications},'' \emph{\ed{IEEE Trans. Veh. Technol.}}, vol.~74, no.~1, pp. 861--876, 2025 \ed.

\bibitem{10974611}
Y.~Guo, F.~Wang, R.~Li, S.~Shi, X.~Li, and D.~Benevides~da Costa, ``{Optical IRS Assisted-Visible Light Positioning in Indoor Non-LOS IoV Scenarios},'' \emph{IEEE Internet Things J.}, vol.~12, no.~14, pp. 27\,686--27\,698, 2025.

\bibitem{Shi:25}
J.~Shi, Y.~Han, Y.~Li, X.~Gui, X.~Fu, and Z.~Li, ``{Single LED visible light positioning scheme based on intelligent reflecting surfaces},'' \emph{Appl. Opt.}, vol.~64, no.~15, pp. 4438--4446, May 2025 \color{black}.

\bibitem{ClosasDirect}
P.~Closas, C.~Fernández-Prades, and J.~Fernández-Rubio, ``{Direct Position Estimation approach outperforms conventional two-steps positioning},'' in \emph{European Signal Processing Conference}, 2009, pp. 1958--1962.

\bibitem{obeed2019optimizing}
M.~Obeed, A.~M. Salhab, M.-S. Alouini, and S.~A. Zummo, ``{On Optimizing VLC Networks for Downlink Multi-User Transmission: A Survey},'' \emph{\ed{IEEE Commun. Surv. Tutorials}}, vol.~21, no.~3, pp. 2947--2976, 2019.

\bibitem{ibne_mushfique2022mirrorvlc}
S.~Ibne~Mushfique, A.~Alsharoa, and M.~Yuksel, ``{MirrorVLC: Optimal Mirror Placement for Multielement VLC Networks},'' \emph{\ed{IEEE Trans. Wireless Commun.}}, vol.~21, no.~11, pp. 10\,050--10\,064, 2022.

\bibitem{komine2004fundamental}
T.~Komine and M.~Nakagawa, ``{Fundamental analysis for visible-light communication system using LED lights},'' \emph{\ed{IEEE Trans. Consum. Electron.}}, vol.~50, no.~1, pp. 100--107, 2004.

\bibitem{Zhang2014Theoretical}
X.~Zhang, J.~Duan, Y.~Fu, and A.~Shi, ``{Theoretical Accuracy Analysis of Indoor Visible Light Communication Positioning System Based on Received Signal Strength Indicator},'' \emph{\ed{J. Lightwave Technol.}}, vol.~32, no.~21, pp. 4180--4186, 2014.

\bibitem{kay1993fundamentals}
S.~M. Kay, \emph{Fundamentals of Statistical Signal Processing: Estimation Theory}.\hskip 1em plus 0.5em minus 0.4em\relax Prentice-Hall, Inc., 1993.

\bibitem{MissEstim2}
P.~Zheng, H.~Chen, T.~Ballal, H.~Wymeersch, and T.~Y. Al-Naffouri, ``{Misspecified Cramér-Rao Bound of RIS-Aided Localization Under Geometry Mismatch},'' in \emph{IEEE International Conference on Acoustics, Speech and Signal Processing (ICASSP)}, 2023, pp. 1--5.

\bibitem{MismEstim1}
C.~Ozturk, M.~F. Keskin, V.~Sciancalepore, H.~Wymeersch, and S.~Gezici, ``{RIS-Aided Localization Under Pixel Failures},'' \emph{\ed{IEEE Trans. Wireless Commun.}}, vol.~23, no.~8, pp. 8314--8329, 2024.

\end{thebibliography}

\end{document}